\documentclass[twocolumn]{aastex631}

\let\tablenum\relax
\usepackage{natbib}
\usepackage{siunitx}
\usepackage{float}
\usepackage{comment}
\usepackage{amsmath,here,afterpage,accents}
\usepackage{longtable}
\usepackage{multirow}
\usepackage{ulem}

\received{}
\revised{}
\accepted{}
\submitjournal{ApJ}

\tighten
\shorttitle{Volume Density Structure of NGC~253}
\shortauthors{Tanaka et al.}
\tighten

\newcommand{\prob}[1]{{\ifmmode{\mathrm{Pr}\left(#1\right)}\else{$\mathrm{Pr}\left(#1\right)$}\fi}}

\newcommand\nHH{\ifmmode{n_{\rm H_2}}\else{$n_{\rm H_2}$}\fi}
\newcommand\NHH{\ifmmode{N_{\rm H_2}}\else{$N_{\rm H_2}$}\fi}
\newcommand\dNHHdv{\ifmmode{{\mathrm{d}N_{\mathrm{H}_2}}/{\mathrm{d}v}}\else{${\mathrm{d}N_{\mathrm{H}_2}}/{\mathrm{d}v}$}\fi}
\newcommand\ff{\ifmmode{\Phi}\else{$\Phi$}\fi}
\newcommand\NHHbeam{\ifmmode{\left<N_{\rm H_2}\right>_\mathrm{beam}}\else{$\left<N_{\rm H_2}\right>_\mathrm{beam}$}\fi}
\newcommand\dNHHbeam{\ifmmode{\frac{\mathrm{d}\NHHbeam}{\mathrm{d}v}}\else{${\mathrm{d}\NHHbeam}/{\mathrm{d}v}$}\fi}
\newcommand\vlsr{\ifmmode{v_\mathrm{LSR}}\else{$v_\mathrm{LSR}$}\fi}
\newcommand\Tkin{\ifmmode{T_{\rm kin}}\else{$T_{\mathrm{kin}}$}\fi}
\newcommand\ncrit{\ifmmode{n_{\rm crit}}\else{$n_{\rm crit}$}\fi}
\newcommand\Eu{\ifmmode{E_{\rm u}}\else{$E_{\rm u}$}\fi}
\newcommand\pcc{\si{\per\cubic\centi\metre}}
\newcommand\psc{\si{\per\square\centi\metre}}
\newcommand\kmps{\si{\kilo\meter\per\second}}
\newcommand\PV{$P$--$V$}
\newcommand\PP{$P$--$P$}
\newcommand\PPV{$P$--$P$--$V$}
\newcommand\Msun{\ifmmode M_\sun\else$M_\sun$\fi}
\newcommand\sigmav{\ifmmode \sigma_v\else$\sigma_v$\fi}
\newcommand\Mgas{\ifmmode M_\mathrm{gas}\else$M_\mathrm{gas}$\fi}
\newcommand\pc{\ifmmode\mathrm{pc}\else{pc}\fi}
\newcommand\fDG{\ifmmode{{f_\mathrm{DG}}}\else{${f_\mathrm{DG}}$}\fi}
\newcommand\fDGx{\ifmmode{{f_\mathrm{DG}}}^*\else{${f_\mathrm{DG}}^*$}\fi}
\newcommand\fDGn{\ifmmode{{f_\mathrm{DG}}^*\left(n\right)}\else{${f_\mathrm{DG}}^*\left(n\right)$}\fi}

\newcommand\formaldehyde{\ifmmode{\mathrm{H_2CO}}\else{$\mathrm{H_2CO}$}\fi}
\newcommand\pformaldehyde{\ifmmode{\mathit{p}\text{-}\mathrm{H_2CO}}\else{{\it p}-$\mathrm{H_2CO}$}\fi}
\newcommand\ammonia{\ifmmode{\mathrm{NH_3}}\else{$\mathrm{NH_3}$}\fi}
\newcommand\HCCCN{\ifmmode{\mathrm{HC_3N}}\else{$\mathrm{HC_3N}$}\fi}
\newcommand\HCOp{\ifmmode{\mathrm{HCO^+}}\else{$\mathrm{HCO^+}$}\fi}
\newcommand\NNHp{\ifmmode{\mathrm{N_2H^+}}\else{$\mathrm{N_2H^+}$}\fi}

\newcommand\xmol[1]{\ifmmode{x_\mathrm{mol}\left(\mathrm{#1}\right)}\else{$x_\mathrm{mol}\left(\mathrm{#1}\right)$}\fi}
\newcommand\alpvt{\ifmmode \alpha_\mathrm{VT}\else$\alpha_\mathrm{VT}$\fi}

\newcommand\revcol{black}

\newcommand\myrev[1]{#1}
\newcommand\erase{\bgroup\markoverwith{\textcolor{\revcol}{\rule[.5ex]{2pt}{0.4pt}}}\ULon}

\makeatletter
  \def\widebar{\accentset{{\cc@style\underline{\mskip10mu}}}}
  \def\Widebar{\accentset{{\cc@style\underline{\mskip8mu}}}}
\makeatother
\defcitealias{Tanaka2018b}{T18}
\defcitealias{Rosenberg2014}{R14}
\defcitealias{Perez-Beaupuits2018a}{P18}
\begin{document}

\title{Volume density structure of the NGC\,253 CMZ through ALCHEMI excitation analysis}

\author[0000-0001-8153-1986]{Kunihiko Tanaka}
\affiliation{Department of Physics, Faculty of Science and Technology, Keio University, 3-14-1 Hiyoshi, Yokohama, Kanagawa 223--8522 Japan}
\email{ktanaka@phys.keio.ac.jp}
  
\author[0000-0003-1183-9293]{Jeffrey G.~Mangum}
\affiliation{National Radio Astronomy Observatory, 520 Edgemont Road, Charlottesville, VA 22903-2475, USA}
 
\author[0000-0001-8504-8844]{Serena Viti}
\affiliation{Leiden Observatory, Leiden University, P.O. Box 9513, 2300 RA Leiden, The Netherlands}
\affiliation{Department of Physics and Astronomy, University College London, Gower Street, London WC1E6BT, UK}

\author[0000-0001-9281-2919]{Sergio Mart\'in}
\affiliation{European Southern Observatory, Alonso de C\'ordova, 3107, Vitacura, Santiago 763-0355, Chile}
\affiliation{Joint ALMA Observatory, Alonso de C\'ordova, 3107, Vitacura, Santiago 763-0355, Chile}

\author[0000-0002-6824-6627]{Nanase Harada}
\affiliation{National Astronomical Observatory of Japan, 2-21-1 Osawa, Mitaka, Tokyo 181-8588, Japan}
\affiliation{Institute of Astronomy and Astrophysics, Academia Sinica, 11F of AS/NTU Astronomy-Mathematics Building, No.1, Sec. 4, Roosevelt Rd, Taipei 10617, Taiwan   }
\affiliation{Department of Astronomy, School of Science, The Graduate University for Advanced Studies (SOKENDAI), 2-21-1 Osawa, Mitaka, Tokyo, 181-1855 Japan}

\author[0000-0001-5187-2288]{Kazushi Sakamoto}
\affiliation{Institute of Astronomy and Astrophysics, Academia Sinica, 11F of AS/NTU Astronomy-Mathematics Building, No.1, Sec. 4, Roosevelt Rd, Taipei 10617, Taiwan   }

\author[0000-0002-9931-1313]{Sebastien Muller}
\affiliation{Department of Space, Earth and Environment, Chalmers University of Technology, Onsala Space Observatory, SE-43992 Onsala, Sweden}

\author[0000-0002-1413-1963]{Yuki Yoshimura}
\affiliation{Institute of Astronomy, Graduate School of Science, The University of Tokyo, 2-21-1 Osawa, Mitaka, Tokyo 181-0015, Japan}

\author[0000-0002-6939-0372]{Kouichiro Nakanishi}
\affiliation{National Astronomical Observatory of Japan, 2-21-1 Osawa, Mitaka, Tokyo 181-8588, Japan}
\affiliation{Department of Astronomy, School of Science, The Graduate University for Advanced Studies (SOKENDAI), 2-21-1 Osawa, Mitaka, Tokyo, 181-1855 Japan}
   
\author[0000-0002-7758-8717]{Rub\'en~Herrero-Illana}
\affiliation{European Southern Observatory, Alonso de C\'ordova, 3107, Vitacura, Santiago 763-0355, Chile}
\affiliation{Institute of Space Sciences (ICE, CSIC), Campus UAB, Carrer de Magrans, E-08193 Barcelona, Spain}

\author[0000-0001-6527-6954]{Kimberly L. Emig}
\altaffiliation{Jansky Fellow of the National Radio Astronomy Observatory}
\affiliation{National Radio Astronomy Observatory, 520 Edgemont Road,Charlottesville, VA  22903-2475, USA}
       
\author{S. M\"uhle}
\affiliation{Argelander-Institut f\"ur Astronomie, Universit\"at Bonn, Auf dem H\"ugel 71, D-53121 Bonn, Germany}

\author[0000-0002-2699-4862]{Hiroyuki Kaneko}
\affiliation{Institute of Science and Technology, Niigata University, 8050 Ikarashi 2-no-cho, Nishi-ku, Niigata 950--2181, Japan}
\affiliation{Joetsu University of Education, 1, Yamayashiki-machi, Joetsu, Niigata 943--8512, Japan}

\author[0000-0001-9016-2641]{Tomoka Tosaki}
\affiliation{Joetsu University of Education, 1, Yamayashiki-machi, Joetsu, Niigata 943--8512, Japan}

\author[0000-0002-2333-5474]{Erica Behrens}
\affiliation{Department of Astronomy, University of Virginia, P.~O.~Box 400325, 530 McCormick Road, Charlottesville, VA 22904-4325}

\author[0000-0002-2887-5859]{V\'ictor M. Rivilla}
\affiliation{Centro de Astrobiolog\'ia (CSIC-INTA), Ctra. de Ajalvir Km. 4, Torrej\'on de Ardoz, 28850 Madrid, Spain}
\affiliation{INAF-Osservatorio Astrofisico di Arcetri, Largo Enrico Fermi 5, 50125, Florence, Italy}
    
\author[0000-0001-8064-6394]{Laura Colzi}
\affiliation{Centro de Astrobiolog\'ia (CSIC-INTA), Ctra. de Ajalvir Km. 4, Torrej\'on de Ardoz, 28850 Madrid, Spain}

\author[0000-0003-0563-067X]{Yuri Nishimura}
\affiliation{Institute of Astronomy, Graduate School of Science, The University of Tokyo, 2-21-1 Osawa, Mitaka, Tokyo 181-0015, Japan}
\affiliation{National Astronomical Observatory of Japan, 2-21-1 Osawa, Mitaka, Tokyo 181-8588, Japan}

\author[0000-0003-3537-4849]{P. K. Humire}
\affiliation{Max-Planck-Institut f\"ur Radioastronomie, Auf dem H\"ugel 69, 53121 Bonn, Germany}

\author[0000-0003-0167-0746]{Mathilde Bouvier}
\affiliation{Leiden Observatory, Leiden University, P.O. Box 9513, 2300 RA Leiden, The Netherlands}

\author[0000-0002-1227-8435]{Ko-Yun Huang}
\affiliation{Leiden Observatory, Leiden University, P.O. Box 9513, 2300 RA Leiden, The Netherlands}

\author[0000-0002-5353-1775]{Joshua Butterworth}
\affiliation{Leiden Observatory, Leiden University, P.O. Box 9513, 2300 RA Leiden, The Netherlands}

\author[0000-0001-9436-9471]{David S.~Meier}
\affiliation{New Mexico Institute of Mining and Technology, 801 Leroy Place, Socorro, NM 87801, USA}
\affiliation{National Radio Astronomy Observatory, PO Box O, 1003 Lopezville Road, Socorro, NM 87801, USA}

 \author[0000-0001-5434-5942]{Paul P.~van der Werf}
 \affiliation{Leiden Observatory, Leiden University, P.O. Box 9513, 2300 RA Leiden, The Netherlands}



\begin{abstract}
We present a spatially-resolved excitation analysis for the central molecular zone (CMZ) of the starburst galaxy NGC~253 using the data from the ALMA Large program ALCHEMI, whereby we explore parameters distinguishing NGC~253 from the quiescent Milky Way's Galactic Center (GC).
Non-LTE analyses employing a hierarchical Bayesian framework are applied to Band 3--7 transitions from nine molecular species to delineate the position--position--velocity distributions of column density ($N_\mathrm{H_2}$), volume density ($n_\mathrm{H_2}$), and temperature ($T_\mathrm{kin}$) at $27$ pc resolution.
Two distinct components are detected: a low-density component with $(n_\mathrm{H_2},\ T_\mathrm{kin})\sim (10^{3.3}\ \mathrm{cm}^{-3}, 85\ \mathrm{K})$ and a high-density component with $(n_\mathrm{H_2},\ T_\mathrm{kin})\sim (10^{4.4}\ \mathrm{cm}^{-3}, 110\ \mathrm{K})$, separated at $n_\mathrm{H_2}\sim10^{3.8}\ \mathrm{cm}^{-3}$.
NGC~253 has $\sim10$ times the high-density gas mass and $\sim3$ times the dense-gas mass fraction of the GC. These properties are consistent with their HCN/CO ratio but cannot alone explain the factor of $\sim 30$ difference in their star formation efficiencies (SFEs), \myrev{contradicting} the dense-gas mass to star formation rate scaling law.
The $n_\mathrm{H_2}$ histogram toward NGC~253 exhibits a shallow declining slope up to $n_\mathrm{H_2}\sim10^6\ \mathrm{cm}^{-3}$, while that of the GC steeply drops in $n_\mathrm{H_2}\gtrsim10^{4.5}\ \mathrm{cm}^{-3}$ and vanishes at $10^5 \mathrm{cm}^{-3}$.
Their dense-gas mass fraction ratio becomes consistent with their SFEs when the threshold $n_\mathrm{H_2}$ for the dense gas is taken at $\sim 10^{4.2\mbox{--}4.6}\ \pcc$. 
The rich abundance of gas above this density range in the NGC~253 CMZ, or its scarcity in the GC, is likely to be the critical difference characterizing the contrasting star formation in the centers of the two galaxies.
\end{abstract}

\section{Introduction\label{section:introduction}}
NGC~253 is an archetypal starburst galaxy in the nearby universe, whose relative closeness \citep[$D$ = 3.5 Mpc;][]{Rekola2005} makes it an ideal target to study the physical and chemical environments for highly active star formation (SF). 
In particular, a comparison between the central molecular zone (CMZ) of NGC~253 and the Galactic center (GC) of Milky Way (MW) provides insights into the conditions distinguishing starburst and quiescent galaxies \citep{Paglione1995,Sakamoto2011,Leroy2015,Krieger2020}.
While they exhibit contrasting SF activities, they are similar in the total stellar mass \citep{Bailin2011ApJ,Licquia2015ApJ} and share many common physical characteristics. 
The molecular gas in their central region, and the giant molecular clouds (GMCs) in their central regions are both in a barred potential and characterized by high temperature and density  \citep{Nagai2007,Rosenberg2014,Gorski2017,Perez-Beaupuits2018a,Tanaka2018b,Mangum2013,Mangum2019}, an elevated degree of turbulence \citep{Tsuboi1999,Shetty2012,Rathborne2014,Tanaka2020,Krieger2020,Harada2022ApJ,Huang2023AAp}, and enhanced cosmic-ray ionization rate \citep[CRIR;][]{OkaTakeshi2005,Goto2008,Belloche2013,Harada2021ApJ,Holdship2021AAp,Holdship2022ApJ,Behrens2022ApJ}.
However, the dense gas content with molecular hydrogen volume density (\nHH) of $\gtrsim 10^4\ \pcc$ of the GC is dominated by quiescent gas lacking cluster formation, in stark contrast to the NGC~253 CMZ, which harbours 14 embedded young massive clusters \citep{Leroy2018}.
The overall star formation rate (SFR) of the CMZ of NGC~253 is two orders of magnitude higher than that of the GC \citep{Ott2005,Kauffmann2013a},  which is approximately an order of magnitude higher than the molecular gas mass ratio of the NGC~253 CMZ to the GC \citep{Dahmen1998}.

The dispersion of star formation efficiencies (SFEs) of molecular clouds and galaxies, where SFE refers to the SFR per unit molecular gas mass, are observationally related to the difference in dense gas fraction (\fDG), which represents the mass fraction of gas immediately responsible for SF \citep{Solomon1992ApJ,Gao2004,Lada2012a}.
The HCN/CO luminosity ratio is the most commonly used measure of \fDG, as HCN~1--0 luminosity was considered to trace dense-gas mass with $\nHH\gtrsim 3\times10^4$\ \pcc{} \citep{Gao2004,Lada2012a}.
The HCN/CO ratio of NGC~253 is a few times that of the GC \citep{Paglione1995,Sakamoto2011}, being qualitatively consistent with the higher SFE of NGC~253.
However, this difference in \fDG\ is still by an order of magnitude smaller than that in their difference in SFE, apparently \myrev{contradicting} the scaling relation between the dense-gas mass and SFR.
This case with the NGC~253 CMZ and the GC can be regarded as a striking example of the $\sim 1$ dex dispersion of the SFR/HCN ratio among extragalactic sources \citep[e.g.,][]{Usero2015,Bigiel2016a,Gallagher2018,Jimenez-Donaire2019,Neumann2023}.
Such dispersion is attributed to the environmental dependence of the threshold \nHH\ for SF based on the turbulent cloud model \citep[e.g.][]{Krumholz2005} and deficiency of the HCN/CO ratio as an accurate measure of the dense-gas fraction \citep[e.g.][]{Kauffmann2017,Barnes2020,Jones2023}.

Accurate measurement of \nHH\ is crucial to addressing the above issues.
Luminosity-based measurements of \nHH\ based on multi-line excitation analysis have been reported toward NGC~253 (\citealt{Rosenberg2014,Perez-Beaupuits2018a}; hereafter R14 and P18, respectively) and the GC \citep{Nagai2007,Tanaka2018b,Mills2018}.  
These studies consistently indicate that both NGC~253 and the GC have multi-component molecular gas consisting of a relatively low-density ($\nHH\sim$ a few $10^3\ \pcc$) component mainly traced by low- to mid-$J$ CO lines and higher-density ($\nHH\sim$ a few $10^{4\mbox{--}5}\ \pcc$) a component visible in high-density tracers such as HCN and high-$J$ CO lines.
Multi-component gas was also detected in measurements of gas kinetic temperature (\Tkin) both for NGC~253 \citep{Mangum2013,Gorski2017} and for the GC \citep{Arai2016} from excitation analysis of the \ammonia\ inversion lines.
However, the \nHH\ values and mass fractions of the dense gas components differ among the authors depending on the analysis method, the tracer lines used, and the spatial resolution, leaving the comparison of the dense gas properties in NGC~253 and the GC not straightforward.

Further spatially resolved measurement of \nHH\ was performed for the GC.
\cite{Tanaka2018b} (hereafter \citetalias{Tanaka2018b}) conducted spatially-resolved excitation analysis using GC survey data of 13 transitions from nine molecular species ($^{13}$CO~2--1, HCN~1--0 and 4--3, H$^{13}$CN~1--0, HCO$^+$~1--0, H$^{13}$CO$^+$~1--0, HNC~1--0, \pformaldehyde~$2_{0,2}$--$1_{0,1}$ and $3_{0,3}$--$2_{0,2}$, CS~2--1, N$_2$H$^+$~1--0, SiO~2--1, and \HCCCN~10-9).
They constructed position--position--velocity (\PPV) cubes of the physical condition parameters and molecular abundances at a spatial resolution of \ang{;;60} = 2.4 pc.
They found that signatures of SF were limited to dense-gas clumps with typical $\nHH \gtrsim 10^{4.6}\ \pcc$ and argued that the scarcity of such high-density clumps is one reason for the suppressed SF in the GC \citep{Tanaka2020}.
For the NGC~253 CMZ, however, the previous \nHH\ measurements (\citetalias{Rosenberg2014} and \citetalias{Perez-Beaupuits2018a}) were performed with the luminosity-based CO spectral line energy distribution (SLED), and hence the spatial distribution of \nHH\ and its correlation with SF activities remain yet to be fully investigated.

By exploiting large-scale molecular line survey data at millimeter--submillimeter wavelengths accessible with Atacama Large Millimeter/submillimeter Array (ALMA), it is now possible to simultaneously delineate the spatial variation of \nHH, \Tkin, and hydrogen column density (\NHH) even in relatively distant sources.
In this paper, we apply the same analysis as \citetalias{Tanaka2018b} to the NGC\,253 CMZ by utilizing the unprecedentedly rich spectral data from the ALMA Comprehensive High-resolution Extragalactic Molecular Inventory (ALCHEMI) large program \citep{Holdship2021AAp,Barrientos2021ExA,Harada2021ApJ,
Martin2021A&A,Haasler2022A&A,Holdship2022ApJ,Behrens2022ApJ,Humire2022A&A,Harada2022ApJ,Huang2023AAp}. 
The ALCHEMI survey \citep{Martin2021A&A} covers the 84--373 GHz frequency range at a \ang{;;1.6} (= 27 pc) spatial and 10~\kmps\ velocity resolution, including all tracers used in the \citetalias{Tanaka2018b} analysis.  The richness of the spectral data with high enough spatial resolution ensures a reliable comparison of the \PPV\ distributions of \NHH, \Tkin, and \nHH\ to those of the GC with less ambiguity due to differences in the analysis settings.
With this analysis, we discuss the fundamental differences that distinguish starburst and quiescent GMCs based on the physical condition parameters, not only on the line intensities.

The remainder of this paper is structured as follows.
Section \ref{section:method} outlines two statistical models for the parameter inference: the Hierarchical Bayesian (HB) method and the standard non-hierarchical Bayesian (NHB) method. 
We also use four different input data sets from ALCHEMI and \citetalias{Tanaka2018b}, which are described in \S\ref{section:data}.   The results of the analysis are presented in \S\ref{section:results}.   
A comparative study with previous works, including a comparison between NGC~253 and the GC, is made in \S\ref{section:discussion}.   The final section (\S\ref{section:summary}) summarizes the important results of this article.

\section{Method\label{section:method}}
\subsection{Outline of the Analysis}

Different molecular lines probe different \Tkin\ and \nHH\ ranges depending on their upper state energies and transitional critical densities.
Physical conditions can be constrained by solving the equations of molecular line excitation of multi-transition data.
GMC complexes generally consist of multiple components with different physical conditions, making the results of the excitation analysis dependent on the choice of input lines when the spatial resolution is insufficient to resolve multi-component gas.   
The analysis is also affected by the statistical modeling employed in the parameter inference.  

In this paper, we perform five different analyses for the NGC~253 CMZ using three input data sets (Low-density, High-density, and Minimal data sets) and two statistical frameworks (HB and NHB).
The Low- and High-density data sets probe molecular gas components with $\nHH\sim10^3~\pcc$ and $\gtrsim 10^4~\pcc$, respectively. We can approximately decompose multiple component molecular gas in the NGC~253 CMZ by applying excitation analyses independently for these two data sets.  
We will perform both the HB and NHB analyses for them to examine the effect of the choice of the statistical framework on the outputs. 
In addition to the main analyses using the Low- and High-density data sets, we perform a supplementary analysis using the `Minimal' data set.
This data set is a common subset of the above two data sets for the NGC~253 and the data set used for the GC in the \citetalias{Tanaka2018b}.
It consists of five molecular lines that are barely sufficient to determine \Tkin\ and \nHH\ simultaneously.
By using of the same data set for the NGC~253 and the GC, we can exclude bias due to the selection of tracer lines in comparing their results.

In addition to the above five analyses for the NGC~253 CMZ, we also perform a re-analysis for the GC using the Minimal data set. 
The GC result is used for comparison with the result with the Minimal data set for the NGC~253 CMZ.

The following subsections describe the model parameters, non-LTE excitation analysis, statistical frameworks, and data sets used in these analyses.

\subsection{Model Parameters and Non-LTE Excitation Analysis \label{subsection:method:parameters}}

Table \ref{table:modelparams} lists the parameters required to calculate model line intensities that can be directly compared to the observed data.
The parameter space includes the physical and chemical conditions, \Tkin, \nHH, \dNHHdv, fractional abundances of the molecules involved ($x_\mathrm{mol}$), and $[^{12}\mathrm{C}]/[^{13}\mathrm{C}]$ isotopic abundance ratio ($R_{13}$).
We assume $R_{13}$ is common for all $^{13}\mathrm{C}$-bearing species except for $\mathrm{^{13}CO}$, though the [$^{12}\mathrm{C}$]/[$^{13}\mathrm{C}$] isotopic ratio may differ among species due to chemical fractionation and selective photo-dissociation effects.  The assumption of common $R_{13}$ is supported by the ALCHEMI compact array data \citep{Martin2021A&A}, which show a consistent [$^{12}\mathrm{C}$]/[$^{13}\mathrm{C}$] isotopic ratio for the high dipole moment molecules used in the present analysis.
The $\mathrm{^{13}CO}$ is treated differently from other $\mathrm{^{13}C}$-bearing species. We fix $\xmol{^{13}CO}$ at the value taken from \cite{Martin2019}, since  $\xmol{^{13}CO}$ cannot be accurately constrained from our input data set as we describe later in \S\ref{subsection:data:lowdensity}. 

\myrev{
The line intensities are calculated by solving the non-LTE rate equations, where the large velocity gradient (LVG) approximation \citep{Goldreich1974} is employed to evaluate the opacity effect.
}
The coefficients of the radiative and collisional transitions are taken from the Leiden Molecular and Atomic Database \citep[LAMDA; ][]{Schoier2005}. 
The collision coefficients of HCN, \pformaldehyde, and \HCCCN\ are not provided for temperatures high enough to cover the full \Tkin\ range of the present analysis, which is 1000~K.   We applied constant extrapolation for the coefficients outside the temperature range given in the tables.
References for the collision rates of individual species are provided in Table \ref{table:datasets}.
The escape probabilities are calculated assuming spherical geometry.
We do not consider excitation through IR-pumping, since the analysis on the HCN and HNC intensity ratio has shown that the IR-pumping is unlikely to be the dominant excitation mechanism in the ALCHEMI data \citep{Behrens2022ApJ}.  We will also examine the effect of the IR-pumping independently later in  this paper ( \S\ref{subsection:results:line2lineDifference}).
The LVG-calculated intensities are converted into the observable intensities by multiplying by the beam filling factor ($\Phi$).
We introduce the $\phi$-parameter, which is related to $\Phi$ as $\Phi = 1-e^{-\phi}$, so that $\Phi$ is constrained to the range 0--1 for $\phi > 0$.
We represent the above parameters at the $i$th spaxel by the parameter vector $\Vec{p}_i = \left(p_{ij}\right)$, where $j$ indexes the individual parameter elements.
All parameters are in the base-10 logarithm scale.
We fix a few parameters to obtain reliable results from currently available observational information, 
of which details are given in \S\ref{subsection:data:fixedParams}.

The observed intensities are also practically dependent on non-statistical errors, which may arise from calibration uncertainties and model deficiencies, such as the breakdown of the ideal LVG approximation. We assume the non-statistical errors to be multiplicative and represent the relative errors by the vector $\Vec{\epsilon}_i = \left(\epsilon_{ik}\right)$, where the subscripts $i, k$ index the spaxels and lines, respectively.

In addition to the above fundamental parameters, we also use a composite parameter representing the beam-diluted column density: $\left<\NHH\right>_\mathrm{beam} \equiv \Phi\cdot\NHH$, which is a direct measure of the gas mass contained within a spaxel.
This paper uses simplified symbols for the primary parameters \dNHHdv, \dNHHbeam, \nHH, \Tkin, $R_{13}$, and $\phi$, \myrev{represented by $N$, $\left<N\right>$, $n$, $T$, $R$, and $\phi$, respectively}.
They are primarily used in subscript indices of vectors and matrices; for instance, $p_{iN}$ denotes the \dNHHdv\ value at $i$th spaxel.

\begin{deluxetable*}{lll}
\tablecolumns{3}
\tablehead{\colhead{parameter}& \colhead{} & \colhead{description}}
\tablecaption{Model Parameters and Hyperparameters \label{table:modelparams}}
\startdata
\multicolumn{2}{l}{$\vec{p}_i$ (parameter vector at $i$th voxel)} \\
&$\log_{10} \dNHHdv$ & hydrogen column density per unit velocity width in unit of  $\mathrm{cm^{-2}} (\mathrm{km\,s^{-1}})^{-1}$ \\
&$\log_{10} \Tkin$ & gas kinetic temperature in unit of K\\
&$\log_{10} \nHH$ & hydrogen volume density  in unit of $\mathrm{cm^{-3}}$ \\
&$\log_{10} x_\mathrm{mol}\left(X\right)$ & fractional abundance of molecule $X$ relative to H$_2$ \\
&$\log_{10} R_{13}$ & $[^{12}\mathrm{C}]/[^{13}\mathrm{C}]$ isotopic abundance ratio \\
&$\log_{10} \phi_i$ & $\phi$-parameter; related to the beam filling factor $\Phi$ by $\Phi = 1-e^{-\phi}$ \\
\\
\multicolumn{2}{l}{$\vec{\epsilon}_i$ (non-statistical error vector at $i$th voxel)} \\
&$\epsilon_{ik}$ & non-statistical relative error of the $k$th line intensity/ratio  \\
\\
\multicolumn{2}{l}{$\theta$ (hyperparameter)}\\
&$\Vec{\mu}$ & location vector of the multivariate log-student prior of $\Vec{p}_i$ \\
&$\Sigma$   & scale matrix of the multivariate log-student prior of $\Vec{p}_i$ \\
&$\Vec{\sigma}$  & scale parameters of the log-normal prior of $\Vec{\epsilon}_i$ 
\enddata
\end{deluxetable*}

\subsection{Hierarchical Bayesian Analysis\label{subsection:method:hierarchical}}

The HB analysis  (\citealt{Kelly2012}, \citetalias{Tanaka2018b}) can handle problems including the non-statistical errors, whose values are difficult to infer with the standard maximum likelihood (ML) analysis.  The inclusion of non-statistical errors often makes the number of degrees of freedom less than zero.   
The HB analysis can also explicitly forbid artificial (anti-)correlation among the parameters by using appropriate hyperprior functions.   
The following outlines the HB framework we utilized in this paper; we refer readers to \citetalias{Tanaka2018b} for details on the method.

Let $\mathcal{V} = \left\{\Vec{v}_i\right\}$, where the elements of vector $\Vec{v}_i = \left\{v_{ik}\right\}$ represent the
$k$th line intensity at the $i$th spaxel.   
The present analysis uses raw line intensities or intensity ratios as $v_{ik}$.
Then the hierarchical posterior probability is given by Bayes' theorem;
\begin{eqnarray}
\prob{\mathcal{P},\mathcal{E},\theta|\mathcal{V}} & = & \prob{\mathcal{V}|\mathcal{P},\mathcal{E}} \cdot\prob{\mathcal{P},\mathcal{E}|\theta}\cdot\prob{\theta}\ \label{eq:bayesTheorem},
\end{eqnarray}
where \prob{\cdot|\cdot}\ denotes the conditional probability density function (PDF), $\mathcal{P}=\left\{\Vec{p}_i\right\}$, and  $\mathcal{E}=\left\{\Vec{\epsilon}_i\right\}$.
The function \prob{\mathcal{P},\mathcal{E}|\theta} is the prior probability, which is the conditional PDF of $\mathcal{P}$ and $\mathcal{E}$ when the hyperparameter $\theta$ is given.  
The hyperparameter $\theta$ consists of the location ($\vec{\mu}$) and scale parameters ($\Sigma$) of the prior function.
The scale parameter $\Sigma$ is further decomposed into the scaling diagonal matrix $S$ and correlation matrix $R$ as $\Sigma = SRS$. 
The PDFs of the hyperparameters are given by the hyperprior function \prob{\theta}.  Here and henceforth, normalization constants of PDFs are omitted whenever possible.
The function forms of $\prob{\mathcal{V}|\mathcal{P},\mathcal{E}}, \prob{\mathcal{P},\mathcal{E}|\mathcal{\theta}}$, and $\prob{\mathcal{\theta}}$ adopted in the present analysis are described in Appendix \ref{appendix:A}.

The final product of the HB inference is the marginal posterior $\prob{\mathcal{P}|\mathcal{V}}$ that describes the probability distribution of the model parameters.
The marginal posterior is obtained by integrating \prob{\mathcal{P},\mathcal{E},\theta|\mathcal{V}} over the nuisance parameters, i.e., (hyper)parameters not of immediate interest;
\begin{eqnarray}
\prob{\mathcal{P}|\mathcal{V}} & = & \int{\mathrm{d}\mathcal{E}}{\mathrm{d}\theta\cdot\prob{\mathcal{P},\mathcal{E},\theta|\mathcal{V}}}\label{eq:marginalposteriorH}.
\end{eqnarray}

\subsection{Non-Hierarchical Analysis}
We compare the results of the HB analysis with those obtained from a more conservative analysis adopting the non-hierarchical posterior function. 
The posterior probability in the NHB framework is
\begin{eqnarray}
\prob{\mathcal{P},\mathcal{E}|\mathcal{V}} & = & \prob{\mathcal{V}|\mathcal{P},\mathcal{E}} \cdot\prob{\mathcal{P},\mathcal{E}} \label{eq:nonHprob}.
\end{eqnarray}
The likelihood function $\prob{\mathcal{V}|\mathcal{P},\mathcal{E}}$ is the same as that used in the HB analysis.
The prior function \prob{\mathcal{P},\mathcal{E}} is not hierarchical, i.e., not parameterized by variable hyperparameters.
The function form \prob{\mathcal{P},\mathcal{E}} is chosen to be non-informative so that the analysis becomes essentially identical to the standard maximum-likelihood method.
The details of the non-hierarchical posterior are given in Appendix \ref{appendix:B}.

\section{Data\label{section:data}}

\subsection{Two-Component Model\label{subsection:method:twoComponentModel}}

\begin{figure}[t]
\plotone{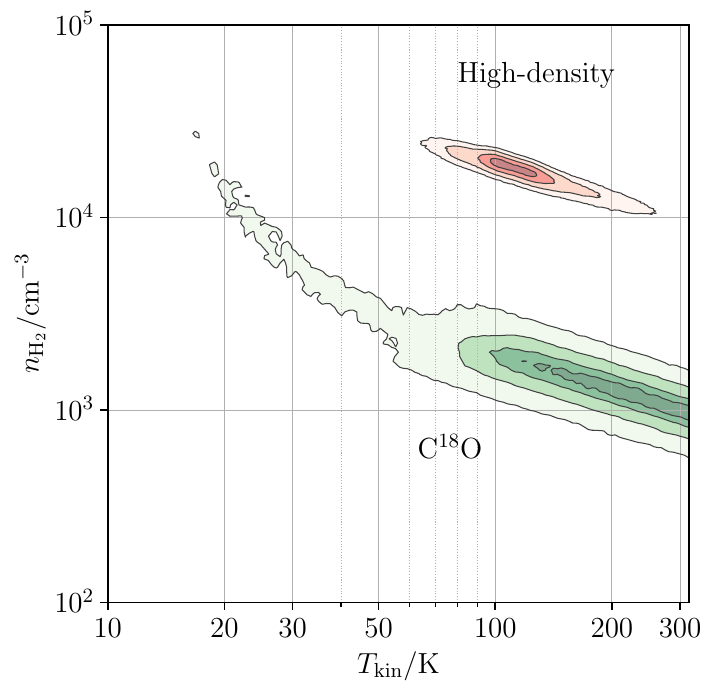}
\caption{Posterior probabilities calculated from the intensity ratios of the $\mathrm{C^{18}O}$ lines (green contours; labeled ``$\mathrm{C^{18}O}$'') and those from the   $\mathrm{H^{13}CN}$, $\mathrm{H^{13}CO^+}$ and \pformaldehyde ~lines (red contours; labeled ``High-density''). 
Contours indicate 20\%, 50\%, 80\%, and 95\% credible intervals. 
The parameters other than \Tkin\ and \nHH\ are fixed at typical values in the GC \citep{Tanaka2018b}:  $\phi = 0.1$, $\frac{\mathrm{d}\NHH}{\mathrm{d}v} = 10^{22}\ \psc\,\left(\kmps\right)^{-1}$, $\xmol{CO}=10^{-4}, \xmol{H^{13}CN} = 10^{-8.1}, \xmol{H^{13}CO^+} = 10^{-9.5}$, and $\xmol{\pformaldehyde} = 10^{-8.3}$.
The two line groups represent two independent physical components whose \nHH\ differ by $\sim$ 1 dex.
\label{fig:highlow}}
\end{figure}

The physical condition parameters deduced from the excitation analysis for a non-uniform medium depend on the critical density (\ncrit) and the upper state energy (\Eu) of the input lines.
Figure \ref{fig:highlow} shows the posterior PDFs calculated using the median intensities from two different data sets chosen from the ALCHEMI survey.
One is calculated from the line intensity ratios among the 1--0, 2--1, and 3--2 transitions of $\mathrm{C^{18}O}$ (labeled ``$\mathrm{C^{18}O}$'' in the figure), and the other is from the median 4--3/3--2 ratios of $\mathrm{H^{13}CN}$ and $\mathrm{H^{13}CO^+}$, and the $3_{2,1}$--$2_{2,0}$/$3_{0,3}$--$2_{0,2}$ ratio of $p$-\formaldehyde\ (labeled ``High-density'' in the figure).
A relative uncertainty of 15\% is conservatively assumed for all line intensities \citep{Martin2021A&A}.
The posterior PDFs were calculated using the non-hierarchical model, where \Tkin\ and \nHH\ are taken as the free parameters and the other parameters are fixed at typical GC values given in the figure caption.
The 95\% credible intervals of the two results do not overlap in the given parameter space, strongly suggesting that the two data sets trace different components of molecular gas, which likely corresponds to multiple components found in the dust and molecular line SEDs (\citetalias{Rosenberg2014}; \citetalias{Perez-Beaupuits2018a}; \citealt{Gorski2017,Mangum2019,Humire2022A&A}).
Similar multi-component gas was also reported for the GC  (\citealt{Nagai2007},\citealt{Arai2016}, \citealt{Krieger2017}, \citealt{Mills2015},  and \citetalias{Tanaka2018b}).

In the present analysis, we separate the input data into two groups, i.e., the Low-density and High-density data sets, and apply a one-zone model for each.
We adopt this simplified approach because a bona-fide multi-component analysis requires too many free parameters to constrain from the available observational information.
The Low-density and High-density data set comprise CO lines and high-density tracer lines, respectively, whose details are given in the next subsections (\S\ref{subsection:data:lowdensity}, \S\ref{subsection:data:highdensity}). 
This method assumes that the lines in the Low- and High-density data strictly trace the low- and high-density components, respectively, such that no line has significant emission from both components.
This is a reasonable approximation for the high-density tracer lines (e.g., \citetalias{Tanaka2018b}).
The CO lines may have intense emission both from the low- and high-excitation gas when they are optically thick, which may cause an overestimate of \nHH\ of the low-density component;  however, the previous CO SLED analyses (\citetalias{Rosenberg2014}; \citetalias{Perez-Beaupuits2018a}) show that the lowest-density component dominates the $J\leq$ 3 transitions of CO.
    Although the CO SLED analyses show a third component with $\Tkin, \nHH\ \sim$ 110--160 K, $10^{5.5\mbox{--}6.4}\ \pcc$ \citepalias{Rosenberg2014,Perez-Beaupuits2018a}, we do not consider this highest-excitation component in this analysis, since its mass is approximately 2 orders magnitude smaller than the second component mass.  
    In addition, we exclude transitions with high \Eu\ ($\Eu/k_\mathrm{B} > 70$ K) from the High-density data set to suppress the contamination from the highest-excitation component (\S\ref{subsection:data:highdensity}). 

It should also be noted that the validity of the one-zone approximation needs verification, in particular for the high-density component.
For example, star-forming dense gas generally consists of multiple components with different physical characteristics probed by different molecular lines, which cannot be spatially resolved with the ALCHEMI beam size of 27 pc. 
The present analysis treats the breakdown of the one-zone model by introducing the error parameter $\vec{\epsilon}$ in the Hierarchical analysis, which enables us to estimate averaged physical conditions over multiple physical components co-existing in a beam.
This also means that the results of the HB analysis may differ from the physical conditions deduced from a fewer number of molecular lines;  we will investigate the species-to-species difference in their excitation conditions in a later section (\S\ref{subsection:results:line2lineDifference}).

In addition to the Low- and High-density data sets, we use a subset consisting of five transitions commonly included in the two data set and the transitions used in the \citetalias{Tanaka2018b} analysis.
We refer to this data set as the ``Minimal data set'' hereafter.
In addition to the main analysis using the High-density data set, we perform the HB analyses utilizing the Minimal data set both for NGC~253 and the GC.
Their results are expected to be more suitable for comparisons between the NGC~253 CMZ and the GC than the main analysis and the \citetalias{Tanaka2018b}, since they are free from the potential bias arising from the selection of tracer lines.

In the following we describe the details of the three data sets.
\myrev{All of the data cubes for for the NGC~253~CMZ are taken from the ALCHEMI survey data (2017.1.00161.L; \citealt{Martin2021A&A}).
The data cubes are convolved into an angular resolution of \ang{;;1.6} at a \ang{;;0.45} grid spacing and velocity channel bins of 20~\kmps\ width. 
We show the integrated intensity maps of the used transitions in Appendix \ref{appendix:integratedIntensityMaps}.}

\subsection{Low-Density data set\label{subsection:data:lowdensity}}
The Low-density data set consists of the $J$=1--0, 2--1, and 3--2 transitions of CO, $^{13}\mathrm{CO}$, and $\mathrm{C^{18}O}$.
Their optically-thin critical densities and upper-state energies are $\mathrm{log}_{10}\,n_\mathrm{crit}/\mathrm{cm^{-3}} \sim 3$--$4$, and $E_\mathrm{u}$/$k_\mathrm{B}\sim 5$--$30$ K, respectively. 
To limit the analysis to emission detected with sufficient signal quality, we applied a signal-to-noise (S/N) cutoff of 3 to all CO and $\mathrm{^{13}CO}$ measurements.   Only spaxels with both CO and $^{13}\mathrm{CO}$ intensities above this S/N cutoff are included the analysis.

The present analysis approximates that the beam-filling factor $\Phi$ is common for all lines at each spaxel;  
however, the $^{12}\mathrm{CO}$ lines are expected to have significantly large $\Phi$ because of their large optical depths of $\sim 10$.  
The breakdown of the approximation of common $\Phi$ may lead to a significant underestimation of the $\mathrm{^{12}CO}$ beam-filling factor, and hence an overestimate of their undiluted intensities.  This is problematic because the undiluted $\mathrm{^{12}CO}~J$=1--0 intensity almost directly determines $T_\mathrm{kin}$ because of its high opacity and low $n_\mathrm{crit}$.  
To avoid this problem, we use the 3--2/1--0 and 3--2/2--1 intensity ratios as the input data for the $^{12}\mathrm{CO}$ lines instead of their absolute intensities.
As a drawback, we cannot determine the isotopic abundances with sufficient accuracy because of the lack of the absolute  $^{12}\mathrm{CO}$ intensities.
Hence, we fix $R_{13}$ at 21 according to \cite{Martin2019} in the analysis using the Low-density data set.   
We also assume a constant $[\mathrm{CO}]/[\mathrm{C^{18}O}]$ isotopic abundance of 130 \citep{Martin2019} for consistency.

\begin{figure*}
    \centering
    \plotone{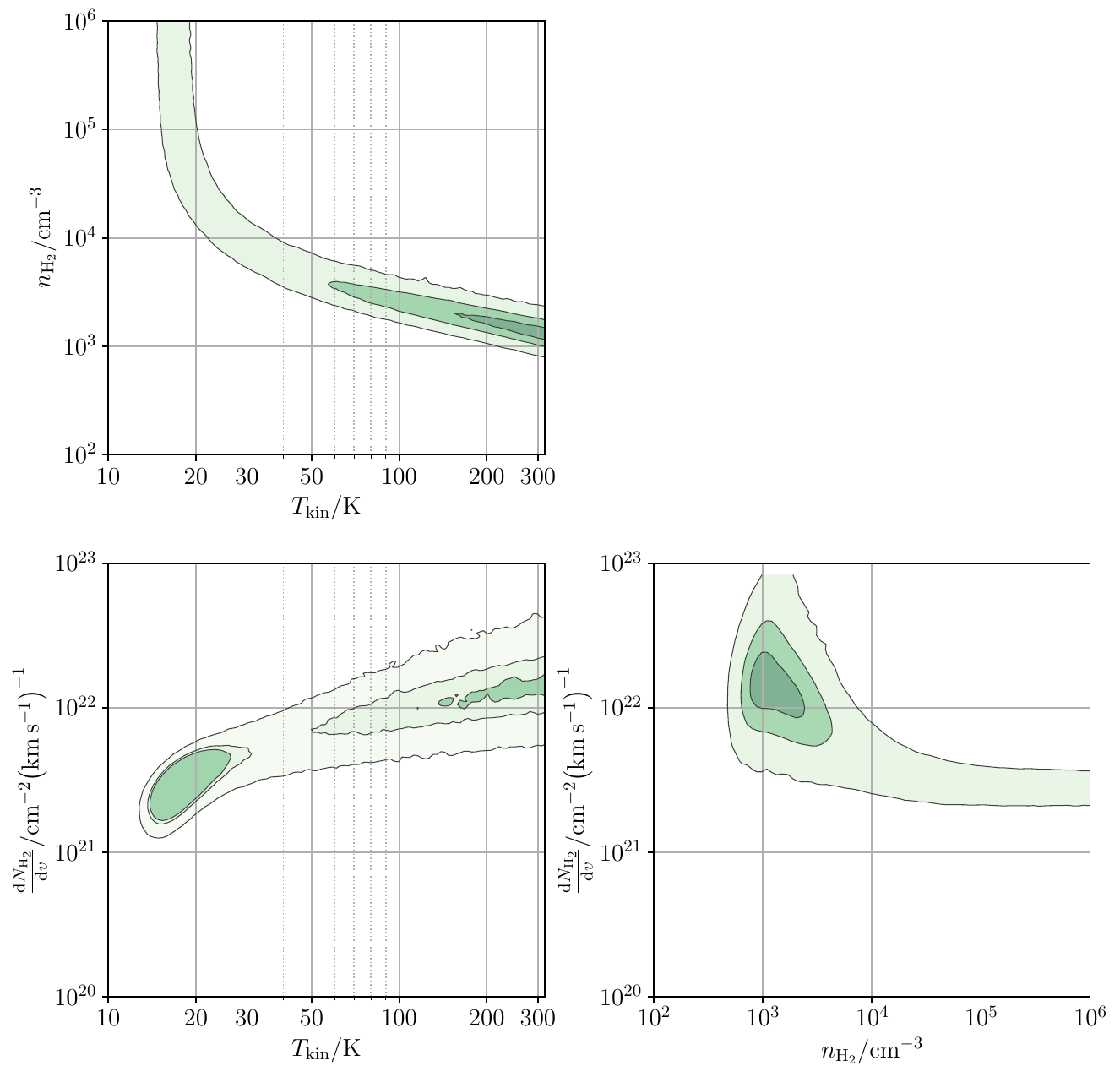}
    \caption{Example of the excitation analysis using the Low-density data set averaged over the central \ang{;;10} radius.  The posterior probability calculated using the NHB analysis is shown in the $\dNHHdv$--$\Tkin$--$\nHH$ parameter space, with  
    contours drawn at 20\%, 50\%, and 90\% credible intervals.
    The molecular abundances are fixed at $x_\mathrm{mol}\left(^{13}\mathrm{CO}, \mathrm{C^{18}O}\right)$ = $10^{-5.32}, 10^{-6.11}$ assuming $\xmol{CO} = 10^{-4}$.
    Relative uncertainty of 15\% is assumed for all line intensities according to \cite{Martin2021A&A}. 
    }
    \label{fig:LVGlow}
\end{figure*}

Figure \ref{fig:LVGlow} shows an example of the excitation analysis using the Low-density data set, 
which includes the marginal posterior $\prob{\mathcal{P}|\mathcal{V}}$ calculated using the NHB method.
The input intensities $\mathcal{V}$ are averages over the central \ang{;;10} radius.
The posterior probability shows a highly elongated shape on the \Tkin--\nHH\ plane, indicating the strong degeneracy of the two parameters.

\begin{figure*}
    \centering
    \plotone{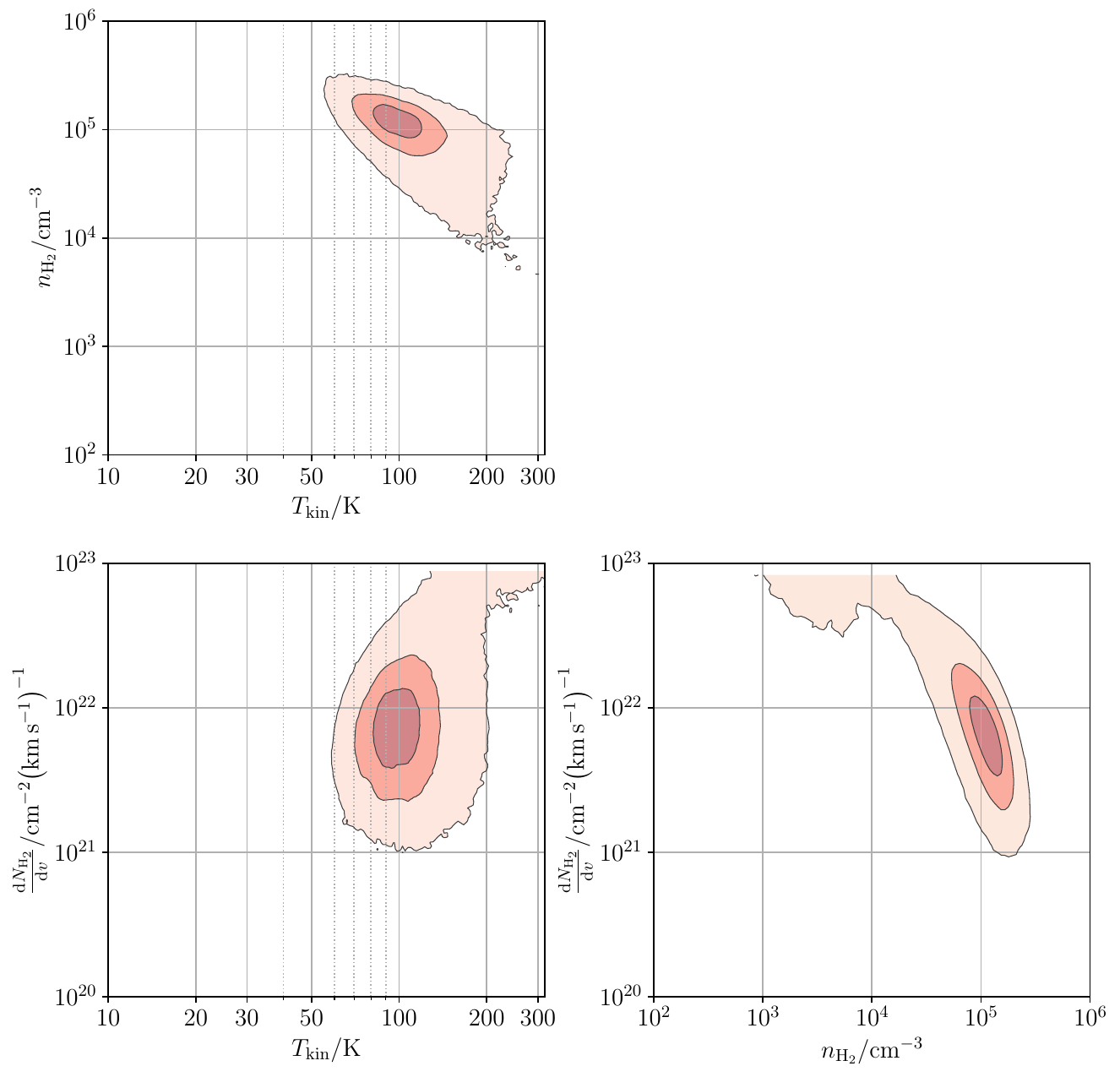}
    \caption{Same as Figure \ref{fig:LVGlow}, but for the NHB analysis with High-density data set.  
    \label{fig:LVGhigh}}
\end{figure*}

\begin{figure*}
    \centering
    \plotone{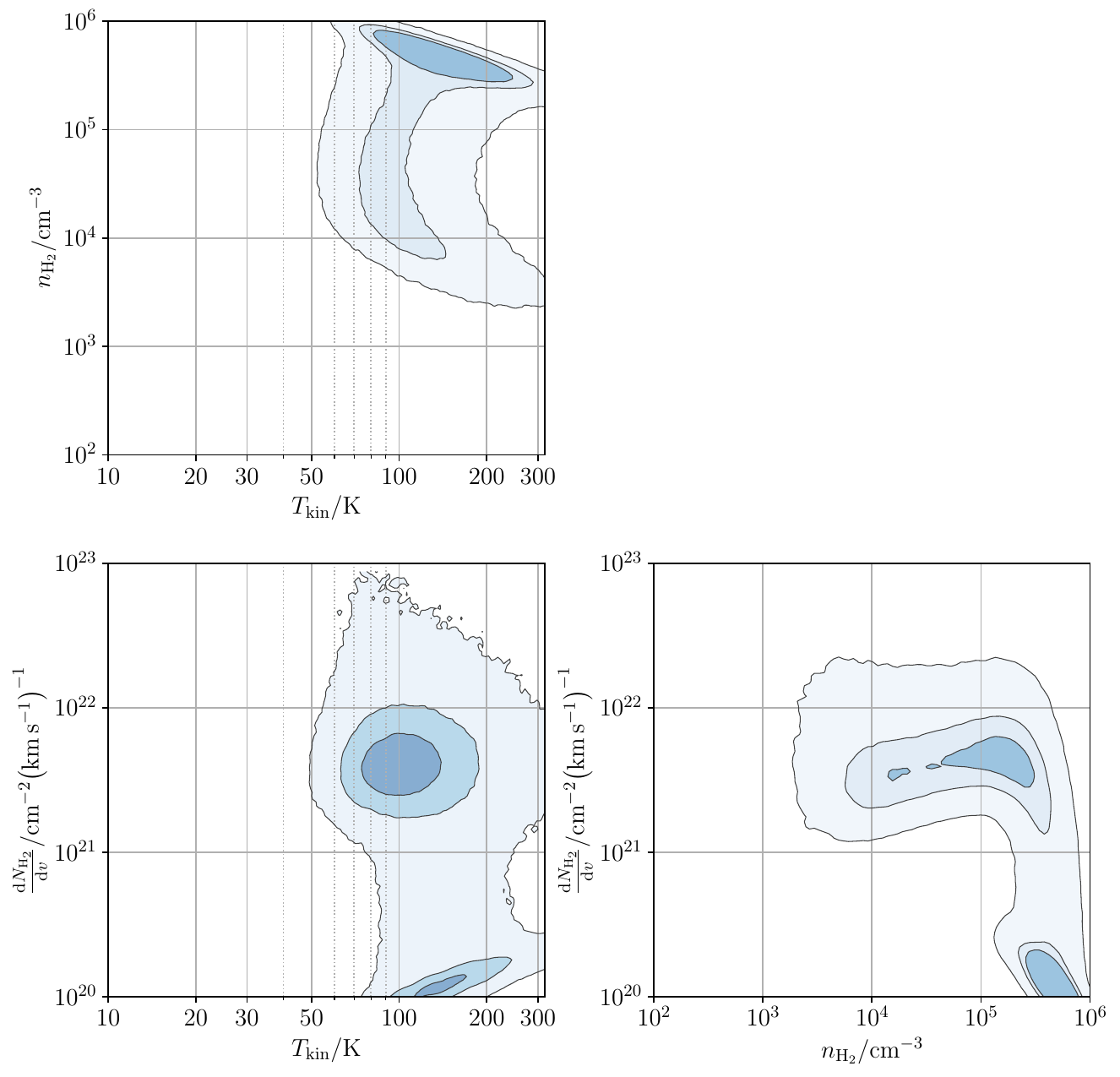}
    \caption{Same as Figure \ref{fig:LVGlow}, but for the NHB analysis with the Minimal data set. 
    }
    \label{fig:LVGmin}
\end{figure*}

\begin{deluxetable*}{llccccll}
\tablecaption{Input data sets\label{table:datasets}}
\tablecolumns{7}
\tablehead{
\colhead{data set} & \colhead{molecule} & \colhead{transitions} & \colhead{frequency} & \colhead{$E_\mathrm{u}/k_\mathrm{B}$} & \colhead{$\mathrm{log}_{10}\,n_\mathrm{crit}$\tablenotemark{$\star$}} & 
\colhead{collisional excitation rate reference} 
\\
\colhead{}&\colhead{}&\colhead{}&\colhead{GHz}&\colhead{K}&\colhead{\pcc} &
\colhead{}}
\tabletypesize{\footnotesize}
\startdata
Low-density & $\mathrm{CO} $ & \tablenotemark{$\dagger$}$ 1$--$0 $ &  115.2712  & {\phn}5.5  &  3.3  & \cite{rates:CO}  \\
 & $            $ & \tablenotemark{$\dagger$}$ 2$--$1 $ &  230.5380  &  16.6  &  3.8 & \\
 & $            $ & \tablenotemark{$\dagger$}$ 3$--$2 $ &  345.7960  &  33.2  &  4.2 & \\
 & $\mathrm{^{13}CO} $ & \tablenotemark{$\dagger$}$ 1$--$0 $ &  110.2014  & {\phn}5.3  &  3.3 & \cite{rates:CO}  \\
 & $            $ & \tablenotemark{$\dagger$}$ 2$--$1 $ &  220.3987  &  15.9  &  3.8 &  \\
 & $            $ & \tablenotemark{$\dagger$}$ 3$--$2 $ &  330.5880  &  31.7  &  4.2 &  \\
 & $\mathrm{C^{18}O} $ & $ 1$--$0 $ &  109.7822  & {\phn}5.3  &  3.3 & \cite{rates:CO} \\
 & $                 $ & $ 2$--$1 $ &  219.5604  &  15.8  &  3.8 &  \\
 & $                 $ & $ 3$--$2 $ &  329.3306  &  31.6  &  4.2 &  \\
\hline
High-density & $\mathrm{HCN} $ & $ 1$--$0 $ & {\phn}88.6316  & {\phn}4.3  &  6.4 & \cite{rates:HCN_HNC} \tablenotemark{\footnotesize $a$} \\
 & $             $ & \tablenotemark{$\ddagger$}$ 3$--$2 $ &  265.8864  &  25.5  &  7.0 & \\
 & $             $ & $ 4$--$3 $ &  354.5055  &  42.5  &  7.3 &  \\
 & $\mathrm{HCO^+} $ & $ 1$--$0 $ & {\phn}89.1885  & {\phn}4.3  &  5.4 & \cite{rates:HCOp_N2Hp} \\
 & $               $ & \tablenotemark{$\ddagger$}$ 3$--$2 $ &  267.5576  &  25.7  &  6.4 &  \\
 & $               $ & $ 4$--$3 $ &  356.7343  &  42.8  &  6.6 &  \\
 & $\mathrm{HNC} $ & \tablenotemark{$\ddagger$}$ 3$--$2 $ &  271.9811  &  26.1  &  6.8 & \cite{rates:HCN_HNC} \tablenotemark{\footnotesize $a$} \\
 & $\mathrm{CS}  $ & \tablenotemark{$\ddagger$}$ 3$--$2 $ &  146.9690  &  14.1  &  5.8 & \cite{rates:CS} \\
 & $             $ & $ 6$--$5 $ &  293.9121  &  49.4  &  6.6 &  \\
 & $\mathrm{SiO} $ & $ 3$--$2 $ &  130.2686  &  12.5  &  5.8 & \cite{rates:SiO} \\
 & $             $ & $ 5$--$4 $ &  217.1050  &  31.3  &  6.4 &  \\
 & $\mathrm{HC_3N} $ & $ 14$--$13 $ &  127.3677  &  45.8  &  5.4 & \cite{rates:HC3N}  \tablenotemark{\footnotesize $b$} \\
 & $               $ & $ 17$--$16 $ &  154.6573  &  66.8  &  5.6 &  \\
 & $\mathrm{N_2H^+} $ & $ 3$--$2 $ &  279.5117  &  26.8  &  6.3 & \cite{rates:HCOp_N2Hp} \\
 & $                $ & $ 4$--$3 $ &  372.6725  &  44.7  &  6.6 &  \\
 & $\pformaldehyde $ & $ 2_{0,2}$--$1_{0,1} $ &  145.6029  &  10.5  &  5.7 & \cite{rates:H2CO}  \tablenotemark{\footnotesize $b$} \\
 & $               $ & \tablenotemark{$\ddagger$}$ 3_{0,3}$--$2_{0,2} $ &  218.2222  &  21.0  &  6.1 &  \\
 & $               $ & $ 3_{2,1}$--$2_{2,0} $ &  218.7601  &  68.1  &  5.6 &  \\
 & $\mathrm{H^{13}CN} $ & $ 3$--$2 $ &  259.0118  &  24.9  &  7.0 & \cite{rates:H13CN} \\
 & $                  $ & $ 4$--$3 $ &  345.3398  &  41.4  &  7.2 &  \\
 & $\mathrm{H^{13}CO^+} $ & $ 3$--$2 $ &  260.2553  &  25.0  &  6.3 & \cite{rates:HCOp_N2Hp} \\
 & $                    $ & $ 4$--$3 $ &  346.9983  &  41.6  &  6.6 &  \\
 & $\mathrm{^{13}CS} $ & $ 3$--$2 $ &  138.7393  &  13.3  &  5.7 & \cite{rates:CS} \\
\hline
Minimal & HCN & \tablenotemark{$\ddagger$}1--0\\
        &     & \tablenotemark{$\ddagger$}4--3\\
        & \pformaldehyde & \tablenotemark{$\ddagger$}$3_{0,3}$--$2_{0,2}$\\
        &                & $3_{2,1}$--$2_{2,0}$\\
        & $\mathrm{^{13}CO}$ & {$\ddagger$}2--1\\
\enddata
\tablenotetext{\star}{
According to the definition of \ncrit\ given in Equation 4
of \cite{Shirley2015}; $\ncrit \equiv A_{i,j} / \sum_{k \neq
i} \gamma_{i,k}$ for the $i \rightarrow j$ level transition, where $A$
and $\gamma$ are the Einstein $A$ coefficient and collisional
excitation rate, respectively.
}
\tablenotetext{\dagger}{Used for the pre-analysis filtering of low-S/N spaxels. The cut-off level is 3.}
\tablenotetext{\ddagger}{Same as $\dagger$, but the cut-off level is 2.}
\tablenotetext{$\it{a}$}{Collision coefficients are provided for temperatures up to 500 K.}
\tablenotetext{$\it{b}$}{Collision coefficients are provided for temperatures up to 300 K.}
\end{deluxetable*}

\subsection{High-Density data set\label{subsection:data:highdensity}}       
The High-density data set is selected from the transitions of the molecules used in the \citetalias{Tanaka2018b} analysis (HCN, \HCOp, HNC, CS, SiO, \HCCCN, $\mathrm{N_2H^+}$, \pformaldehyde, and their $\mathrm{^{13}C}$ isotopologues) using the following criteria:
\begin{enumerate}
\item \myrev{\ncrit\ $\gtrsim 10^5\ \pcc$ in the optically thin limit} and the upper state energy $E_\mathrm{u}/k_\mathrm{B} \lesssim 70\ \mathrm{K}$, \label{enum:crit1}
\item negligible confusion from adjacent lines, \label{enum:crit3}
\item spatial distribution widespread over the entire CMZ (not applied to $\mathrm{^{13}C}$ isotopologues), \label{enum:crit4}
\item no prominent self-absorption dips toward the TH2 position, \label{enum:crit5} and
\item more than one transition or isotopologue line that satisfy the criteria \ref{enum:crit1}--\ref{enum:crit5} in the ALCHEMI data. \label{enum:crit2}
\end{enumerate}
The selected lines are shown in Table \ref{table:datasets}.
The criterion \ref{enum:crit1} is used to select high-density tracers while avoiding potential contamination from local high-temperature spots such as embedded hot cores.
The criterion \ref{enum:crit5} is necessary to avoid confusion with self-absorption features. 
TH2 is an intense non-thermal continuum source \citep{Turner1985,Ulvestad1997}, toward which deep self-absorption dips are detected in low frequency bands of the ALCHEMI data.
All optically thin transitions of the molecules in the High-density data set in Band-3 do not satisfy criterion \ref{enum:crit5}.
As an exception to the above criteria, we include HNC 3--2 in the data set to use it as a proxy of $\mathrm{H_2}$ though it does not satisfy criterion \ref{enum:crit2}.
Low-S/N measurements were excluded in the same manner as in the Low-density data set.
The lines used in the pre-analysis filtering of the low-emission spaxels are marked in Table \ref{table:datasets};  relatively abundant species in Bands 4 and 6 were chosen for the filtering.  
The cutoff S/N level is set to be 2, though the cut off level of 3 is used for the Low-density data set.  We adopt a lower cutoff level so that the High-density data set has a \PPV\ coverage close to the Low-density data set while keeping sufficient quality.

The problem of the coupling between $\Tkin$ and the beam-filling factor $\Phi$ is not critical in the High-density data set, as the High-density set does not contain high-$\Phi$ and low-\ncrit\ lines such as CO~$J$=1--0.  With the High-density data set, \Tkin\ is mainly determined by the \pformaldehyde\ $3_{2,1}$--$2_{2,0}$/$3_{0,3}$--$2_{0,2}$ intensity ratio \citep{Mangum2019}, which is unaffected by $\Phi$ when $\Phi$ is common for the two transitions.
Hence, the absolute intensities are used as the input for all lines in the High-density set.
Since the absolute intensities of HCN, $\mathrm{HCO^+}$, CS, and their $^{13}\mathrm{C}$ isotopologues are available, we do not fix the $R_{13}$ value and treat it as a probability variable in the analysis with the High-density data set.

Figure \ref{fig:LVGhigh} shows an example of excitation analysis using the High-density data set averaged over the central $\pm\ang{;;10}$ region.
The optimal \Tkin\ and \nHH\ values are relatively well determined compared with the Low-density analysis (Figure \ref{fig:LVGlow}), owing to the inclusion of the two $p$-$\formaldehyde$ lines, whose intensity ratio only weakly depends on \nHH.
However, parameter degeneracy is still identified in the \NHH--\nHH\, and \Tkin--\nHH\ planes, which can cause artificial anti-correlation between \PPV\ distributions of these parameters.

\subsection{Minimal data set\label{subsection:data:minimal}}
From the lines commonly included in the High-density data set and \citetalias{Tanaka2018b} data sets, we selected five lines ($\mathrm{^{13}CO}$~2--1, HCN~1--0 and 4--3, and \pformaldehyde~$3_{0,3}$--$2_{0,2}$ and $3_{2,1}$--$2_{2,0}$; listed in Table \ref{table:datasets}).
The \pformaldehyde\ $3_{2,1}$--$2_{2,0}$/$3_{0,3}$--$2_{0,2}$ intensity ratio is a good \Tkin\ probe as mentioned above, which effectively resolves the \Tkin--\nHH\ degeneracy in the HCN~4--3/1--0 ratio. 
Thus, this data set is expected to allow physical condition measurements of the high-density component with low degeneracy. 
Note that this is not the unique selection of the smallest data set required for a non-degenerate solution in general; 
however, we refer to this data set as ``Minimal'' since it comprises the minimum necessary transitions that can be constructed from the intersection of the High-density data set and the \citetalias{Tanaka2018b}\ data set.

All transitions qualify the same criteria for the High-density data set except for $^{13}\mathrm{CO}$ 2--1, which does not satisfy the criteria 1 and 2.
We include $^{13}\mathrm{CO}$ 2--1 to use it as an approximate \NHH\ tracer independent of the HCN and \formaldehyde\ intensities, similarly to the \citetalias{Tanaka2018b} analysis.
Although the $^{13}\mathrm{CO}$ 2--1 line may not be an ideal tracer of the high-density component due to its low $\ncrit$ of $\sim 10^{3.8} \pcc$, it is still useful for an approximate measure of \NHH\  on the assumption that the high-density and low-density components have similar large-scale mass distribution patterns in the \PPV\ space.
Indeed, \citetalias{Tanaka2018b} found good correlations between intensities of representative high-density tracer lines (HNC~1--0 and $\mathrm{HCO^+}$~1--0) and $^{13}\mathrm{CO}$~2--1-based column density for the GC. 
We will verify correlation between \NHH\ of the low- and high-density components for the NGC~253 CMZ, based on the analyses using the Low- and High-density data sets in a later section (\S\ref{subsection:results:LowVsHigh}).
More accurate \NHH\ tracers, such as higher-$J$ $^{13}\mathrm{CO}$ lines or $^{13}\mathrm{C}$ isotopologues of high-density tracer lines, are not available for the GC or do not satisfy criterion 5.

The Minimal data set for the GC is prepared using the data compiled from \cite{Jones2012}, \cite{Ginsburg2016}, and \citetalias{Tanaka2018b}. 
All GC data are re-gridded to the same angular and velocity griddings and resolutions as those in \citetalias{Tanaka2018b}, i.e., $\ang{;;30}\times\ang{;;30}\times10~\kmps$ and $\ang{;;60}\times\ang{;;60}\times10~\kmps$, respectively.
The angular resolution of \ang{;;60} corresponds to 2.4~pc at the distance to the GC \citep[8.18 kpc; ][]{GravityCollaboration2019}.
We refer readers to \citetalias{Tanaka2018b} for a detailed data description.

Figure \ref{fig:LVGmin} shows an example of the joint posterior probability of \NHH, \Tkin, and \nHH\ calculated using the averaged intensities of the Minimal data sets over the central \ang{;;10} region.
The posterior probability shows two local maxima.
The lower-\NHH\ solution at $\dNHHdv\sim 10^{20}\ \mathrm{cm}^{-2}\left(\kmps\right)^{-1}$ is not consistent with the High-density result (Figure \ref{fig:LVGhigh}), and hence the higher-\NHH\ peak at $\dNHHdv\sim 10^{21.5}\ \mathrm{cm}^{-2}\left(\kmps\right)^{-1}$ is likely to represent the true physical conditions of the high-density component.
The higher-\NHH\ solution shows $\Tkin \sim 100\ \mathrm{K}$, which is approximately consistent with the High-density analysis, as \Tkin\ values are predominantly determined by the \pformaldehyde\ lines for both analyses.
The \nHH\ value is relatively poorly determined, whose 50\% interval spans over more than an order of magnitude around the higher-\NHH\ solution.
Hence, analysis using the Minimal data set is less accurate than that using the High-density data set, reflecting the limited number of the input transitions in the former.
We also note that the HCN and \pformaldehyde\ lines may be biased to relatively high-temperature gas likely associated with turbulence (\citetalias{Tanaka2018b}).  In this sense, the High-density data set better represents the entire high-density component that includes both turbulent and quiescent regions than the Minimal data set.

\subsection{Fixed Parameters\label{subsection:data:fixedParams}}
We have to fix the abundance of one molecular species in each data set to use it as a proxy for $\mathrm{H_2}$.
We assume $\xmol{CO} = 10^{-4}$ in the analysis with Low-density and Minimal data sets, i.e., the same assumption as in \citetalias{Tanaka2018b}.
We use HNC to measure \NHH\ for the High-density data set as HNC is considered as a ``neutral'' species in the GC, in the sense that HNC~1--0 is the best correlated with the dust 500~\micron\ flux in the Band-3 lines of the high dipole moment molecules analyzed by \citetalias{Tanaka2018b} (HNC, HCN, $\mathrm{HCO^+}$, CS, SiO, \HCCCN, $\mathrm{N_2H^+}$, and \pformaldehyde).
\citetalias{Tanaka2018b} has also shown that HNC is one of the species with the least spatial variation in the fractional abundance ($\sim 0.1$ dex in rms at a 27 pc resolution) in the GC.
For the NGC~253 CMZ, \cite{Behrens2022ApJ} reported a factor of $\sim 2$ variation in the [HCN]/[HNC] abundance ratio among the major GMCs.  
This [HCN]/[HNC] variation is similar to that in the GC (\citetalias{Tanaka2018b}), which would allow us to assume that \xmol{HNC} variation in the NGC~253 CMZ is also similar to that in the GC.
We assume $\xmol{HNC} = 4\times10^{-8}$ for the NGC~253 CMZ, which is estimated by multiplying the typical \xmol{HNC}\ in the GC, $2\times10^{-8}$ (\citetalias{Tanaka2018b}), by the NGC~253-to-GC ratio of the HNC~1--0/$\mathrm{^{13}CO}$~2--1 intensity ratio, 2.
We will verify this assumption by comparing the derived dense gas mass to the previous CO SLED results (\S\ref{subsubsection:discussion:massEstimates}).
Breakdown of the constant \xmol{HNC}\ approximation is a potential source of systematic errors in \NHH, but it does not critically affect the excitation analysis since the line intensities depend on the column densities of individual molecules, not on \NHH.

We fix $R_{13} = 21$ in the analysis using the Low-density data set as already mentioned in \S\ref{subsection:data:lowdensity}.
The same $R_{13}$ value is applied to the Minimal data sets both for NGC~253 and the GC, which is used to translate $N\left(\mathrm{^{13}CO}\right)$ into \NHH.
Although the GC has a slightly higher $R_{13}$ value of $\sim 25$ (\citealt{Langer1990,Langer1993}; \citetalias{Tanaka2018b}), we ignore the difference to make the comparison between the two galaxies easier.
We have to consider this simplification and the uncertainty in \xmol{CO}\ when we compare \NHH\ of NGC~253 and the GC.
We also fix the [$\mathrm{C^{18}O}$]/[$\mathrm{^{12}CO}$] abundance in NGC~253 at the value in \cite{Martin2019}, i.e., $\xmol{\mathrm{C^{18}O}} = 7.8\times10^{-7}$, for the purpose of consistency.

\section{Results\label{section:results}}

\subsection{Analysis Runs\label{subsection:results:analysisRuns}}

\begin{deluxetable*}{lllll}
\tablecolumns{6}
\tablehead{\colhead{Run Name} & \colhead{data set} & \colhead{Statistical Modeling} & \colhead{Fixed Abundance} & \colhead{$R_{13}$} }
\tablecaption{Analysis runs\label{table:analysisruns}}
\startdata
High-HB &  High-Density  & HB      & $\xmol{HNC} = 4\times10^{-8}$  & Variable \\
High-NHB &  High-Density  & NHB   & $\xmol{HNC} = 4\times10^{-8}$  & Variable \\
Low-HB  &  Low-Density   & HB     & $\xmol{CO} = 10^{-4}$, $\xmol{C^{18}O} = 7.8\times10^{-7}$    & 21 \\
Low-NHB  &  Low-Density   & NHB  & $\xmol{CO} = 10^{-4}$, $\xmol{C^{18}O} = 7.8\times10^{-7}$    & 21 \\
Min-HB  &  Minimal       & HB     & $\xmol{CO} = 10^{-4}$    & 21 \\
GC-HB  &  Minimal (MW)  & HB     & $\xmol{CO} = 10^{-4}$    & 21 
\enddata
\end{deluxetable*}

We ran six analyses using the two statistical modelings (HB and NHB) and three data sets (High-density, Low-density, and Minimal; abbreviated as High, Low, and Min, respectively) described in the previous sections.
Table \ref{table:analysisruns} tabulates the combination of the statistical modelings and data sets used in individual analysis runs.
We conducted both HB and NHB analyses for each of the High- and Low-density sets to verify the effect of the prior probability.
Only the HB analysis was applied to the Minimal sets of the ALCHEMI and MW data; the NHB analyses was not necessary for them, since their validity can be verified through comparison with the results with the High-density data set and the \citetalias{Tanaka2018b} results.
We refer to individual analysis runs by combination of the abbreviations of the names of the data set and the statistical framework used, e.g., ``High-HB'' for the HB analysis with the High-density data set, and so on.
Table \ref{table:analysisruns} also summarizes the fixed parameters in individual analysis runs described in \S\ref{subsection:data:fixedParams}.

For each analysis run, the marginal posterior function $\prob{\mathcal{P}|\mathcal{V}}$ (Equations \ref{eq:marginalposteriorH} and \ref{eq:marginalposteriorNonH}) was numerically calculated by employing the Markov-Chain Monte Carlo (MCMC) method, where the hybrid Monte Carlo method \citep{Duane1987} was employed;   
details of the sampling method are provided in \citetalias{Tanaka2018b}.
Examples of the MCMC results including trace plots and corner plots of the posterior function are provided in Appendix \ref{appendix:C}.

The marginal posteriors of the individual $\mathcal{P}$ elements, i.e., all parameters at all spaxels, are calculated based on the MCMC results after convergence.
The marginal posteriors are reduced into \PPV{} cubes of individual parameters by adopting the medians and 50\% credible intervals as the representative value and uncertainty, respectively.
Poorly determined values with the 50\% credible intervals above 0.3, which corresponds to the relative uncertainty of $> 41\%$ in the linear scale, are excluded from the final cubes.

Table \ref{table:fluxCoverage} shows the flux coverage of the Low-HB and High-HB analyses, which we define as the fraction of molecular line fluxes included in the region where the HB analyses were successful (i.e., credible interval of \NHHbeam\ is less than 0.3 dex) to the total fluxes.
The total fluxes were measured over the spaxels where the CO~1--0 intensity is above $5\sigma$. 
We find that $\gtrsim 90\%$ of the fluxes of the primary gas mass tracers (the $\mathrm{^{13}CO}$ and HNC transitions) and 77--89\%\ of the fluxes of the CO transitions and the $J\ge$3--2 transitions of major high-density tracers (HCN, $\HCOp$, and HNC) were covered by the present analysis, with the lower excitation transitions having the lower flux coverages.
Table \ref{table:fluxCoverage} also shows that $\sim35$\% of the total flux of HCN~1--0 and \HCOp~1--0, which are frequently used dense-gas mass tracers, are missed by the High-HB analysis.
\begin{deluxetable}{llcc}
\tablecolumns{4}
\tablehead{\colhead{Analysis} & \colhead{Molecule} & \colhead{Transition} & \colhead{Flux Coverage} }
\tablecaption{Flux Coverage\label{table:fluxCoverage}}
\startdata
Low-HB & CO & 1--0  & 0.85 \\
       &    & 2--1  & 0.84 \\
       &    & 3--2  & 0.87  \\
       & $\mathrm{^{13}CO}$ & 1--0  &  0.92 \\
       &    & 2--1  & 0.94  \\
       &    & 3--2  & 0.97  \\
High-HB & HCN & 1--0  & 0.65 \\
       &    & 3--2  & 0.83 \\   
       &    & 4--3  & 0.89 \\
       & $\mathrm{HCO^+}$ & 1--0 & 0.64 \\
       &    & 3--2  & 0.83  \\
       &    & 4--3  & 0.89 \\
       & HNC  & 3--2  &  0.89 \\   
       & CS & 3--2  & 0.77 \\   
       &     & 6--5  & 0.88 \\  
\enddata
\end{deluxetable}

\myrev{We also evaluated the optical depths of the transitions used in the analysis from the parameters obtained from the Low-HB and High-HB analyses.  Detailed descriptions of the calculation and the results are provided in Appendix \ref{appendix:tau}}.

\subsection{Hierarchical Versus Non-hierarchical Analysis \label{subsection:results:HBvsNHB}}
\begin{deluxetable*}{@{\extracolsep{9pt}}lccccccccc}
\tablecaption{Analysis Results: Statistical Properties\label{table:stats}}
\tablecolumns{10}

\tablehead{
\colhead{} & \multicolumn{6}{c}{$\tilde{\mu}_i$\tablenotemark{$\dagger$}} & \multicolumn{3}{c}{$\tilde{R}_{ij}$} \\
\cline{2-7}  \cline{8-10} 
\colhead{$i$ or $i\,j$}&\colhead{$N$}&\colhead{$\left<N\right>$}&\colhead{$n$}&\colhead{$T$}&\colhead{$R$}&\colhead{$\phi$}&\colhead{${Nn}$}&\colhead{${N\phi}$}&\colhead{${Tn}$} \\
\colhead{unit}&\colhead{$\mathrm{cm}^{-2}\left(\mathrm{km\,s^{-1}}\right)^{-1}$}&\colhead{$\mathrm{cm}^{-2}\left(\mathrm{km\,s^{-1}}\right)^{-1}$}&\colhead{$\mathrm{cm}^{-3}$}&\colhead{$\mathrm{K}$}&\colhead{}&\colhead{}&\colhead{}&\colhead{}&\colhead{}}
\startdata
Low-NHB&$21.28$&$20.23$&$3.16$&$1.86$&$1.32$&$-0.92$&$0.11$&$-0.73$&$-0.68$\\
Low-HB&$21.37$&$20.19$&$2.95$&$1.86$&$1.32$&$-1.18$&$0.84$&$-0.02$&$0.48$\\
High-NHB&$20.69$&$19.91$&$4.62$&$2.20$&$1.72$&$-0.69$&$-0.72$&$-0.07$&$-0.42$\\
High-HB&$21.52$&$20.56$&$4.07$&$1.99$&$1.78$&$-0.95$&$0.13$&$-0.15$&$0.76$\\
Min-HB&$21.10$&$20.18$&$3.86$&$1.96$&$1.32$&$-0.91$&$0.46$&$-0.01$&$0.68$\\
GC-HB&$21.25$&$20.68$&$3.99$&$1.92$&$1.32$&$-0.50$&$0.56$&$-0.03$&$0.12$\\
\enddata
\tablecomments{Median values $\tilde{\mu}_i$ and correlation coefficients $\tilde{R}_{ij}$ of the parameters.
$N$, $\left<N\right>$, $n$, $T$, and $R$ denote 
\dNHHdv, \dNHHbeam, \nHH, \Tkin, and $R_{13}$, respectively.
}
\tablenotetext{\dagger}{In base-10 logarithm scale}
\end{deluxetable*}

In addition to the \PPV\ distributions of the individual parameters, 
we also calculated their medians ($\tilde{\mu}_i$) and the covariance matrix ($\tilde{R}_{ij}$) at every MCMC steps, whereby constructing their marginal posteriors.
Similarly to $\mathcal{P}$, the median and 50\% credible intervals of the marginal posteriors are adopted as the representative values and uncertainties of these statistical properties.
Table \ref{table:stats} shows $\tilde{\mu}_i$ of \dNHHdv, \dNHHbeam, \nHH, \Tkin, $\phi$, and $\tilde{R}_{ij}$ for important parameter pairs.
Note that they are similar to, but not equal to the hyperparameters $\vec{\mu}$ and $R$, which are the location and the correlation coefficient parts of the scale matrix $\Sigma$ of the log-Student prior function (see Appendix \ref{appendix:A} for their definitions),
as the actual parameter distribution is not determined only by the truncated log-student prior distribution, but by the whole posterior function.

Table \ref{table:stats} shows that the NHB results (High-NHB and Low-NHB) show large negative correlation coefficients between several parameters: $\tilde{R}_{Nn}$ and $\tilde{R}_{Tn}$ in the High-NHB result, and $\tilde{R}_{N\phi}$ and $\tilde{R}_{Tn}$ in the Low-NHB, reflecting their degeneracy in the excitation equations.
These correlation coefficients are positive or only slightly negative in the High-HB and Low-HB analyses compared with their respective NHB counterparts, indicating that the HB analyses suppressed the anti-correlation from the parameter degeneracy.

Another significant difference is identified in $\tilde{\mu}_i$ for \dNHHdv, \nHH, and $\phi$ of the High-HB and High-NHB results; in particular, $\tilde{\mu}_N$ and $\tilde{\mu}_n$ are 0.83 dex higher and 0.55 dex lower in High-HB than in High-NHB, respectively.
This inconsistency likely originates from the \NHH--\nHH\ degeneracy in the excitation equations with the High-density set, which is exemplified in Figure \ref{fig:LVGhigh}.
A similar but smaller difference in \dNHHdv\ and \nHH\ is also present between the Low-HB and Low-NHB results.
However, the combination of the low \dNHHdv\ and high \nHH\ in the High-NHB result yields too small a spatial scale of $\NHH/\nHH = 3.8\times10^{-3}\ \left(\frac{\Delta v}{\kmps}\right)\mathrm{pc}$, where $\Delta\,v$ is the typical velocity width.
If we assume a typical cloud radius ($r$) of 0.1--10~pc, this $\NHH/\nHH$ yields that a typical velocity dispersion (\sigmav) of 26--2600 \kmps, which is obviously too large compared with the $r$--\sigmav\ relation for the NGC~253~CMZ \citep{Krieger2020}.
The same estimate using the High-HB result yields $\NHH/\nHH = 9.1\times10^{-2}\ \left(\frac{\Delta v}{\kmps}\right)\mathrm{pc}$.
This size scale gives \sigmav\ of 1--100\ \kmps\ for $r$ of 0.1--10~pc, which better fits the $r$-\sigmav\ relation.

The above comparison indicates that the HB analysis was able to provide more reasonable \PPV\ distributions of the physical conditions than the standard NHB analysis.   
Therefore, we adopt Low-HB and High-HB as the final results for the low- and high-excitation components in the rest of this article.
However, the validity of the hierarchical framework needs verification, as 
the hyperprior function enforcing non-negative correlations between \NHH, \Tkin, \nHH, and $\phi$ possibly creates artificial parameter distributions in the HB results.  
The Low-HB results need particularly careful verification, since the Low-HB analysis relies on the prior and hyperprior probabilities to resolve the severe parameter degeneracy shown in Figure \ref{fig:LVGlow}.
The effect of the logistic hyperprior is discussed in section \S\ref{subsection:results:hyperPrior}.

\subsection{Distributions of \NHH, \nHH, and \Tkin\label{subsection:results:maps}}

Figures \ref{fig:mapLowH}, \ref{fig:mapHighH}, and \ref{fig:mapMinH} show the \PPV\ distributions of the Low-HB, High-HB, and Min-HB results, respectively.
The results of the non-hierarchical analyses for the NGC~253 CMZ (Low-NHB and High-NHB) and the HB analysis for the GC (GC-HB)  are presented in \myrev{Appendix \ref{appendix:D}}.
The cubes of \NHHbeam, \nHH, and \Tkin\ are projected onto the \PP\ and \PV\ 2-D images.
The \PP\ image of \NHHbeam\ shows velocity integration of \dNHHbeam, and the other \PP\ and \PV\ images show the \NHHbeam-weighted averages of the non-blank spaxels along the respective projection axes \footnote{\NHHbeam-weighted average of parameter $X$ means $\frac{\sum \Phi_i N_i X_i}{\sum \Phi_i N_i}$, where $\Phi_i$, $N_i$, and $X_i$ stand for the $i$th spaxel values for $\Phi$, \dNHHbeam, and $X$, respectively, along the projection axis.}.
The position axis of \PV\ images is taken in the major axis direction indicated in the figure, whose position angle is $55^\circ$ \citep{Leroy2015}.
All \PP\ and \PV\ images of the Low-HB and High-HB results show smooth continuous distribution though the analysis is on the per spaxel basis.
The Min-HB images are less smooth than the Low-HB and High-HB images, having a non-negligible number of discontinuous pixels near emission peaks; this discontinuity is created by the blank spaxels whose 50\% credible interval exceeds 0.3 due to insufficient decoupling of the model parameters.

The \NHHbeam\ distributions basically follows the intensity distributions of the \NHH\ tracers ($\mathrm{^{13}CO}$, $\mathrm{C^{18}O}$, and HNC lines).
The high--\NHHbeam\ region is concentrated in the central $\pm\ang{;;10}$ region along the major axis of the CMZ, which harbors stellar clusters and star-forming dense clumps \citep{Sakamoto2011,Leroy2018,Mills2021} (referred to as the ``central starburst region'' hereafter).
The highest \NHHbeam\ is found toward the position of GMC6 identified by \cite{Leroy2015} (denoted by an open triangle in Figures \ref{fig:mapLowH} and \ref{fig:mapHighH}) in the Low-HB and High-HB results,
whereas no particular \NHHbeam\ peak is present toward the TH2 position (denoted by a cross in Figures \ref{fig:mapLowH}, \ref{fig:mapHighH}, and \ref{fig:mapMinH}) at the resolution of the ALCHEMI data in all results.
The central starburst region has also higher \Tkin\ and \nHH\ than the outer region.
The \Tkin\ enhancement in the central starburst region is approximately consistent with the spatial distribution of the CRIR measured by \cite{Behrens2022ApJ}, which could support the picture that the cosmic-ray heating dominates the molecular gas heating in the NGC\,253 CMZ \citep{Behrens2022ApJ}.

We may consider the possibility that the co-enhancement of \nHH\ and \Tkin\ in the high-\NHH\ region is artificially induced by the logistic hyperprior function for $R$, which explicitly forbids anti-correlation between these parameters. 
In particular, the Low-HB result shows a remarkably tight \NHH--\nHH\ correlation coefficient of 0.84, which is close to the maximum allowed by the hyperprior function.  
This high correlation coefficient possibly indicates an insufficient decoupling of \dNHHdv\ and \nHH\ in the excitation analysis, since the hierarchical hyperprior gives higher values for solutions with higher \NHH--\nHH\ correlation when the two parameters degenerate in the excitation equations.
However, the overall \nHH\ distribution in the Low-HB results is similar to that in the High-HB results as we will see in detail in the following subsections (\S\ref{subsection:results:high-nT},\ref{subsection:results:LowVsHigh}).
Since the High-density data set does not suffer the parameter degeneracy as we have seen in \S\ref{subsection:data:highdensity}, this similarity between the Low-HB and High-HB results suggests that the artificial effect on the \nHH\ and \Tkin\ result is limited, though the correlation coefficient between \NHH\ and \nHH\ is possibly overestimated for the Low-HB result.
In a later subsection (\S\ref{subsection:results:hyperPrior}), we will also show that the effect of the hierarchical prior function on the analysis results is not critical at least for the High-HB results based on analysis of the PDF of the $R$ matrix.

\subsection{High-temperature Features\label{subsection:results:high-nT}}

\begin{figure*}
\epsscale{1.0}
\plotone{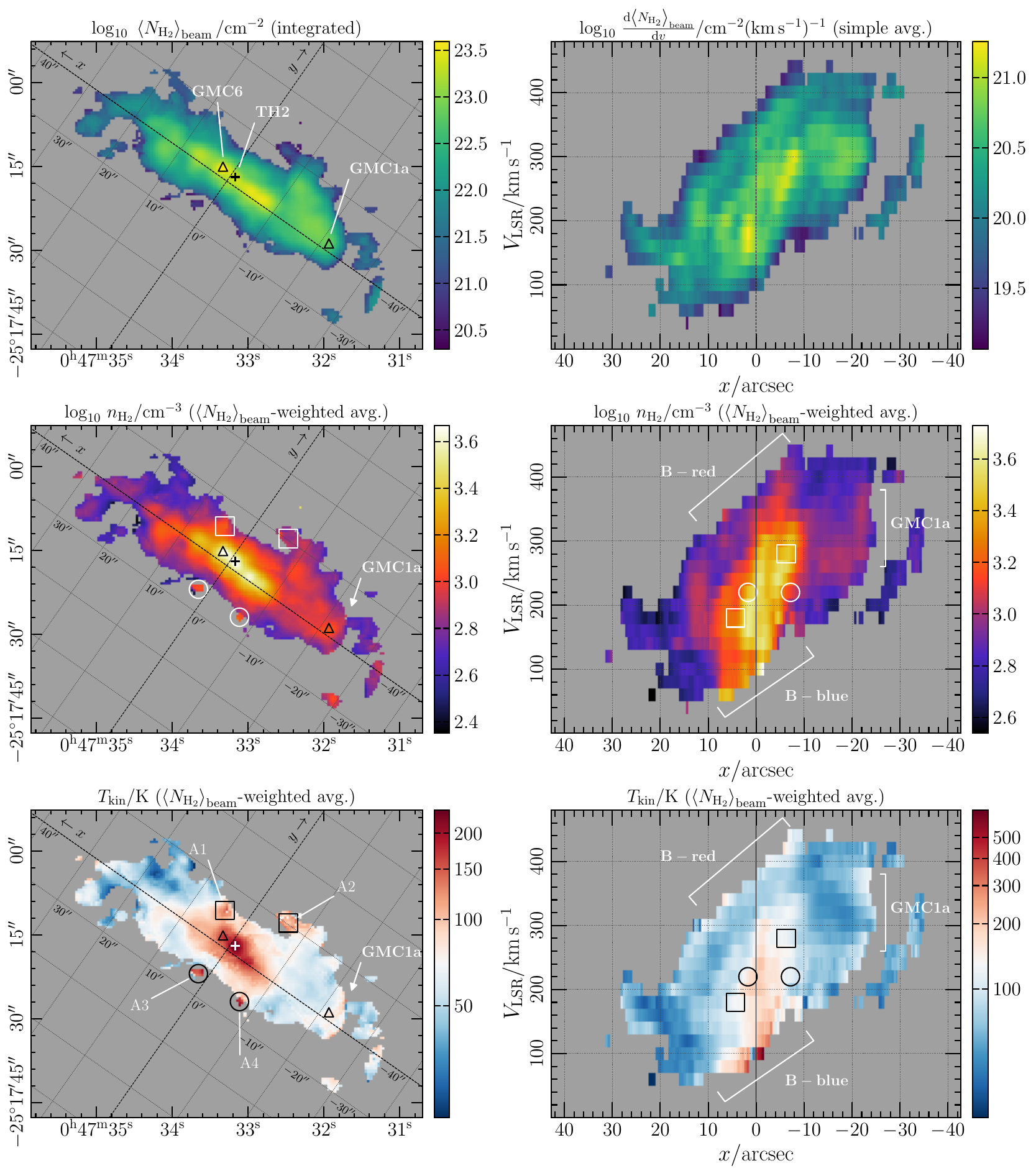}
\caption{Position--position (left column) and position--velocity (right column) images of \NHHbeam, \nHH, and \Tkin\ of the Low-HB run (from top to bottom). The relative coordinate from the phase center of the ALCHEMI observation $(\alpha_\mathrm{ICRS}, \delta_\mathrm{ICRS}) =  (0^\mathrm{h}47^\mathrm{m}33^\mathrm{s}.26, -25^\circ17'17''.7)$ \citep{Martin2021A&A} along the major and minor axes is overlaid on the \PP\ maps with a \ang{;;10} grid spacing.  The position angle of the major axis is $55^\circ$.
 The \PP\ and \PV\ images of \NHHbeam\ show the velocity-integration and latitude-average of $\frac{\mathrm{d}\NHH}{\mathrm{d}v}$ over non-blank \PPV\ spaxels, respectively.   The other \PP\ and \PV\ images show \NHH-weighted averages along the respective projection axes. 
 The positions of TH2 and two GMCs (GMC1a and GMC6; their positions are taken from \citet{Huang2023AAp} and \citet{Leroy2015}, respectively)  are denoted by cross and open triangle marks, respectively, in the \PP\ images. 
The labels A1 through A4 show the \PPV\ positions of the high-\Tkin\ features described in \S\ref{subsection:results:high-nT}.  The off-plane high-\Tkin\ spots are denoted by open and closed rectangles for the northern (A1 and A2) and southern (A3 and A4) spots, respectively, both in the \PP\ and \PV\ images.
 \label{fig:mapLowH}}
\end{figure*}

\begin{figure*}
\epsscale{1.1}
\plotone{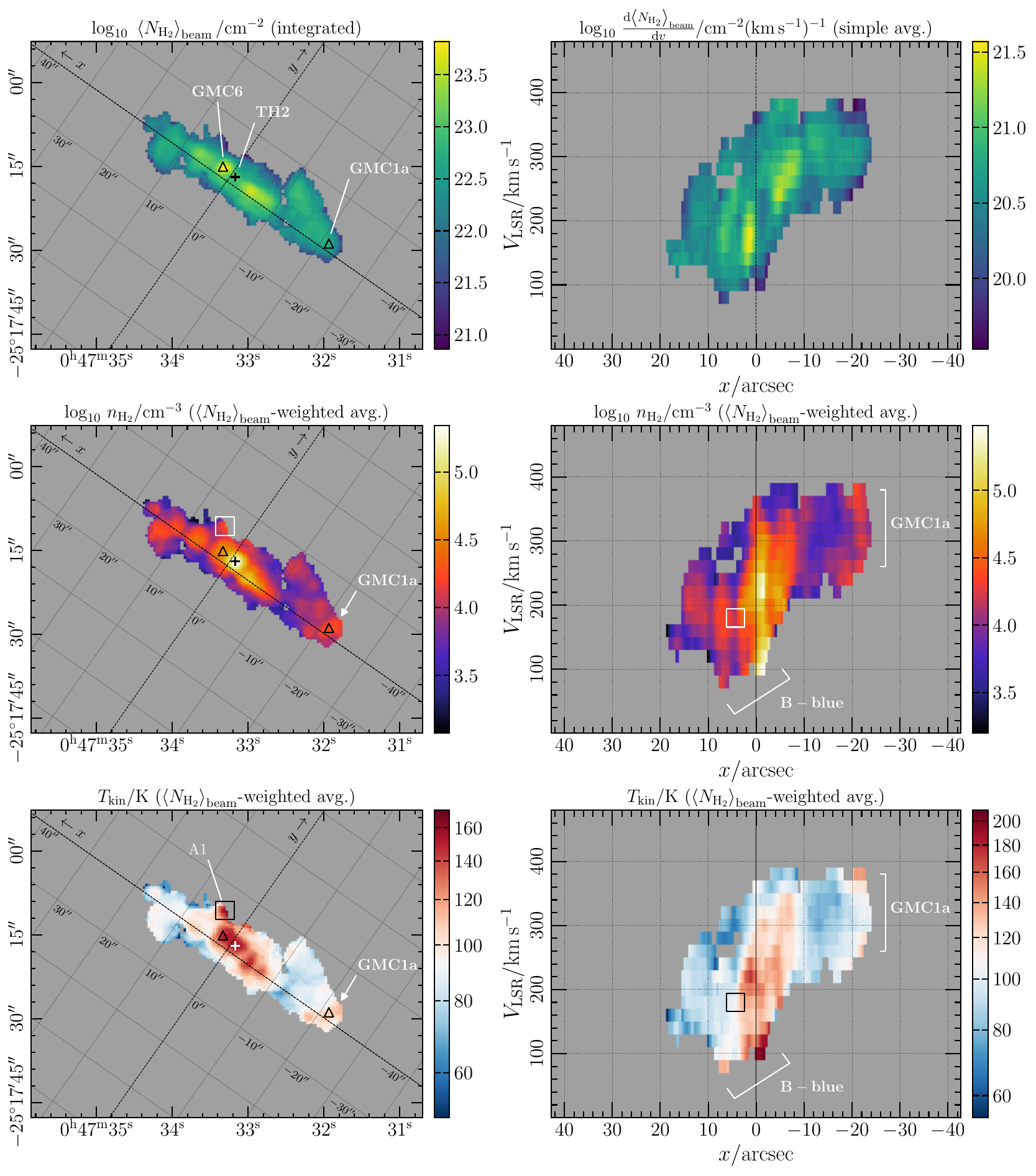}
\caption{Same as Figure \ref{fig:mapLowH}, but for the High-HB result. \label{fig:mapHighH}}
\end{figure*}

\begin{figure*}
\epsscale{1.1}
\plotone{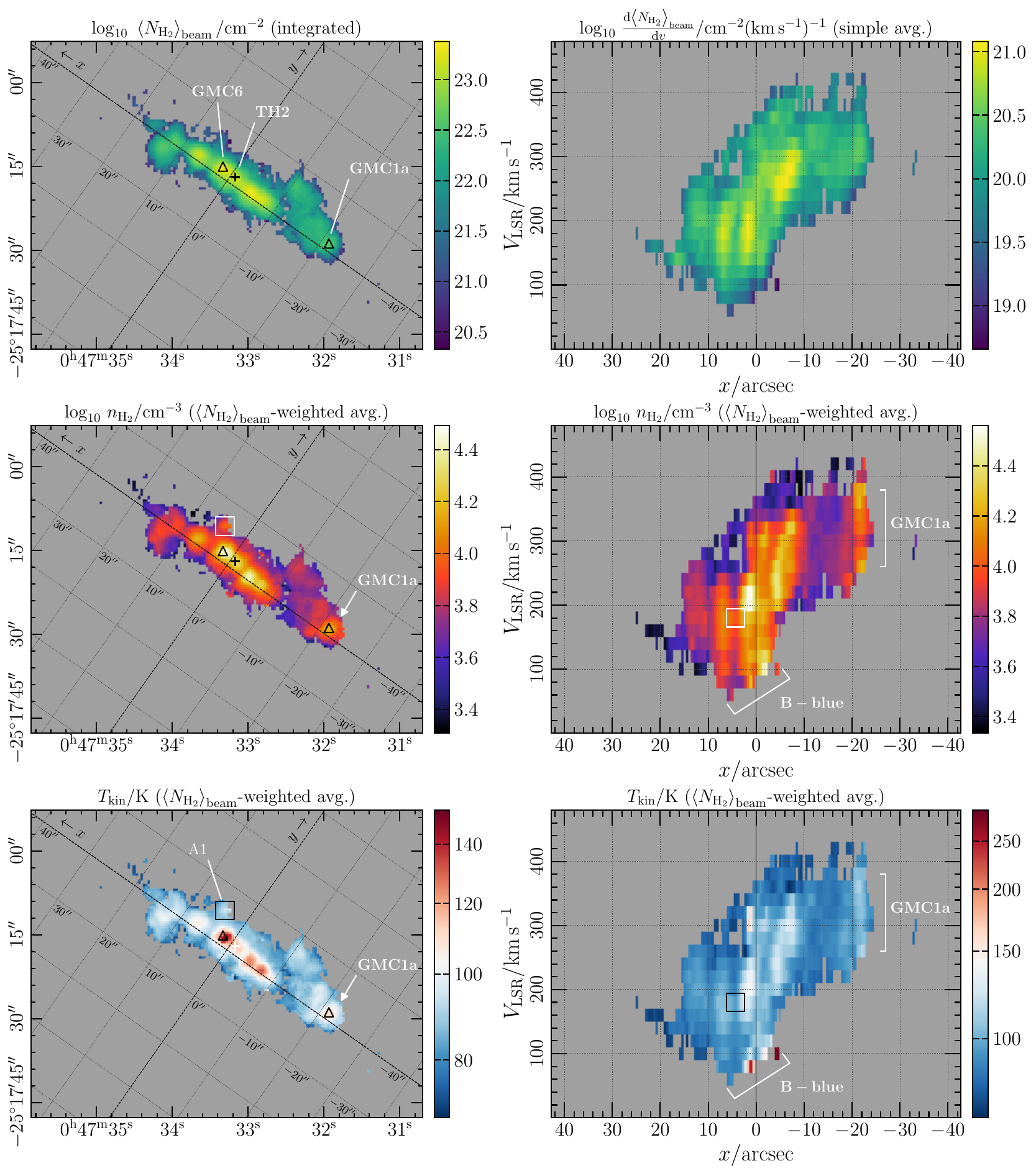}
\caption{Same as Figure \ref{fig:mapLowH}, but for the Min-HB result. \label{fig:mapMinH}}
\end{figure*}

Both in the Low- (Figure \ref{fig:mapLowH}) and High-density (Figure \ref{fig:mapHighH}) results, high-\Tkin\ regions are predominantly found in the high-\NHHbeam\ spaxels toward the central starburst region. 
This overall \Tkin\ distribution is reasonably explained as due to cosmic-ray heating as discussed in \S\ref{subsection:results:maps}.
Meanwhile, we have detected several high-\Tkin\ features outside the central high-\NHH\ region either in space or velocity, which are not associated with known heating sources.
In the following we briefly describe these high-\Tkin\ features commonly found in the Low-HB, High-HB, and Min-HB results, which are labeled as A1--A4, B (red and blue), and GMC1a in Figures \ref{fig:mapLowH}--\ref{fig:mapMinH}.
The \PPV\ positions of A1--A4 and GMC1a are shown in both the \PP\ and \PV\ images, whereas B-red and -blue are shown only in the \PV\ images.   

The \PP\ image of \Tkin\ in Figure \ref{fig:mapLowH} shows four high-\Tkin\ spots that are approximately $\pm$\ang{;;5}--\ang{;;10}\ from the major axis of the CMZ, which are  labeled A1--4.
Spot A1 also appears in the High-HB and Min-HB maps (Figures \ref{fig:mapHighH}--\ref{fig:mapMinH}), while the other three spots are outside the spatial coverage of the High-density and Minimal data sets.
Their projected locations, apparently at high latitudes from the CMZ plane, may indicate that they are shocked gas at the interface between the molecular gas and the large-scale outflow \citep{Bolatto2013}.  
Indeed, A2 and A4 coincide with the SW and NW streamers, respectively, which consists the western edge of the molecular outflow.

Another prominent high-\Tkin\ feature is found in the lowermost LSR velocities ($\sim$ 100--200 \kmps) in the central starburst region in the \PV\ diagram of the Low-HB and High-HB results, which is labeled B-blue in the figure. 
Its red-shifted counterpart, B-red, is also found in the Low-HB \PV\ diagram of \Tkin\ in 300--450\ \kmps\ velocities.
Their spatial distributions are consistent with that of the main velocity component of the central starburst region in 200--300\ \kmps\ velocities, suggesting that this high-velocity warm gas is associated with the shocked GMCs detected in the SiO and HNCO analyses \citep{Huang2023AAp}.

Finally, a high-\Tkin\ spot is detected toward GMC1a \citep{Huang2023AAp}, which is near the southwestern end of the CMZ.   
This feature is clearly visible in the \PP\ and \PV\ images of \Tkin\ of the High-HB and Min-HB results, where a pair of high-velocity gas with $\Tkin\sim$ 100--150 K is detected at the low- and high-velocity ends.
The GMC1a was detected as the high-\Tkin\ spot A7 in the \ammonia\ inversion line study by \cite{Gorski2017}.
The $\mathrm{H_2}$ and HNCO column densities from the High-HB result and \cite{Huang2023AAp}, respectively, shows that GMC1a has a remarkably high [HNCO]/[$\mathrm{H_2}$] ratio of $10^{-7.3}$, which is by more than an order of magnitude enhancement from the values in the GMCs in the central starburst region. 
\cite{Gorski2017} showed that the GMC1a/A7 is adjacent to an expanding shell-like structure of dense gas accompanied by class-I methanol masers.
Hence, the GMC1a could be mechanically heated by SNR shock or shocks induced by cloud--cloud collisions associated with the expanding shell.
The dynamical time scale of the expanding shell estimated from the radius (30 pc) and the expanding velocity ($\sim60$ \kmps), $0.5\times10^5$ year, is approximately consistent with the shock time scale estimated by \cite{Huang2023AAp}. 

\subsection{Low-HB versus High-HB analyses \label{subsection:results:LowVsHigh}}

\begin{figure*}
\plotone{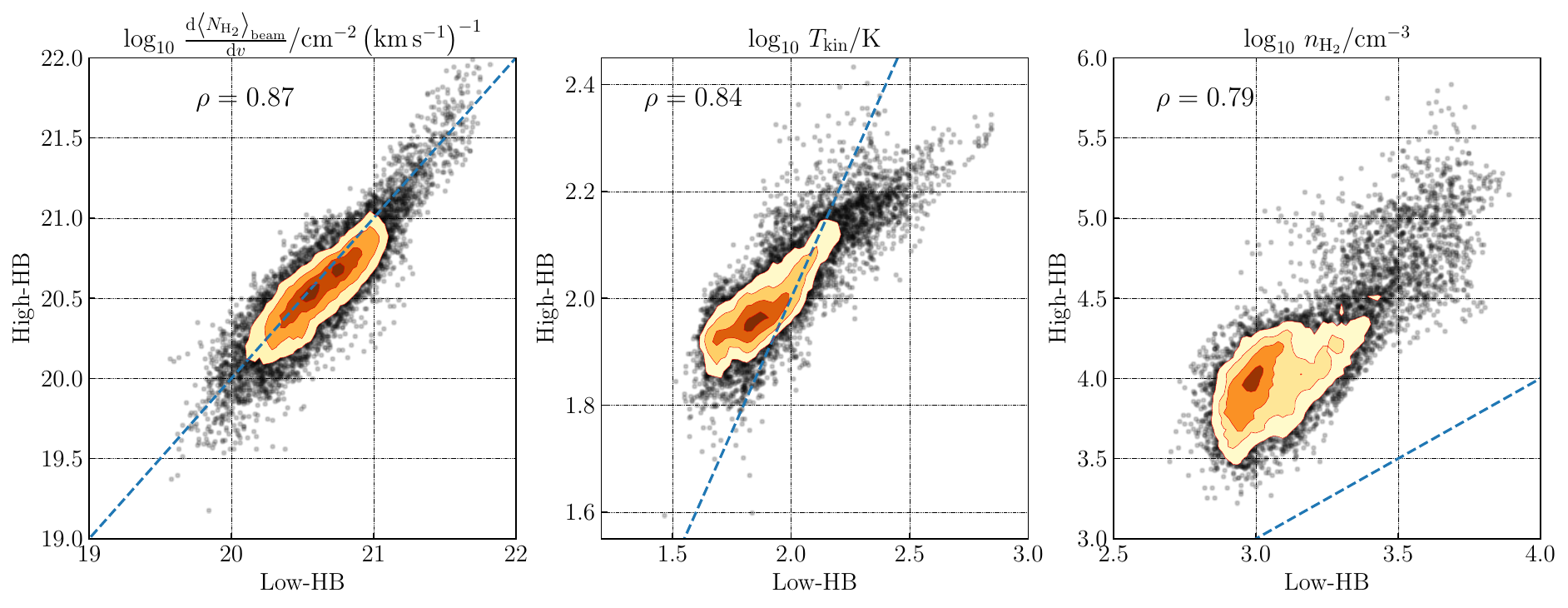}
\caption{Spaxel-by-spaxel scatter plots of the parameters (\NHHbeam, \Tkin, and \nHH) from the High-HB results vs. those from the Low-HB results.  The colored contours indicate 2-D histograms at relative frequencies of 5, 20, 50, and 70\%.   The correlation coefficient ($\rho$) is given in each panel. 
The dashed blue lines are for the same parameter values from the Low-HB and High-HB analyses. 
\label{fig:HighLowComparison}}
\end{figure*}

Figure \ref{fig:HighLowComparison} compares the \NHHbeam, \Tkin, and \nHH\ of the Low-HB and High-HB results in scatter plots.
All parameters show good spaxel-by-spaxel correlation with the correlation coefficient of $\gtrsim 0.8$, confirming the similarity in their \PPV\ distributions we saw in \S\ref{subsection:results:maps}.
The consistency between the independent results from different data sets corroborates the validity of the HB analysis.
The \nHH\ scatter plot shows that \nHH\ is higher for the High-HB results than for the Low-HB results at all spaxels. 
This confirms the assumption that the low- and high-density components do not spatially overlap in space, although they appear mixed in the 27-pc resolution.
The overall similarity of the parameter behaviors indicates that the low-density and high-density components trace the consistent molecular cloud structures, rather than being two isolated structures.

\begin{figure*}
\epsscale{1.}
\plotone{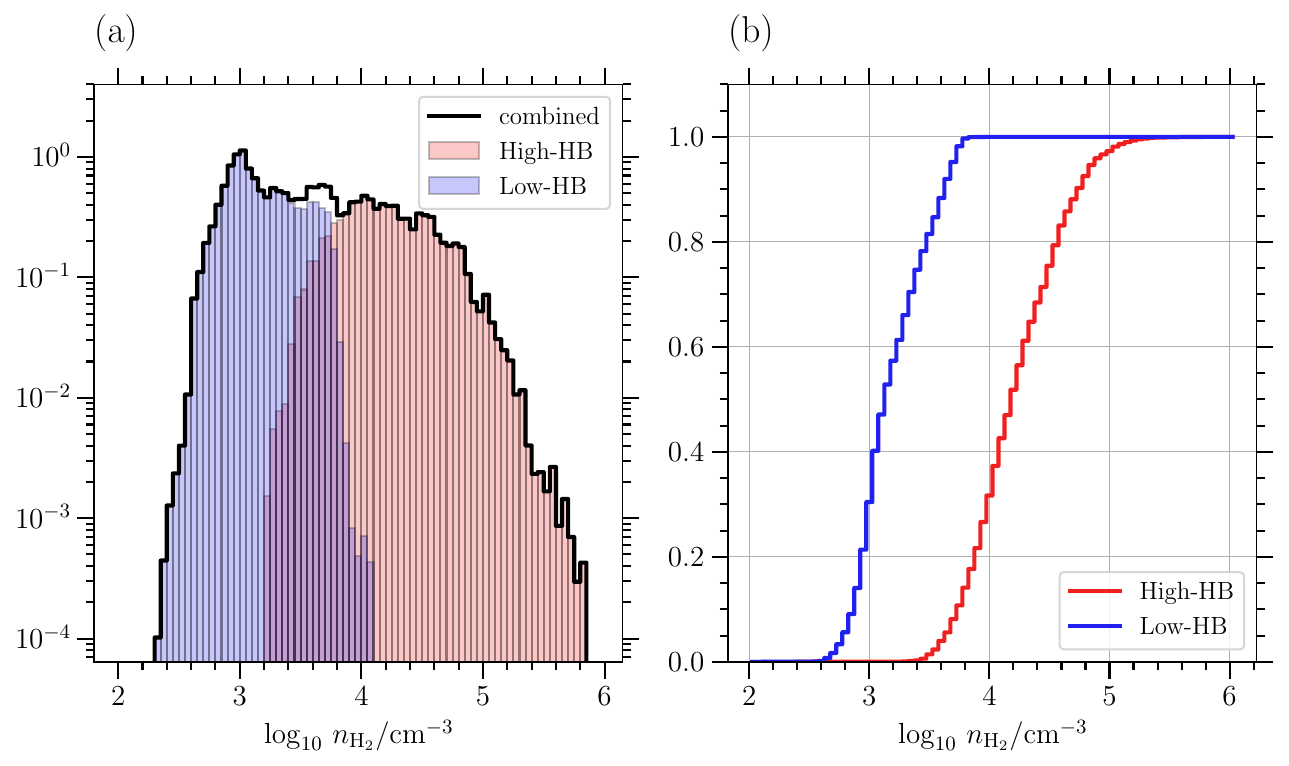}
\caption{(a) Frequency histograms of \nHH\ from the Low-HB (blue hatched) and High-HB (red hatched) results.  The combined histogram using the both results is shown in thick black lines.  All histograms are weighted by \NHHbeam\ and normalized by a common factor so that the integral of the combined histogram is unity.  (b) Cumulative forms of the same \nHH\ histograms as \myrev{(a)} but without the combined frequency.  Normalization is independently applied to the Low-HB and High-HB plots so that the respective maxima are unity. \label{fig:nHistogram} }
\end{figure*}

Figure \ref{fig:nHistogram}a compares the histograms of \nHH\ for the Low-HB and High-HB results along with the combined histogram for the two results, where \myrev{the} frequencies are weighted by \NHHbeam\ and normalized by a common factor so that the integral of the combined histogram is unity.    
The figure shows that the mass fractions of the low- and high-density components are comparable at $\nHH\sim 10^{3.8}\ \pcc$, above which \nHH\ bins the high-density component dominates the gas mass.
The cumulative \nHH\ histograms for the Low-HB and High-HB results are presented in Figure \ref{fig:nHistogram}b.
The gas mass with $\nHH > 10^{3.8}$\ \pcc\ constitutes 87\%\ of the mass traced by the High-HB result.
Hence, we can consider $\nHH = 10^{3.8}\ \pcc$ as the boundary between the two components.

\subsection{High-HB versus Min-HB analyses \label{subsection:results:HighVsMin}}

\begin{figure*}
\plotone{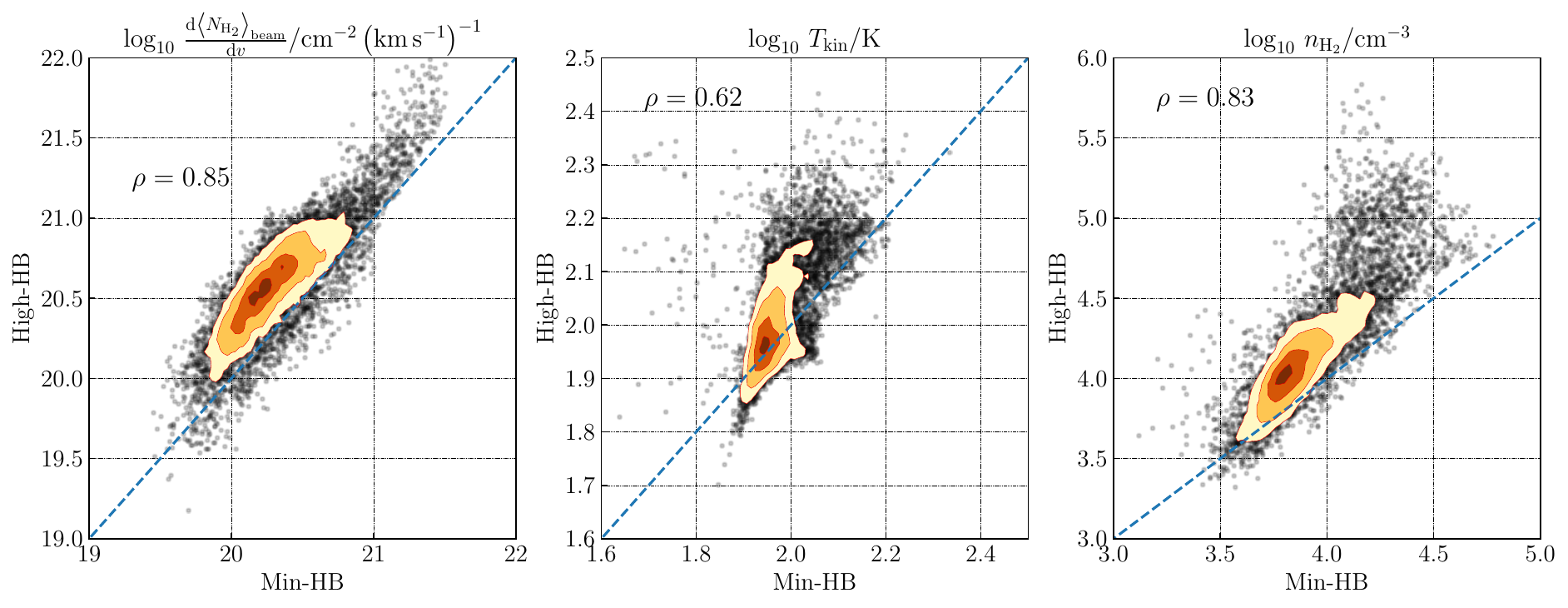}
\caption{Same as Figure \ref{fig:HighLowComparison} but for the High-HB results vs. Min-HB results  \label{fig:HighMinComparison}}
\end{figure*}

Figure \ref{fig:HighMinComparison} compares \NHHbeam, \Tkin, and \nHH\ of the High-HB and Min-HB results in scatter plots.
The two analyses target the same high-density component, and hence their results are ideally expected to agree with each other.
The \NHHbeam\ scatter plot follows a tight linear correlation between the High-HB and Min-HB though they use different \NHH\ tracers (HNC~3--2 and $\mathrm{^{13}CO}$~2--1, respectively). 
This tight correlation confirms that $\mathrm{^{13}CO}$~2--1 can be used to {\it qualitatively} measure the \NHH\ variation in the high-density component at the ALCHEMI resolution, as we assumed in \S\ref{subsection:data:minimal}.
On the other hand, the High-HB \NHHbeam\ values are consistently $\sim$ 0.2--0.3 dex higher the Min-HB values.  
This inconsistency indicates that $\mathrm{^{13}CO}$~2--1 intensity is inaccurate as a {\it quantitative} \NHH\ tracer for the high-density component.
We leave this inconsistency unsolved at present, but will compare the masses based on the High-HB and Min-HB results with previous CO SLED and dust SED measurements in a later section (\S\ref{subsubsection:discussion:massEstimates}).

The \Tkin\ scatter plot shows a relatively large dispersion, and the major part of the \Tkin\ values are distributed in a narrow range from $10^{1.9}$ to $10^{2.1}$\,K in the Min-HB result.
The \nHH\ scatter plot holds a good correlation, but the Min-HB analysis consistently yields $\sim 0.2$ dex lower values than the High-HB result.   
The \nHH\ range is also slightly narrower in the Min-HB result.
We note that the narrower variable ranges of \Tkin\ and \nHH\ in the Min-HB result may result from the stronger effect of the multivariate student prior, since the Min-HB results use a fewer number of observable intensities and hence the relative importance of the prior probability is larger.  
The multivariate student prior is a decreasing function of $|\Sigma|$, causing narrower parameter distributions (i.e., smaller variances or larger co-variances) preferred if other conditions are equal.
Hence, the High-HB result is likely to describe the dense-gas properties more reliably than the Min-HB result, since the former solution is more constrained by the observable information than by the prior probability.
Indeed, a significant number of spaxels near the emission peaks were filtered out from the final physical condition maps in the Min-HB analysis due to too broad credible intervals as we saw in \S\ref{subsection:results:maps} (Figure \ref{fig:mapMinH}); this indicates that the Minimal data set is insufficient to fully solve the parameter degeneracy in the excitation analysis.

The two analyses for the GC, the GC-HB results and the results from \citetalias{Tanaka2018b}, are compared in Appendix \ref{appendix:D}.

\subsection{\myrev{Molecular Abundances and [$^{12}$C]/[$^{13}$C] Ratios} \label{subsection:results:xmols}}
\begin{figure*}
    \centering
    \epsscale{1.1}
    \plotone{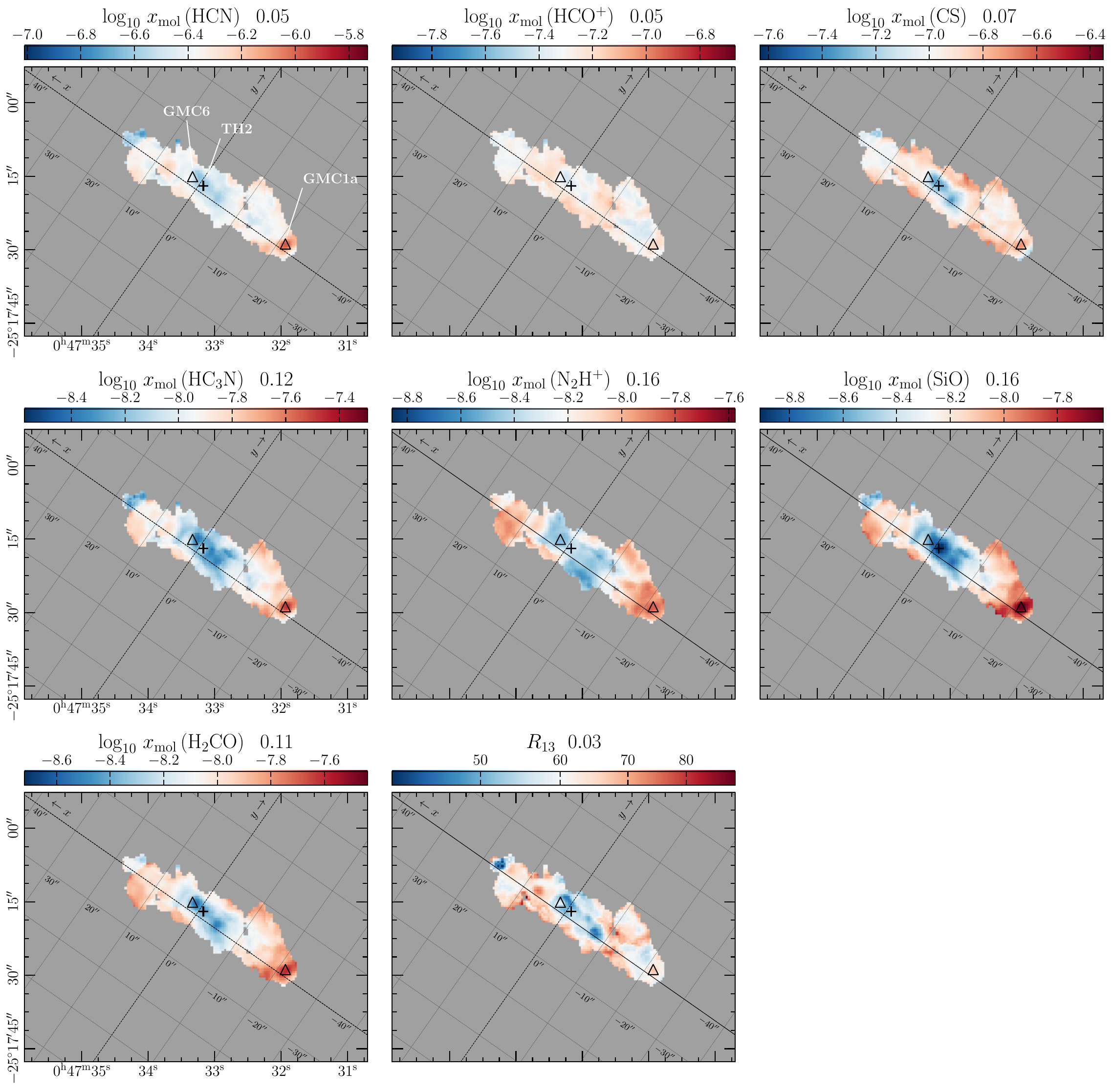}
    \caption{Velocity integrated \PP\ distributions of the molecular abundances ($x_\mathrm{mol}$) and the $\mathrm{[^{12}C]/[^{13}C]}$ isotopic abundance ($R_{13}$) from the High-HB result.  A common color bar range of 1.28 dex is used for all $x_\mathrm{mol}$ panels.  The median absolute deviation is shown in each panel captions. \label{fig:molabundance}}
    \label{fig:xmol}
\end{figure*}

As byproducts of the High-HB analysis, we obtained the $x_\mathrm{mol}$ distributions of HCN, HCO$^+$, HNC, CS, $\mathrm{HC_3N}$, $\mathrm{N_2H^+}$, SiO, \pformaldehyde, and $R_{13}$ in the high-density component.
Figure \ref{fig:molabundance} shows the \PP\ images of the velocity-integrated molecular abundances and $R_{13}$. 

HCN, \HCOp, and CS show relatively small abundance variations across the CMZ.
The other four species, $\mathrm{HC_3N}$, $\mathrm{N_2H^+}$, SiO, \pformaldehyde,  have similar spatial distributions, which increase outside the central starburst region and reach maximum at the GMC1a near the southwestern edge of the CMZ.   
This trend is most clearly visible in SiO, which is an established tracer of fast shocks.   
\HCCCN\ and \pformaldehyde\ are also tracers of slow shocks or known to be abundant in SF outflow sources \citep{Jorgensen2004,Tafalla2010} and in widespread shocked clouds in the GC (\citealt{Requena-Torres2006}; \citetalias{Tanaka2018b}).
Hence, the outwardly increasing distribution of their abundances may indicate  relative importance of the shock chemistry in the outer region of the CMZ compared to the central starburst region, where UV and CR sources concentrate.  
The compact peak of \xmol{SiO}\ and \xmol{\HCCCN}\ at GMC1a, in addition to the reported high HNCO abundance \citep{Huang2023AAp}, supports the suggestion that mechanical heating is important at GMC1a (\S\ref{subsection:results:high-nT}).
On the other hand, SF-related shocked gas in the inner GMCs \citep{Huang2023AAp} is not identified in the shock tracer distributions.
Our result, however, does not necessarily indicate low abundances of shock tracer species in the inner GMCs, since it is likely that the shocked gas in the innermost region is not fully visible due to the selection criteria for the High-density data set, which exclude high-excitation transitions locally distributed in the inner GMCs.  
We will discuss this issue later (\S\ref{subsubsection:discussion:prevAnalysis:otherALCHEMI}).

The [$^{12}\mathrm{C}$]/[$^{13}\mathrm{C}$] isotopic ratio $R_{13}$ is $\sim45$--50  toward the central portion and $\sim 60$ for the GMCs in the outskirt regions.
These values are close to those measured using CN \citep{Henkel1993,Martin2010,Tang2019},  CCH \citep{Martin2010,Martin2021A&A}, and $\mathrm{CH_3OH}$ \citep{Martin2021A&A} transitions.
The overall increasing gradient from the center to outskirts is similar to that seen in the $\mathrm{^{12}C^{18}O}$/$\mathrm{^{13}C^{18}O}$ ratio \citep{Martin2019}.
However, our $R_{13}$ value is approximately a factor of 2--3 higher than those obtained from the $\mathrm{^{12}C^{18}O}$/$\mathrm{^{13}C^{18}O}$ ratio \citep{Martin2019} and from individual fittings of the ALCHEMI ACA data of the HCN, HCO$^+$, and CS transitions \citep{Martin2021A&A}, which range from 21 to 26.
The inconsistency between the ACA result and ours, which use the same molecular transitions from the ALCHEMI data, arises from the one-zone modeling adopted in the present analysis.  
The High-HB analysis yields a lower \Tkin\ than the individual fittings of the HCN, \HCOp\ and CS transitions toward the central portion of the NGC~253 CMZ, as \Tkin\ is primarily determined by the \pformaldehyde\ $3_{21}$--$2_{20}$/$3_{03}$--$2_{02}$ ratio in the High-HB analysis.
The lower \Tkin\ leads to higher optical depths of the $^{12}\mathrm{C}$ main isotopologues, and hence to a higher $R_{13}$ value than the results of the individual fittings.
We leave this inconsistency open to question in this paper, since our analysis is not specifically designed to accurately measure $R_{13}$.
We just suggest that it is likely that $R_{13}$ has to be relatively high value of 45--60 when we adopt the one zone model.

\subsection{Verification of the Results of the Hierarchical Analysis\label{subsection:results:verification}}
We verify the validity of the HB analyses by investigating the effects of the assumptions in the statistical and physical modelings.
In the following we discuss the effects of the logistic hyperprior, uncertainty in the $\Phi$ measurements in the Low-HB result, and the species-to-species variation in the excitation temperatures.

\subsubsection{Effect of the logistic hyperprior\label{subsection:results:hyperPrior}}
The present analysis uses logistic hyperprior to enforce $R_{Nn} > 0$, $R_{N\phi} > 0$, $R_{NT} >0$, and $|R_{ij}| < 0.9$ for all ($i$, $j$) in the HB runs.
Except for the non-negative \NHH--\nHH\ correlation, these constraints are adopted for a computational reason rather than based on the physics of molecular clouds.
These arbitrary assumptions can create non-physical features in the analysis results.

\begin{figure*}
\plottwo{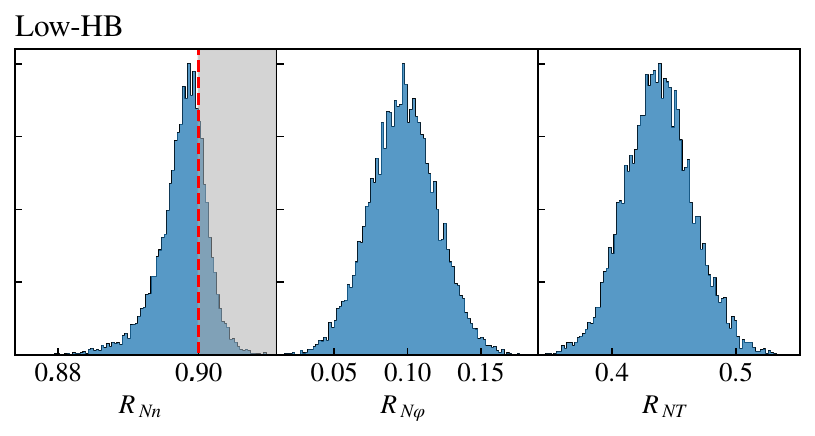}{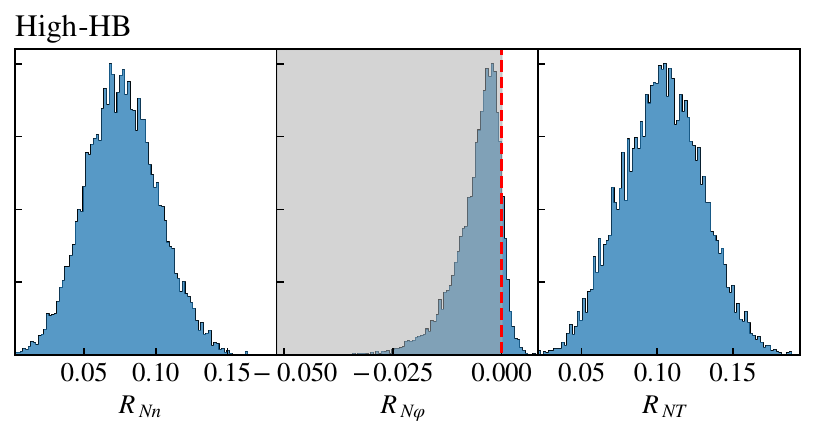}   
\caption{Marginal posteriors of $R_{ij}$ elements in the Low-HB (left panel) and High-HB (right panel) runs. The subscript indices $N$, $n$, and $T$ are for $\frac{\mathrm{d}{\NHH}}{\mathrm{d}v}$, \nHH, and \Tkin, respectively. The gray hatched area and red, vertical broken line denote the parameter ranges forbidden by the logistic hyperprior and their boundaries, respectively. \label{fig:hyperparameters}}
\end{figure*}

Figure \ref{fig:hyperparameters} shows the marginal posteriors of the hyperparameters $R_{Nn}$, $R_{N\phi}$ and $R_{NT}$ in the Low-HB and High-HB runs.
If their non-negative values are enforced by the logistic hyperprior, the function shapes are expected to be truncated or skewed at the boundaries imposed by the hyperprior.
Such a skewed shape is obtained for the $R_{Nn}$ PDF of Low-HB. 
This indicates that the \NHH--\nHH\ degeneracy was not sufficiently solved even in the HB framework.
As mentioned in \S\ref{subsection:results:maps}, the multivariate student prior prefers solutions with higher correlation coefficients (or smaller dispersions) if other conditions are equal.
Therefore, the HB analysis tends to create a tight correlation if the input data are insufficient to solve the parameter degeneracy.

The PDF of $R_{N\phi}$ in the High-HB result is skewed near $0$, indicating that \NHH\ and $\phi$ would be anti-correlated if the logistic hyperprior were absent. 
The median $\tilde{R}_{N\phi}$ (Table \ref{table:stats}) is also slightly negative even with the explicit constraint by the logistic hyperprior.
 It is difficult to determine whether this anti-correlation is real; 
artificial \NHH--$\Phi$ anti-correlation is easily created in solving the excitation equations, since all observable line intensities are approximated in the form $\propto \Phi\cdot\NHH$. 
For example, an overestimation in \NHH\ due to deficiency in the physical or statistical modeling should be compensated by decreasing $\Phi$.
However, we may consider a positive \NHH--$\Phi$ correlation is physically more reasonable, since higher-\NHH\ regions are likely to be more crowded with dense gas.  
For example, larger sizes of the sub-beam structures such as clumps and filaments should increase both $\Phi$ and \NHH, which are more likely to cause a positive \NHH--$\Phi$ correlation than a negative correlation.
If positive $R_{N\Phi}$ is the case in the real clouds in the high-density component, the High-HB analysis possibly overestimates \NHH\ (and hence underestimates \nHH\ and/or \Tkin) in the high-\NHH\ regions.

All other PDFs in Figure \ref{fig:hyperparameters} do not have skewed or truncated posterior PDFs, indicating that their corresponding ${R}_{ij}$ values are not enforced by the logistic hyperprior.
In particular, it is noteworthy that the moderately positive $R_{NT}$ values obtained for all HB analyses are likely to reflect the true physical conditions in the CMZ, rather than being enforced by the statistical modeling.

\subsubsection{Uncertainty in the Beam-Filling Factor\label{subsection:results:beamFillingFactor}}
The median $\Phi$ value in the Low-HB result, $0.064$, is lower than that in the High-HB result, 0.10, although we reasonably expect a wider spatial distribution for the lower-density component.
Indeed, the CO SLED analysis by \citetalias{Perez-Beaupuits2018a} shows that $\Phi$ of the first component is $\sim 5$ times that in the higher-density components measured for the central starburst region.
The smaller $\Phi$ in the Low-HB result may indicate an unsuccessful decoupling of \NHH\ and $\Phi$  due to the lack of the absolute intensities of optically thick lines in the Low-density data set.

We simulated the effect of underestimating $\Phi$ on the excitation analysis by performing NHB analyses using the Low-density data set for a range of fixed values of $\Phi$. 
The input data were the median values of the line intensities and intensity ratios.
The results are shown in Figure \ref{fig:beamFillingFactors}(a), where the medians of the marginal posteriors of \dNHHbeam, \nHH, and \Tkin\ are plotted as a function of $\Phi$ in the unit of $\tilde{\mu}_\Phi$ of the Low-HB result ($\Phi_\mathrm{LowHB} = 0.064$).  The error bars are for the median absolute deviations of the marginal posteriors.
Both the median \nHH\ and \Tkin\ do not significantly differ from the Low-HB results, and their dependence on $\Phi$ is within $\sim 0.2$ dex over the $\Phi/\Phi_\mathrm{LowHB}$ range of 1--5.  
This relative insensitivity of the \Tkin\ and \nHH\ estimates to the assumed $\Phi$ value is due to the fact that the Low-density set consists of absolute intensities of relatively optically thin lines ($\mathrm{^{13}CO}$ and $\mathrm{C^{18}O}$ transitions) and intensity ratios of optically thick lines ($\mathrm{^{12}CO}$ transitions), none of which depend strongly on $\Phi$.
The small variations in \Tkin\ and \nHH\ from $\Phi/\Phi_\mathrm{Low-HB}$ of 2 to 5 are due to the weak dependence of the intensity ratios among the $^{12}\mathrm{CO}$ transitions on \NHH. 
The $\mathrm{^{13}CO}$ transitions become moderately optically thick when $\Phi/\Phi_\mathrm{Low-HB}$ $\leq$ 2, causing the increase and decrease in \Tkin\ and \nHH, respectively, from $\Phi/\Phi_\mathrm{Low-HB}$ of 2 to 1. 
The maximum deviation from the $\Phi = \Phi_\mathrm{Low-HB}$ case is found at $\Phi/\Phi_\mathrm{Low-HB}$ = 2, at which \Tkin\ and \nHH\ are approximately 0.3 dex lower and 0.3 dex higher than the Low-HB values, respectively.
These variations due to the varying $\Phi$ are small compared to the width of the posterior resulting from the uncertainties in the input intensities, and therefore their effect on the final result should not be significant.

The assumption of $\Phi$ has a small effect of \NHHbeam, since the $\mathrm{^{13}CO}$ transitions become moderately optically thick for small $\Phi$.  The \NHHbeam\ is a factor of $\sim 1.5$ lower than the value with $\Phi = \Phi_\mathrm{Low-HB}$  when $\Phi/\Phi_\mathrm{Low-HB} = 5$.
This could be a source of the uncertainty in the mass estimate of the low-density component, along with the uncertainties in \xmol{CO} and $R_{13}$. 

\subsubsection{Uncertainty in the [${^{12}C}$]/[${^{13}C}$] Isotopic Abundance \label{subsection:results:R13}}
\begin{figure*}
\epsscale{0.9}
\plotone{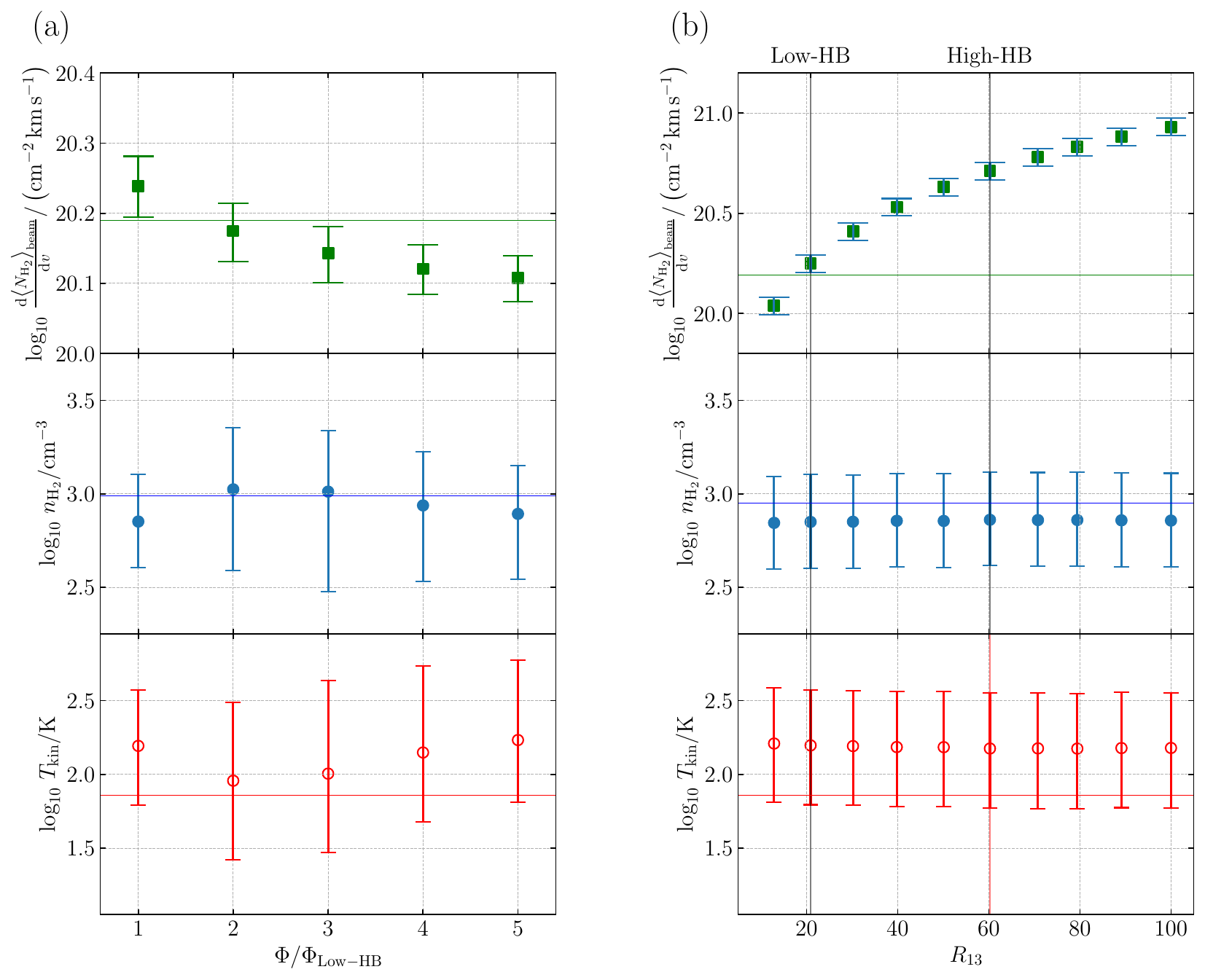}
\caption{(a) Simulated effect of varying beam-filling factor $\Phi$ on the analysis with the Low-density set.  The plot shows the median (\dNHHbeam, \nHH, \Tkin) of the non-hierarchical posterior PDF calculated with the median intensities and intensity ratios of the Low-density set, where $\Phi$ is fixed.  The error bars show the median absolute deviations of the marginal posteriors.  The horizontal axis is $\Phi$ in the unit of the $\tilde{\mu}_\Phi$ of the Low-HB analysis ($\Phi_\mathrm{LowH} = 0.064$).
The median \dNHHbeam, \nHH, and \Tkin\ ($\tilde{\mu}_{\left<N\right>}$, $\tilde{\mu}_{n}$, and $\tilde{\mu}_{T}$) from the Low-HB analysis are shown by horizontal lines.
(b)  Same as (a), but for varying fixed $R_{13}$.
The $R_{13}$ value assumed in the Low-HB analysis and $\tilde{\mu}_{R}$ obtained from the High-HB analysis are indicated by the vertical lines. \label{fig:beamFillingFactors}}
\end{figure*}

The Low-HB analysis adopted $R_{13}$ of 21 \citep{Martin2019}, but the reported [$\mathrm{^{12}C}$]/[$\mathrm{^{13}C}$] ratio measured using the CO isotopologues ranges from 14 to 100 (\citealt{Martin2021A&A} and references therein). 
Figure \ref{fig:beamFillingFactors}(b) shows the variation of the median \Tkin\ and \nHH\ values calculated using the NHB analysis for varying fixed values of $R_{13}$, using the same analysis setup as in the previous subsection (\S\ref{subsection:results:beamFillingFactor}), except that the beam filling factor $\Phi$ is fixed at 0.064 to examine the effect of varying $R_{13}$ independently from the effect of $\Phi$. 
The [$^{13}\mathrm{CO}$]/[$^{12}\mathrm{C}^{18}\mathrm{O}$] abundance ratio is fixed at 6.1.
This result shows that the \Tkin\ and \nHH\ values do not significantly vary within the reported $R_{13}$ range in the literature.   
This insensitivity of the excitation analysis to $R_{13}$ is readily understood as a consequence of the omission of the absolute intensity scales of the $\mathrm{^{12}CO}$ transitions in the present analysis.  
The $R_{13}$ assumption has a relatively minor effect  on the intensity ratios between the CO 1--0, 2--1, and 3--2 transitions via the photon-trapping effect, similarly to the effect of $\Phi$ on the  Low-HB analysis.  Their effects are sufficiently small compared with the credible intervals from the uncertainty of the input data.

The assumption for $R_{13}$ affects estimates of \NHH\ and mass of the low-density component, as \NHH\ is approximately linearly scaled by ${R_{13}}^{-1} N_\mathrm{^{13}CO}$.
We will examine the validity of the derived total mass by comparing it with the masses based on the CO SLED and the dust SED in a later section (\S\ref{subsubsection:discussion:massEstimates}).

\subsubsection{Species-to-Species Variation in Excitation \label{subsection:results:line2lineDifference}}
\begin{figure*}
    \centering
    \plotone{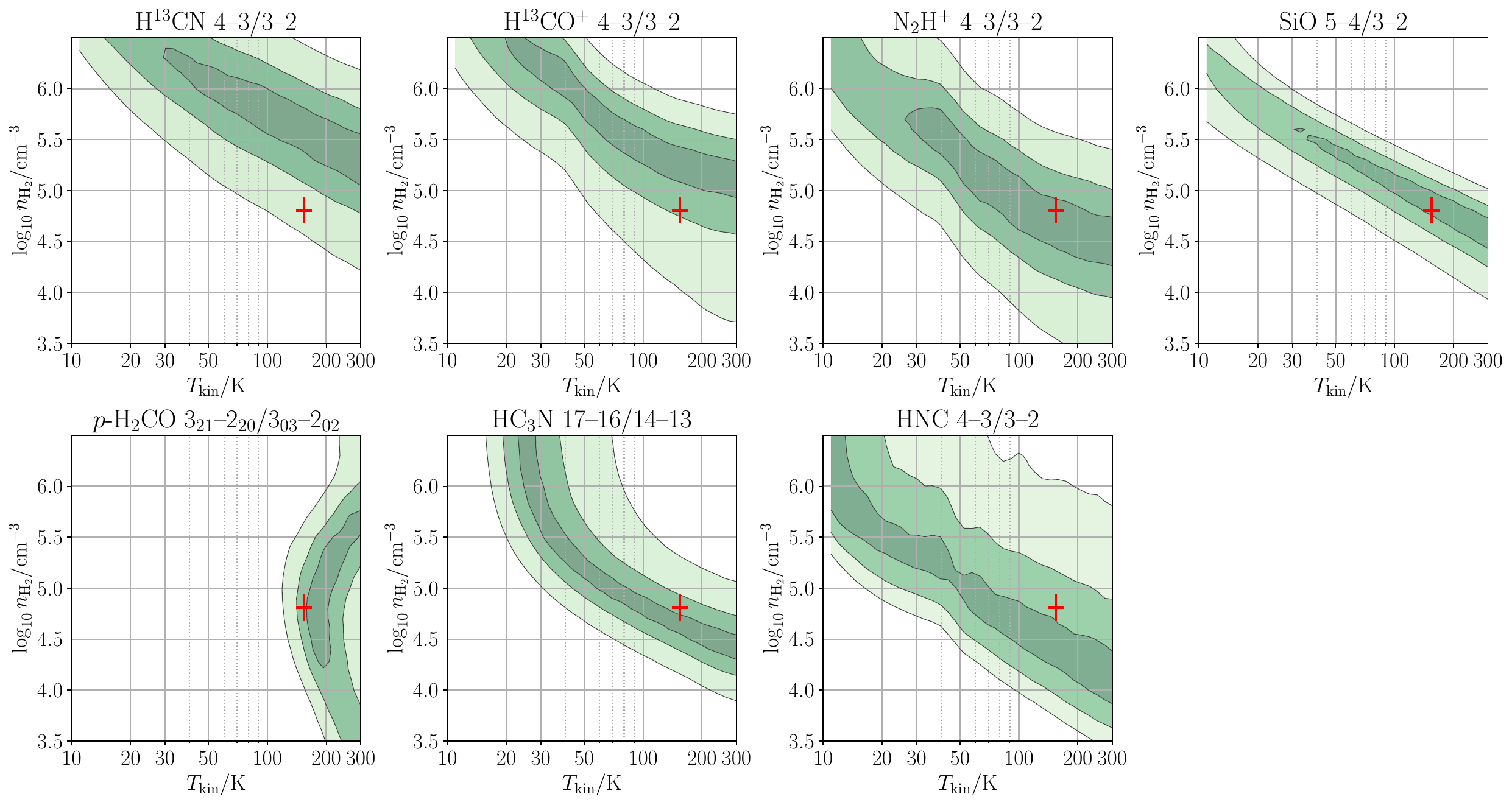}
        \caption{ Comparison of the simple LVG results using different molecular species. 
            The colored contours indicate the (\Tkin, \nHH) ranges calculated from the intensity ratios of individual molecular lines using simple LVG analysis.  The transitions used in the analysis are indicated in the panel captions.  The contours show 30, 50, and 80\% credible intervals, calculated by assuming 15\% relative uncertainties for all input line intensities.  The \Tkin\ and \nHH\ values and their 50\%-percentile errors from the High-HB analysis are indicated by cross marks.\label{fig:lvgVShighH_molPeak}}
\end{figure*}

\begin{figure*}
    \centering
    \plotone{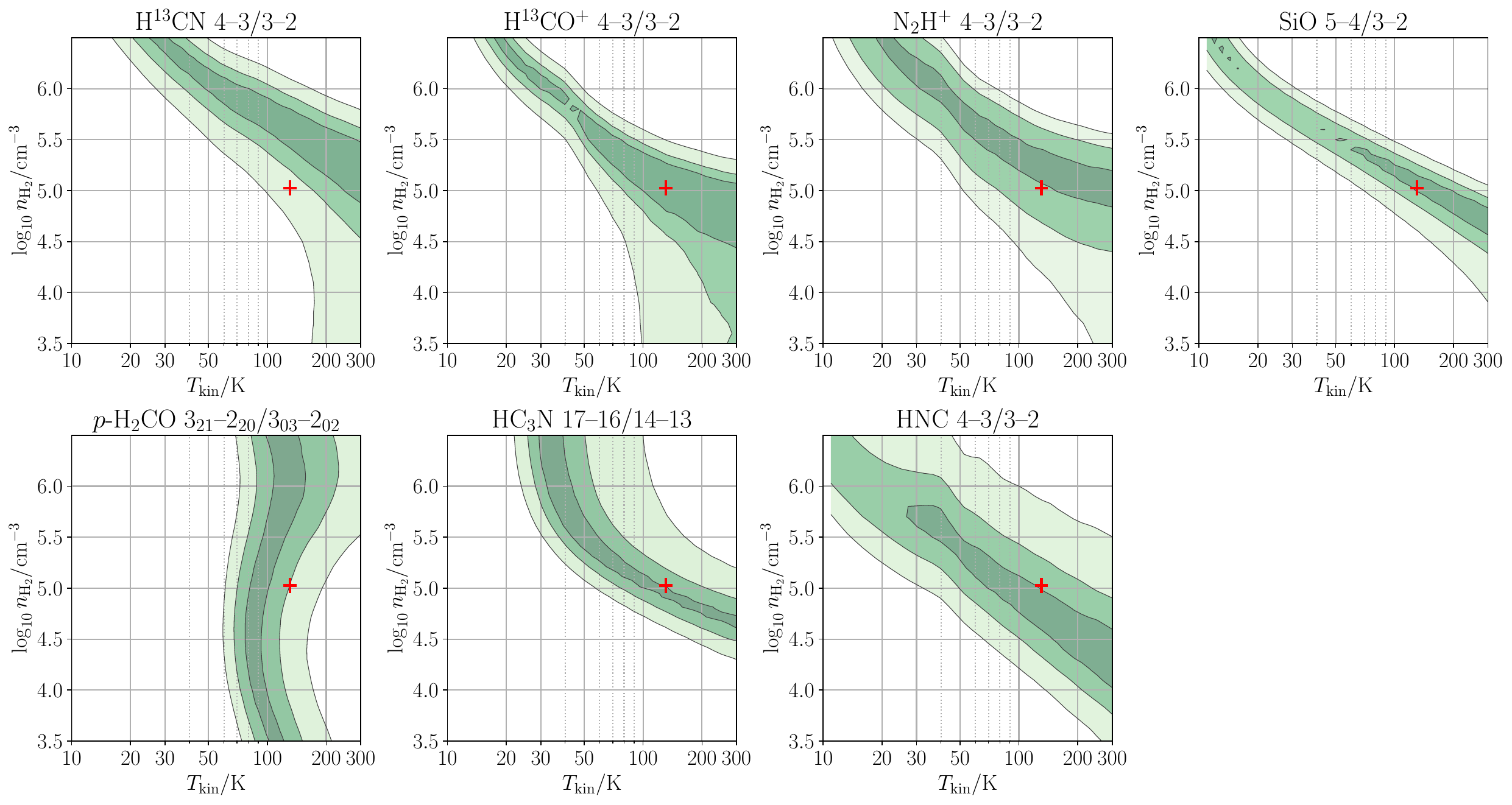}
        \caption{Same as Figure \ref{fig:lvgVShighH_molPeak} but at the TH2 position.}
    \label{fig:lvgVShighH_contPeak}
\end{figure*}

As mentioned in \S\ref{subsection:data:highdensity}, different lines in the High-density data set may be in different excitation conditions depending on their physical and chemical characteristics.  
To see this species-to-species variation, we performed non-hierarchical analyses using single molecular species toward the GMC6 and TH2 positions.
We selected two transitions from six different species ($\mathrm{H^{13}CN}$, $\mathrm{H^{13}CO^+}$, SiO, $\mathrm{N_2H^+}$, $\mathrm{HC_3N}$, \pformaldehyde, and HNC), and ran independent analyses using their intensity ratio for each species assuming a 15~\%\ relative uncertainty for each line intensity.
Spaxels at 200~\kmps\ and 275~\kmps\ are used for the GMC6 and TH2 positions, respectively.
Figures \ref{fig:lvgVShighH_molPeak} and \ref{fig:lvgVShighH_contPeak} show the posterior PDFs projected onto the \Tkin--\nHH\ plane, with the 50~\%\ credible intervals of (\Tkin, \nHH) of the High-HB results. 
The transitions used are indicated by the panel captions.
The posterior PDFs are calculated over the \Tkin\ and \nHH\ ranges of [10 K, 300 K] and [$10^{3.5}$ \pcc,  $10^{6.5}$ \pcc], respectively.   
\myrev{The variation ranges of the molecular column densities (=$\dNHHdv\cdot\xmol{X}$) are conservatively assumed to be $\pm 0.5$ dex from the values in the High-HB result, since the photon trapping effect affects the intensity ratios for optically thick cases.  
Nevertheless, we note that the changes in the intensity ratios due to the column density variation is at most comparable to the relative uncertainties of the intensities, 15\%, even for the optically thickest transitions compared in Figures \ref{fig:lvgVShighH_molPeak} and \ref{fig:lvgVShighH_contPeak}, i.e., HNC 3--2 and 4--3.
} 

Comparison among the six analyses presented in Figures \ref{fig:lvgVShighH_molPeak} and \ref{fig:lvgVShighH_contPeak} shows that the $\mathrm{H^{13}CN}$ lines yield consistently higher \Tkin\ and/or \nHH\ values than the High-HB result toward both positions.  The analysis with $\mathrm{H^{13}CO^+}$ lines also results in slightly higher \Tkin/\nHH\ toward GMC6. 
The 50\%-credible intervals for other molecules are consistent with each other and with the High-HB result.
This consistency is noteworthy because they consist of tracers of a wide variety of physical environments: 
a fast shock tracer (SiO), mild shock tracers ($\mathrm{HC_3N}$ and \pformaldehyde;  \citealt{Requena-Torres2006,Jorgensen2004,Tafalla2010};\citetalias{Tanaka2018b}), and quiescent gas tracers ($\mathrm{HCO^+}$ and $\mathrm{N_2H^+}$).  
The lack of a clear species-to-species difference of the physical conditions (except for HCN) may indicate that the variety in the chemical/physical environments is smoothed-out in the spatial scale of our analysis (\ang{;;1.6} = 27 pc).

The higher excitation of the $\mathrm{H^{13}CN}$ lines could be understood as the consequence of their outstandingly high \ncrit.   The \ncrit\ of $\mathrm{H^{13}CN}$~3--2 and 4--3 are $\sim 10^{7.0}\ \pcc$ and $10^{7.2}\ \pcc$, respectively, whereas those of the highest transitions of other species used in the High-HB analysis are mostly in a relatively narrow range from $10^{6.4}\ \pcc$ to $10^{6.6}\ \pcc$ (except for $\mathrm{HC_3N}$~17--16, whose \ncrit\ is $10^{5.6}\ \pcc$);  it is possible that the $\mathrm{H^{13}CN}$ lines are more affected by the highest-excitation component in the CO SLED \citep{Perez-Beaupuits2018a} than other lines.
We may also consider the IR-pumping effect for the HCN excitation \citep{Sakamoto2010,Mills2013,Veach2013,Aalto2015}, as vibrationally-excited HCN, HNC, and \HCCCN\ lines are detected by ALCHEMI and previous observations \citep{Mangum2019,Rico-Villas2020, Martin2021A&A}.
However, HNC and \HCCCN\ do not show higher excitation than other molecular species.  Since the IR-pumping effect is stronger for HNC than for HCN \citep{Behrens2022ApJ}, this indicates that the IR-pumping effect does not dominate the molecular excitation on the spatial scale of the ALCHEMI data.

\subsubsection{Summary of the Verification of the Analysis\label{subsection:results:caveats}}

The following summarizes caveats on the reliability of the Low-HB and High-HB results described in this and the previous subsections. 
\begin{itemize}
    \item The High-HB analysis is likely to provide more accurate physical conditions of the high-density component than the Min-HB analysis; in particular, the narrow variable range of \Tkin\ in the Min-HB result is possibly due to insufficient constraints from input data (\S\ref{subsection:results:HighVsMin} ,\ref{subsection:results:hyperPrior}).
    \item The strong \NHH--\nHH\ correlation in the Low-HB results is at least partly due to the parameter degeneracy in the excitation analysis (\S\ref{subsection:results:hyperPrior}).   The Low-HB result is also likely to underestimate $\Phi$ by a factor of a few, which may cause $\lesssim$ 0.3 dex underestimate in \nHH\ and $\lesssim0.3$ dex overestimate in \Tkin (\S\ref{subsection:results:beamFillingFactor}).
    \item The assumption for the fixed $R_{13}$ value in the Low-HB analysis does not significantly affect the output \nHH\ and \Tkin\ for the $R_{13}$ range of 14--100 (\S\ref{subsection:results:R13}). \myrev{The High-HB analysis derived the $R_{13}$ values ranging from $\sim$45 to $\sim$$60$, which is 2--3 times higher than the Low-HB value and the values in \cite{Martin2021A&A}. This high $R_{13}$ values result from our adoption of a one-zone approximation, which yielded higher optical depths for the HCN, \HCOp, and CS transitions than \cite{Martin2021A&A}, who performed independent fittings for each species (\S\ref{subsection:results:xmols}).  }
    \item Local \NHH--\Tkin\ (and \nHH--\Tkin) anti-correlation among high-velocity spaxels for the Low-HB and High-HB results is possibly an artifact created by insufficient resolution of the parameter degeneracy.  Nonetheless, the similarity between the Low-HB and High-HB \Tkin\ images may suggest that their overall \Tkin\ distributions are real (\S\ref{subsection:results:maps}).
    \item The High-HB analysis is likely to underestimate $\Phi$ (and hence overestimate \NHH) in high-\NHH\ regions, which possibly causes an underestimation of \nHH\ and/or \Tkin\ there (\S\ref{subsection:results:hyperPrior}).
   \item The IR-pumping effect is unlikely to significantly affect the High-HB results (\S\ref{subsection:results:line2lineDifference}). 
    \item The absolute \NHH\ and \NHHbeam\ values linearly depend on the assumed abundances of $\mathrm{CO}$ and HNC (\xmol{^{12}CO} = $10^{-4}$ and \xmol{HNC} = $4\times 10^{-8}$ in the present analysis), which requires caution when comparing them with other measurements of \NHH\ and gas mass (\S\ref{subsection:data:fixedParams},\ref{subsection:results:R13}).   The uncertainties in $\Phi$ and $R_{13}$ in the Low-HB analysis are additional sources of uncertainty in the mass estimate for the low-density component (\S\ref{subsection:results:beamFillingFactor},\ref{subsection:results:R13}).  We will show that the present analysis provides molecular gas mass consistent with previous results based on CO SLED and dust SED analyses in \S\ref{subsubsection:discussion:massEstimates}). 
\end{itemize}

\section{discussion\label{section:discussion}}
\subsection{Comparison with Excitation Analyses in Previous Studies\label{subsection:discussion:prevAnalysis}}

Measurements of \Tkin\ and \nHH\ toward the NGC~253 CMZ were performed in previous studies through analysis of \formaldehyde\ transitions \citep{Mangum2019}, CO SLEDs (\citetalias{Rosenberg2014,Perez-Beaupuits2018a}) and \ammonia\ inversion transitions \citep{Takano2000,Ott2005,Takano2005,Mangum2013,Gorski2017}, as well as in other ALCHEMI studies \citep{Holdship2021AAp,Behrens2022ApJ,Humire2022A&A,Huang2023AAp}.
This subsection compares our results of \nHH\ and \Tkin\ measurements with those previous results.

\subsubsection{\pformaldehyde\ results\label{subsubsection:discussion:prevAnalysis:h2co}}
A GMC-scale \Tkin\ measurement toward the CMZ of NGC~253 was performed using multi-transition \formaldehyde\ lines by \cite{Mangum2019}.
Their results show that warm gas with \myrev{\Tkin\ ranging from $50$ to $\gtrsim150$ K is extended on $\gtrsim\ang{;;5}$ scales}, which is primarily traced by the \pformaldehyde~$3_{21}$--$2_{20}$ and $3_{03}$--$2_{02}$ transitions.
The High-HB results in the present study are well consistent with this warm component, as a natural consequence that the High-density data set contains the same \pformaldehyde\ lines as in \cite{Mangum2019}.
The widespread distribution of the low-excitation \pformaldehyde\ lines in \cite{Mangum2019} and the ALCHEMI data supports our assumption that they are not localized to specific environments such as hot cores and outflows and hence can be used as \Tkin\ probes on GMC scales, similar to the situation in the GC \citep{Ao2013,Ginsburg2016}.
It is also noteworthy that this warm gas is the lowest-\Tkin\ component identified by the multi-transition \formaldehyde\ study.
Their relatively low upper-state energies ($E_\mathrm{u}/k_\mathrm{B}$ = 21 K and 68 K for $3_{21}$--$2_{20}$ and $3_{03}$--$2_{02}$, respectively) confirm that colder gas than detected with the High-HB analysis, if any, is unlikely to be a major dense gas component of the CMZ of NGC~253.

\subsubsection{CO SLED analyses\label{subsubsection:discussion:prevAnalysis:co}}
Two CO SLED analyses toward the CMZ of NGC~253 were reported in the literature:
\citetalias{Rosenberg2014} using $\mathrm{^{12}CO}$ lines up to the 13--12 transition and $\mathrm{^{13}CO}$ lines up to the 6--5 transition toward the central \ang{;;32.5} region, and \citetalias{Perez-Beaupuits2018a} using $\mathrm{^{12}CO}$ lines up to the 19--18 transition toward the central \ang{;;40} region.
The two analyses consistently show the presence of three CO components.
The low- and high-density components in the present analysis should approximately correspond to the first and second CO components, respectively.
The results for the lowest-density component of the two CO SLED analyses are approximately consistent with each other and with the Low-HB result, as reasonably expected, since all three results are primarily determined by up to the 3--2 transitions of CO. The \Tkin, \nHH\ values of the first component are 60 K, $10^{3.5}\ \pcc$ in \citetalias{Rosenberg2014}, and $90\pm10$ K, $10^{3.2\pm0.1}\ \pcc$ in \citetalias{Perez-Beaupuits2018a}. 
The mass-weighted average of the Low-HB result over the central \ang{;;40} diameter region is $\Tkin, \nHH$ = 120 K, $10^{3.4}\ \pcc$.  The higher \Tkin\ in the Low-HB result than the CO SLED could be explained by the contribution from the second CO component.

For the second CO component, \citetalias{Rosenberg2014} provides the best-fit \Tkin\ and \nHH\ values of 40 K and $10^{4.5}\ \pcc$, respectively, whereas \citetalias{Perez-Beaupuits2018a} provides \Tkin, \nHH = $50\pm3$ K, $(3.2\pm0.8)\times10^5\ \pcc$.
The High-HB result yields \Tkin, \nHH = 100 K, $10^{4.4}\ \pcc$ for the central \ang{;;40} region, which does not critically differ from the \citetalias{Rosenberg2014} result;  although the \Tkin\ of the High-HB result is more than a factor of 2 higher than the best-fit \citetalias{Rosenberg2014} result, it is within the uncertainty due to \Tkin--\nHH\ degeneracy of their solution (presented in Figure 2 in their paper).
The difference between the \citetalias{Perez-Beaupuits2018a} and High-HB results is greater than the uncertainties estimated by \citetalias{Perez-Beaupuits2018a}, though no degeneracy plot is provided in their paper.
However, the observed $^{12}\mathrm{CO}$ 10--9/9--8 and 11--10/10--9 brightness temperature ratios in \citetalias{Perez-Beaupuits2018a}, which are $0.64\pm0.09$ and $0.60\pm0.09$, respectively, are similar to those calculated from the High-HB results, which are 0.77 and 0.65, respectively;  
those three CO transitions are dominated by the second CO component in \citetalias{Perez-Beaupuits2018a}.
Hence, we may consider the \Tkin--\nHH\ degeneracy in the CO SLED analysis as a possible cause of the inconsistency between the \citetalias{Perez-Beaupuits2018a} and High-HB results, too.
The higher-\Tkin\ solution is more preferred by the multi-transition \pformaldehyde\ study \citep{Mangum2019}, as discussed above.
The \pformaldehyde\ results in Figures \ref{fig:lvgVShighH_molPeak} and \ref{fig:lvgVShighH_contPeak} \myrev{indicate} that \Tkin\ of the high-density gas cannot be substantially lower than $\sim 100$ K.
The High-density data set and the CO SLEDs by \citetalias{Rosenberg2014} and \citetalias{Perez-Beaupuits2018a} can be consistently fitted by assuming that a warm ($\Tkin\sim 10^2$ K) and moderately dense ($\nHH\sim$ a few $10^4\ \pcc$) components dominates the second CO component.
Therefore, the High-HB result should more accurately represent the physical conditions of the second CO component, or the high-density component in the present analysis.

\subsubsection{\ammonia\ Inversion Transition Study\label{subsubsection:discussion:prevAnalysis:nh3}}
Measurements of \Tkin\ were also performed by using the \ammonia\ inversion lines \citep[][and references therein]{Takano2000,Ott2005,Takano2005,Mangum2013,Gorski2017}, most of which show multi-temperature components.
The \ammonia\ analysis consistently shows a warm gas component with $\Tkin\sim$ 70--150 K, though the number of the components and their \Tkin\ values differ depending on transitions used, calculation methods, and position/velocities.
Cross comparison between the \ammonia\ studies and the present analysis is not straightforward since \nHH\ is poorly constrained by the \ammonia\ inversion lines while the low- and high-density components in the present analysis are degenerate in the \Tkin\ range.
We confirm that the \Tkin\ $\sim$ 80--100 K of the Low-HB and High-HB results are approximately consistent with that of this warm \ammonia\ component.
The cold \ammonia\ gas with \Tkin\ of 20--50 K reported in a few studies \citep{Takano2000,Takano2005,Mangum2013} is not consistent with either typical \Tkin\ values of the low- or high-density components.
This cold gas component might represent the regions offset from the dynamical center by $\gtrsim 10\arcsec$ along the major axis, where $\Tkin\sim 10^{1.5\mbox{--}1.7}$ K in the Low-HB result (see Fig. \ref{fig:mapLowH}).

\subsubsection{Other ALCHEMI Studies\label{subsubsection:discussion:prevAnalysis:otherALCHEMI}}
\begin{figure*}
\epsscale{1.1}
\plotone{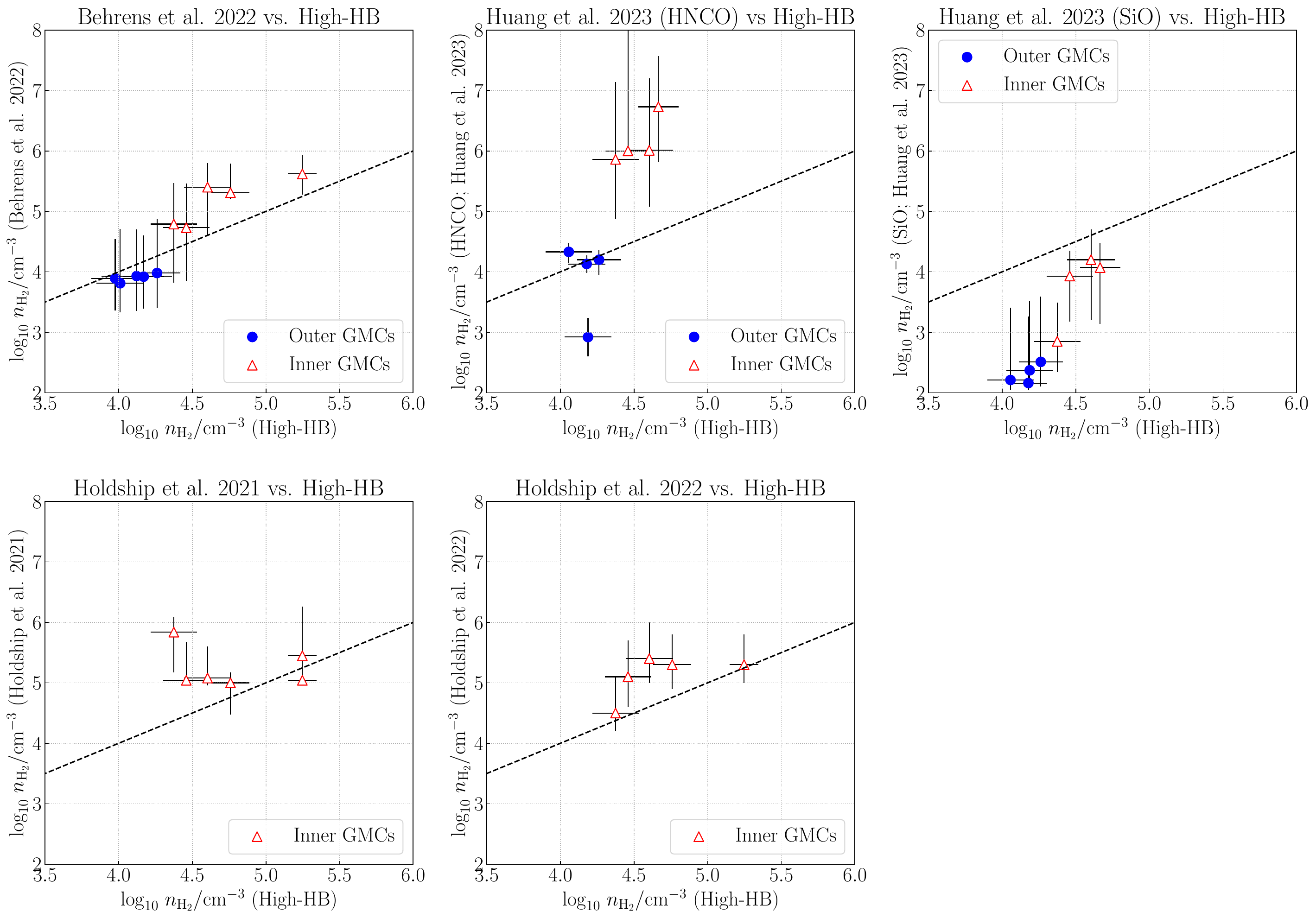}
\caption{Scatter plots of \nHH\ at the GMC positions from the High-HB results vs. those from other ALCHEMI analyses \citep{Behrens2022ApJ,Holdship2021AAp,Huang2023AAp}.  The HNCO and SiO results by \cite{Huang2023AAp} are presented in separate panels. The GMCs inside and outside the central starburst region (the inner and outer GMCs in the text, respectively) are denoted by open rectangles and filled circles, respectively.   The broken lines are the $y=x$ lines.  
\label{fig:compOtherALCHEMI}}
\end{figure*}

Measurements of \nHH\ were performed using various molecular transitions in previous ALCHEMI studies toward representative GMC positions \citep{Holdship2021AAp,Behrens2022ApJ,Humire2022A&A,Huang2023AAp}.
Figure \ref{fig:compOtherALCHEMI} presents scatter plots comparing the \nHH\ measurements from the High-HB result with those from the analyses of HCN and HNC \citep{Behrens2022ApJ}, SiO and HNCO \citep{Huang2023AAp}, $\mathrm{C_2H}$, \citep{Holdship2021AAp} and $\mathrm{H_3O^+}$ and SO \citep{Holdship2022ApJ}.
They all adopt the non-hierarchical Bayesian framework as the statistical modeling method.  \cite{Holdship2022ApJ} and \cite{Behrens2022ApJ} use chemical modeling in addition to the radiative transfer modeling for the parameter inference.   
The inner GMCs in the central starburst region and the outer GMCs in the outskirts of the CMZ are denoted by different symbols in Figure \ref{fig:compOtherALCHEMI}.
Although many of these previous analyses estimate \Tkin\ simultaneously with \nHH,
we do not compare the \Tkin\ results, since their error bars are too large to make meaningful comparisons with the High-HB result.

Figure \ref{fig:compOtherALCHEMI} shows that majority of the \nHH\ values measured by \cite{Behrens2022ApJ}, \cite{Holdship2021AAp}, and \cite{Holdship2022ApJ} agree with the High-HB result within the error bars, 
except for a few GMCs with $\gtrsim 0.5$--1 dex higher \nHH\ than the High-HB results.
Hence, their analyses are likely to probe the approximately same \nHH\ range as the High-HB analysis.
The HNCO result by \cite{Huang2023AAp} for the outer GMCs also approximately agrees with the High-HB \nHH\ values.
However, the HNCO-based \nHH\ values are significantly higher than the High-HB values for the inner GMCs.
This inconsistency for the inner GMCs can be attributed to a larger contribution from the highest-excitation component in the HNCO analysis in \cite{Huang2023AAp}.
They used transitions with up to $\Eu/k_\mathrm{B}$ of 126 K, whereas the High-density data set excludes transitions with $\Eu/k_\mathrm{B} > 70$ K to avoid confusion from the highest-excitation component.
Indeed, intense emission from the high-$J$ ($J\ge$12--11) HNCO transition with $E_\mathrm{u}/k_\mathrm{B} \gtrsim 80$ K are limited to the vicinity of the inner GMC where hot molecular gas associated with the starburst is expected, whereas the lower-$J$  ($J\le$10--9)  transitions are widespread over the entire CMZ \citep{Huang2023AAp}.
Thus, the above three \nHH\ measurements are consistent with the High-HB result except for the overlap of the central highest-excitation component in the HNCO results, despite the difference in the tracer lines and the parameter inference method.

On the other hand, Figure \ref{fig:compOtherALCHEMI} shows that the SiO result by \cite{Huang2023AAp} gives significantly lower \nHH\ than the High-HB and other ALCHEMI results across all GMCs. Their \Tkin\ ranges from 100 to 700 K, which is substantially higher than the High-HB result.
\cite{Humire2022A&A} reported a detection of methanol emission from gas with $\Tkin, \nHH \sim 25\,\mathrm{K}, 10^{8.4}\ \pcc$ near the northeastern end of the CMZ.
These low-\nHH--high-\Tkin\ or high-\nHH--low-\Tkin\ gas components that are visible in particular species are difficult to detect with the High-HB analysis, which assumes a non-negative \nHH--\Tkin\ correlation and the one-zone LVG modeling. 
However, it is likely that such gas components occupy a minor fraction of the dense gas mass in the NGC~253 CMZ, as discussed above (\S\ref{subsubsection:discussion:prevAnalysis:h2co},\ref{subsubsection:discussion:prevAnalysis:co}).

Comparison between the present analysis and \cite{Huang2023AAp} allows for an assessment of the widespread shock chemistry in the NGC253~CMZ.
\cite{Huang2023AAp} discuss multiple origins of the shocked gas, which can be classified into SF-related (starburst scenario and sporadic SF episode scenario) and non-SF-related (cloud--cloud collision scenario) origins.
The above multi-component HNCO emission confirms the coexistence of such multi-component shocked gas. 
The low-$J$ HNCO transition represents the same high-density component as that analyzed in the High-HB analysis.
The high-density tracer molecules (SiO, \pformaldehyde, and \HCCCN) in this widespread shocked gas increase with increasing galactic radius (\S\ref{subsection:results:xmols}).
This spatial distribution suggests that they originate from shocked gas of non-SF origin, i.e., shock associated with cloud--cloud collisions or SN-driven shocks.
The high-$J$ HNCO transitions represent the highest-excitation component in the CO SLEDs.
Their spatial distribution is restricted to the inner GMCs, indicating that this component is associated with active SF there. 

\subsection{NGC~253 Versus GC\label{subsection:discussion:NGC253vsGC}}
In the following, we compare the primary molecular gas properties of the NGC~253 CMZ and the GC, such as the masses, SFRs, and dense-gas mass fraction (\fDGx).
The dense-gas {\it mass} fraction is defined as the mass ratio of the high-density component to the total mass; the asterisked symbol is used to underline the difference from the commonly-used definition based on the HCN/CO luminosity ratio \citep[e.g.][]{Gao2004}.
In addition to these integral quantities, we will use the spatially resolved information to explore the difference characterizing the centers of the two galaxies.
We introduce a new parameter, the modified dense-gas fraction \fDGn, in \S{\ref{subsubsection:discussion:fDGx}}.    
This parameter represents \fDGx\ redefined as a function of the threshold \nHH\ defining the ``dense gas''.
We mainly refer to the Low-HB, High-HB, and \citetalias{Tanaka2018b}\ results in the discussion, but also use the Min-HB and GC-HB results where the precise comparison between the two regions is necessary.

\subsubsection{Molecular Gas Mass and Dense-gas Mass Fraction\label{subsubsection:discussion:massEstimates}}
\begin{deluxetable*}{lllccl}
\tablecaption{Molecular Gas Properties and Star Formation Rates of the NGC~253 CMZ and  the GC }
\tablecolumns{6}
\tablehead{
\multicolumn{3}{c}{}&\colhead{NGC~253 CMZ}& \colhead{GC ($|l|<\ang{1;;}$, $|b| < \ang{0.5;;}$)} & \colhead{notes} 
}
\startdata
\multirow{3}{*}{\begin{minipage}{2.5cm}Low-density\\ component\end{minipage}} & $\bar{n}_\mathrm{H_2}$ &(\pcc) &  $10^{3.3}$ &  $10^3$  & \textit{a} \\
                             & $\bar{T}_\mathrm{kin}$ &(K)   & 85  & 30 & \textit{a} \\
                             & \Mgas &$(10^8\,\Msun)$  & 1.6 & 0.29 & \textit{ac} \\ \vspace{0.5ex}\\
\multirow{3}{*}{\begin{minipage}{2.5cm}High-density\\ component\end{minipage}} & $\bar{n}_\mathrm{H_2}$ &(\pcc) &  $10^{4.4}$ / $10^{4.1}$ &  $10^{4.2}$ / $10^{4.0}$ & \textit{b} \\
                             & $\bar{T}_\mathrm{kin}$ &(K)   & 109 / 106  & \phn83 / 93  & \textit{b} \\
                             & \Mgas &$(10^8\,\Msun)$  & 1.1 / 0.59 & 0.040 / 0.029  & \textit{bc} \\  \vspace{0.5ex}\\
Total & $M_\mathrm{gas}$ & $(10^8\,\Msun)$ & 2.7/2.2 & 0.33/0.32 & b \\ \vspace{0.5ex}\\
\multicolumn{2}{l}{\fDGx} & & 0.40 / 0.27 & 0.13 / 0.09 & \textit{b} \\
\multicolumn{2}{l}{$\mathrm{SFR}$}&{$(\Msun\,\mathrm{yr}^{-1})$}  & 2.8    & 0.012 & \textit{d} \\
\multicolumn{2}{l}{$\mathrm{SFE}$}&{$(10^{-8}\,\mathrm{yr}^{-1})$}  &1.0 / 1.3 & 0.036 / 0.038 & \textit{b} \\
\multicolumn{2}{l}{$\mathrm{SFE_{DG}}$}&{$(10^{-8}\,\mathrm{yr}^{-1})$} & 2.5 / 4.7 & 0.30 / 0.41 & \textit{b}
\enddata
\tablenotetext{a}{From the Low-HB results and \cite{Tanaka2021} for the NGC~253 CMZ and the GC, respectively. }
\tablenotetext{b}{Two estimates from different analyses are given for each of the NGC253~CMZ and the GC, denoted as $X$/$Y$;  $X$ is from the High-HB and \citetalias{Tanaka2018b} results and $Y$ is from the Min-HB and GC-HB results.  }
\tablenotetext{c}{\Mgas\ of the GC is derived from the dust and molecular line data in \cite{Marsh2017}, \cite{Ginsburg2016}, and T18.  See \S\ref{subsubsection:discussion:massEstimates} for details. }
\tablenotetext{d}{From free-free emission at $\sim30$ GHz, taken from \cite{Ott2005} and \cite{Longmore2013a} for NGC~253 and the GC, respectively.}
\label{table:masses}
\end{deluxetable*}

Table \ref{table:masses} tabulates the mass estimates of the low- and high-density components for the NGC~253 CMZ and the GC, along with their respective typical \nHH\ and \Tkin.
The averaged \nHH\ and \Tkin\ of the Low-HB and High-HB results are shown for the NGC~253 CMZ.
The \Tkin\ and \nHH\ values are taken from \cite{Tanaka2021} for the low-density component of the GC, and calculated based on the \citetalias{Tanaka2018b} result for the high-density component.

The low- and high-density component masses for the NGC~253 CMZ are obtained by integrating \NHHbeam\ for the Low-HB and High-HB results, respectively.
The total molecular gas mass, i.e., the sum of the low- and high-density gas masses, is $2\times10^8\ \Msun$ on the assumption that the overlap between the gas traced by the Low-HB and High-HB analyses is negligible.
This is consistent with the mass estimates from multi-component dust SED analysis ($3\times10^8\ \Msun$ in the central $\ang{;;30}\times\ang{;;16}$ region from \citealt{Weiss2008} and $4.5\times10^8\ \Msun$ within the central \ang{;;40} diameter from \citetalias{Perez-Beaupuits2018a}) considering uncertainties in the CO and HNC abundances and dust-to-gas ratio.

For the GC, the low- and high-density gas masses were calculated using the $\mathrm{^{13}CO}~$2--1 flux \citep{Ginsburg2016} and \NHHbeam\ maps by \citetalias{Tanaka2018b}, respectively;  the $^{13}\mathrm{CO}$-2--1-to-mass scaling factor was calculated assuming \Tkin = 30 K, \nHH = $10^3\ \pcc$, and $[\mathrm{^{13}CO}]/[\mathrm{H_2}] = 10^{-5.34}$ \citep{Tanaka2018b}.
The difference in the spatial coverage between \citetalias{Tanaka2018b} and the $^{13}$CO~2--1 data were corrected by extrapolating them to the entire GC ($|l|\leq \ang{1;;}, |b|\leq \ang{0.5;;}$) on the assumption of constant dust-to-gas ratio each for the low- and high-density components.
The HiGAL column density map \citep{Marsh2017} was used as the reference dust mass distribution.
The sum of the low-density and high-density gas masses, $3.3\times10^7\ \Msun$, was confirmed to be the same as that obtained by integrating the HiGAL map.

The dense-gas mass fractions \fDGx\ of the NGC~253 and the GC are also compared in the table.
The boundary of \nHH\ separating the low- and high-density components is $\sim 10^{3.8}\ \pcc$ for the NGC~253 CMZ (\S\ref{subsection:results:LowVsHigh}).
The \nHH\ boundary for the GC can be assumed as $\sim 10^{3.5}\ \pcc$, since the average \nHH\ of the GC is $\sim 0.3$ dex lower than that of the NGC~253 CMZ for both the low- and high-density components.
This assumption is supported by comparison of the \nHH\ frequency histograms based on the measurements using CO lines \citep{Nagai2007} and high-density tracer lines \citepalias{Tanaka2018b} for the GC. 
The \fDGx\ value of 0.40 from the High-HB result is consistent with those based on the CO SLED and dust SED, which are $0.29\pm0.09$ and $0.4\pm0.2$, respectively (\citetalias{Perez-Beaupuits2018a}), which supports our assumption that \xmol{CO} = $10^{-4}$ and \xmol{HNC} = $4\times10^{-8}$. 

The above results confirm the previous results showing that the NGC~253 CMZ is approximately an order-of-magnitude larger than the GC in mass both for the low- and high-density components \citep{Paglione1995,Sakamoto2011,Leroy2015}.
The dense gas mass fraction \fDGx\ in NGC~253 CMZ is $\sim 3$ times that in the GC.

Table \ref{table:masses} also shows mass estimates and averaged \Tkin\ and \nHH\ calculated using the Min-HB and GC-HB results, which yield dense-gas masses that are a factor of 0.54 smaller and 1.4 larger than the High-HB and GC-HB results for the  NGC~253 CMZ and the GC, respectively. 
The inconsistency between the dense gas masses from the  High-HB and Min-HB results should mainly originate from the different \NHH\ tracers used in them (HNC~3--2 and $\mathrm{^{13}CO}$~2--1 in the High-HB and Min-HB analysis, respectively) as mentioned in \S\ref{subsection:results:HighVsMin}.  
The \fDGx\ value of $0.27$ for the NGC253~CMZ is lower than that based on the High-HB result, but still consistent with that from the CO SLED and dust SED \citepalias{Perez-Beaupuits2018a}.
The \fDGx\ ratio between the Min-HB and GC-HB results is 1.7, which is a factor of 2 smaller than  that from the High-HB and \citetalias{Tanaka2018b}\ results. 

\subsubsection{HCN/dense-Gas-Mass Conversion Factor \label{subsubsection:discussion:conversionFactor}}
The mass estimate with the High-HB result yields the $\mathrm{^{12}CO}$-to-$\mathrm{H_2}$ conversion factor $X(\mathrm{CO})$ of $0.7\times10^{20}\ \psc/\left(\mathrm{K}\,\kmps\right)$ for all of the CO 1--0, 2--1, and 3--2 transitions.
This is lower than the canonical value of $2\times10^{20}\ \psc/\left(\mathrm{K}\,\kmps\right)$ as expected for starburst galaxies \citep{Bolatto2013}, and within the range of $X\left(\mathrm{CO}\right)$ estimated for the NGC~253 CMZ, $\left(0.3\right.$--$\left.4\right)\times10^{20}\ \psc/\left(\mathrm{K}\,\kmps\right)$ \citep{Maruersberger1996AAp,Martin2010,Weiss2008}.
The HCN~1--0-to-$\mathrm{H_2}$ conversion factor for the high-density component, $X(\mathrm{HCN})$, is $2\times10^{20}\ \psc/\left(\mathrm{K}\,\kmps\right)$.
This $X(\mathrm{HCN})$ value translates into the HCN luminosity-to-dense-gas-mass conversion factor $\alpha(\mathrm{HCN})$ of $4\ \Msun\pc^{-2}\left(\mathrm{K}\,\kmps\right)^{-1}$, which is a factor of 2.5--3.5 lower than frequently values in extragalactic surveys (10--14 $\Msun\pc^{-2}\left(\mathrm{K}\,\kmps\right)^{-1}$; \citealt{Gao2004,Onus2018,Neumann2023}).

It is noteworthy that the above conversion factors for the NGC~253 CMZ are consistent with those calculated for the GC using the \citetalias{Tanaka2018b} result: $X(\mathrm{CO})$ = $0.6\times10^{20}\ \psc/\left(\mathrm{K}\,\kmps\right)$, and $X(\mathrm{HCN})$ = $2\times10^{20}\ \psc/\left(\mathrm{K}\,\kmps\right)$.
This consistency confirms that \fDG\ defined as the HCN/CO luminosity ratio is indeed a good measure of \fDGx\ based on the mass estimates.
However, we note that the High-HB result contains a wide range of \nHH\ from $10^{3.2}$ \pcc\ to $10^{5.8}$\ \pcc\ as shown in Figure \ref{fig:nHistogram}, and $\sim$70\%\ of the total mass is constituted by the spaxel with $\nHH < 3\times10^4$ \pcc. 
Furthermore, the actual HCN~1--0 emitting gas in the NGC~253 CMZ contains significant amount of even lower-\nHH\ gas, as 35~\% of the HCN~1--0 flux originates outside the coverage of the High-HB analysis as we have seen in \S\ref{subsection:results:analysisRuns}. 
Therefore, our \nHH\ measurement is consistent with the recent arguments against HCN~1--0 luminosity being a reliable tracer of molecular gas mass with $\nHH\sim 3\times10^4\ \pcc$
\citep[e.g.][]{Onus2018,Jones2023}.

\subsubsection{Star Formation Rates and Star Formation Efficiencies \label{subsubsection:discussion:SFRandSFE}}

We show the SFRs of the NGC~253 and MW CMZs taken from the literature, which are measured using the free--free emission in the $\sim 30$-GHz band \citep{Longmore2013a,Ott2005} in Table \ref{table:masses}.
Two SFEs were calculated by dividing the SFR by the mass of either the total molecular gas or the dense molecular gas; we denote the latter as $\mathrm{SFE}_\mathrm{DG}$.
The SFE of NGC~253 is $\sim 30$ times that of the GC, whose difference is substantially greater than the factor of 2--3 difference in \fDGx\ between them.
The $\mathrm{SFE}_\mathrm{DG}$ of NGC~253 is more than an order of magnitude higher than that of the GC.  This difference is significant even when the uncertainty in the mass estimates, which is a factor of $\sim$ 2--3 in \citetalias{Perez-Beaupuits2018a}, is considered.

These results do not conform with a simple picture that the SFR is scaled by the mass of the dense-gas component, at least between the NGC~253 CMZ and the GC.  
A similar result is reported for the comparison between the GC and the circumnuclear ring of M83 \citep{Callanan2021}; their difference in the dense-gas mass based on the HCN~1--0 luminosity is a factor of $\sim 2$, in spite of the an order of magnitude higher SFR of the M83 circumnuclear ring.

\begin{figure*}
\epsscale{1.1}
\plotone{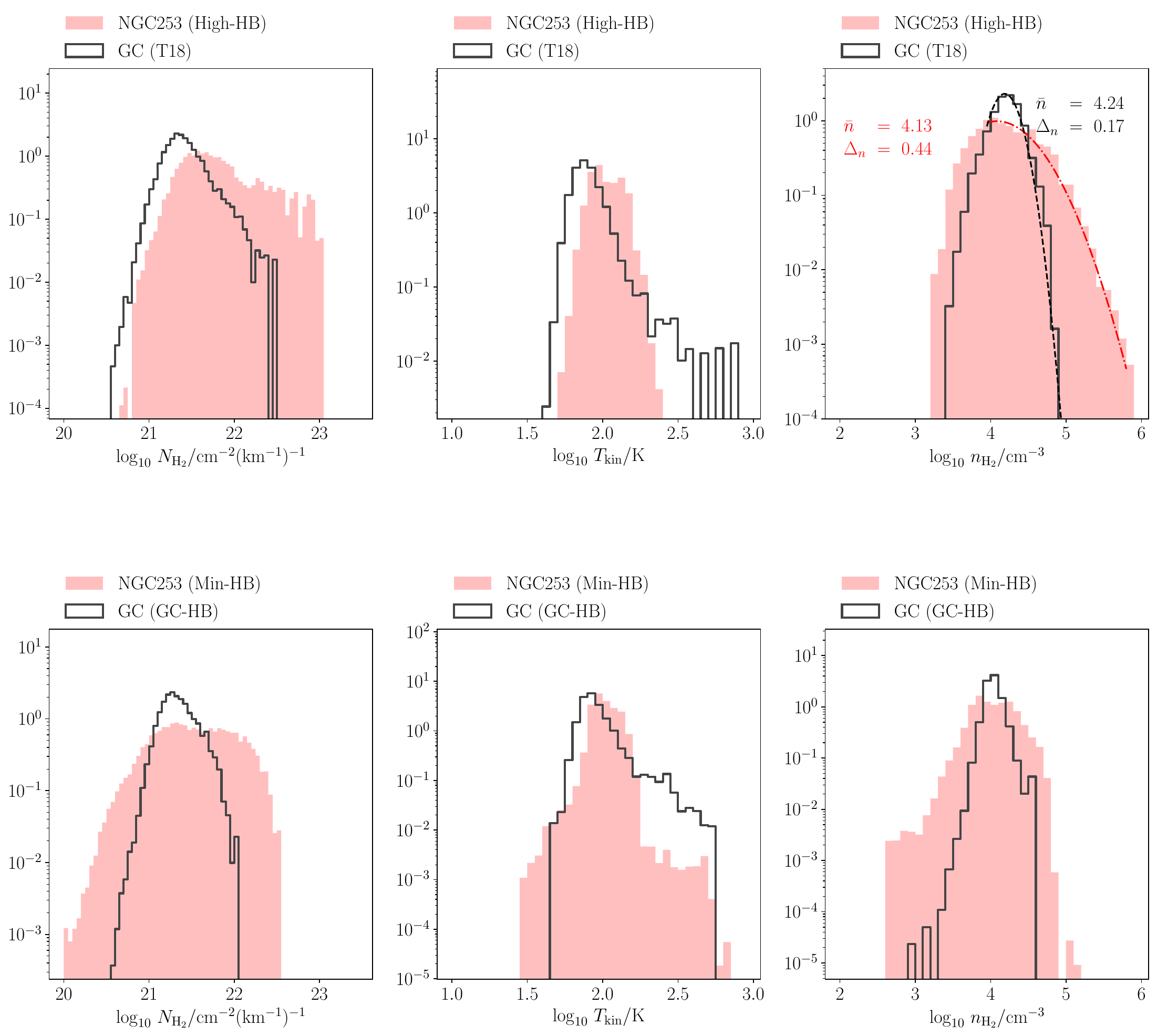}
\caption{Frequency histograms of \NHH, \Tkin, and \nHH\ for NGC~253 (red hatched) and the GC (open).  All frequencies are weighted by \NHHbeam\ and normalized so that the integral is unity.  The upper row uses the High-HB and \citetalias{Tanaka2018b} results for NGC~253 and the GC, respectively, and the lower uses the Min-HB and GC-HB results.   The results of log-normal fitting are shown for the \nHH\ histograms of the High-HB and T18 results.  \label{fig:hist253vsGC}}
\end{figure*}

\subsubsection{Frequency Distributions of \NHH, \nHH, and \Tkin \label{subsubsection:discussion:NnThistograms}}

Figure \ref{fig:hist253vsGC} compares the histograms of \NHHbeam-weighted frequencies of \NHH, \Tkin, and \nHH\ of the high-density components of NGC~253 and the GC.
The comparison is made using two analysis pairs: High-HB vs. \citetalias{Tanaka2018b} (upper row of the figure) and Min-HB vs. GC-HB (lower row of the figure).
In the High-HB vs. \citetalias{Tanaka2018b} comparison, the \NHH\ histogram of the NGC~253 CMZ has a shallower declining slope from the peak at $\sim 10^{21.5}\,\psc\left(\kmps\right)^{-1}$ to the high \NHH\ end at $\sim10^{23}\,\psc\left(\kmps\right)^{-1}$.
A similar trend is identified in the \nHH\ histograms.
The \nHH\ frequency of NGC~253 has a wider distribution toward high \nHH\ values up to $10^6\ \pcc$, whereas that of the GC declines more steeply and vanishes at $\nHH\sim10^5\ \pcc$.
We fit the \nHH\ histograms from the High-HB and T18 analyses with a log-normal function, $f\left(\nHH\right) \propto \exp\left[-\left(\mathrm{log}_{10}\nHH/\pcc - \bar{n}\right)^2/\left(2\Delta_n^2\right)\right]$, for $\nHH \ge 10^4\ \pcc$ and show the results in Figure \ref{fig:hist253vsGC}.
The scaling parameter $\Delta_n$ is a factor of $\sim 2.5$ larger for the High-HB histogram than for the T18 histogram, although the two have similar mean of $10^{4.1\mbox{--}4.2}\ \pcc$.
The Min-HB vs. GC-HB comparison leads to the same conclusion;  both \NHH\ and \nHH\ distributions show broader extensions toward high values for NGC~253 than the GC. 
Though the \nHH\ distributions in the Min-HB and GC-HB results are systematically shifted from the High-HB and \citetalias{Tanaka2018b} histograms by $\sim -0.4$ dex, respectively, it does not affect their relative difference.
The presence of broader high-\NHH\ and high-\nHH\ tails in NGC~253 is noteworthy because the approximately order-of-magnitude lower linear resolution for NGC~253 limits the sensitivity to high-\NHH/-\nHH\ peaks due to stronger beam smearing.  The actual difference between the NGC~253 CMZ and the GC could be larger than detected when compared at the same spatial scales.

On the other hand, the \Tkin\ histogram of the GC shows higher frequencies for $\Tkin\gtrsim10^{2.2}\ \mathrm{K}$ than the NGC~253 histogram in both the High-HB vs. \citetalias{Tanaka2018b} and Min-HB vs. GC-HB comparisons.
This lower frequency of high-\Tkin\ spaxels in the NGC~253 CMZ is consistent with the heavier beam-smearing for the NGC~253 CMZ, and therefore it is unclear whether they differ significantly in \Tkin\ when compared on the same size scale.
The lack of obvious \Tkin\ enhancement in NGC~253 may indicate the enormous difference between NGC~253 and the GC in SFR does not significantly affect the heating budget of the high-density component.

\subsubsection{Modified Dense-Gas Mass Fraction and SF Threshold Density \label{subsubsection:discussion:fDGx}}

\begin{figure*}[ttt]
\plotone{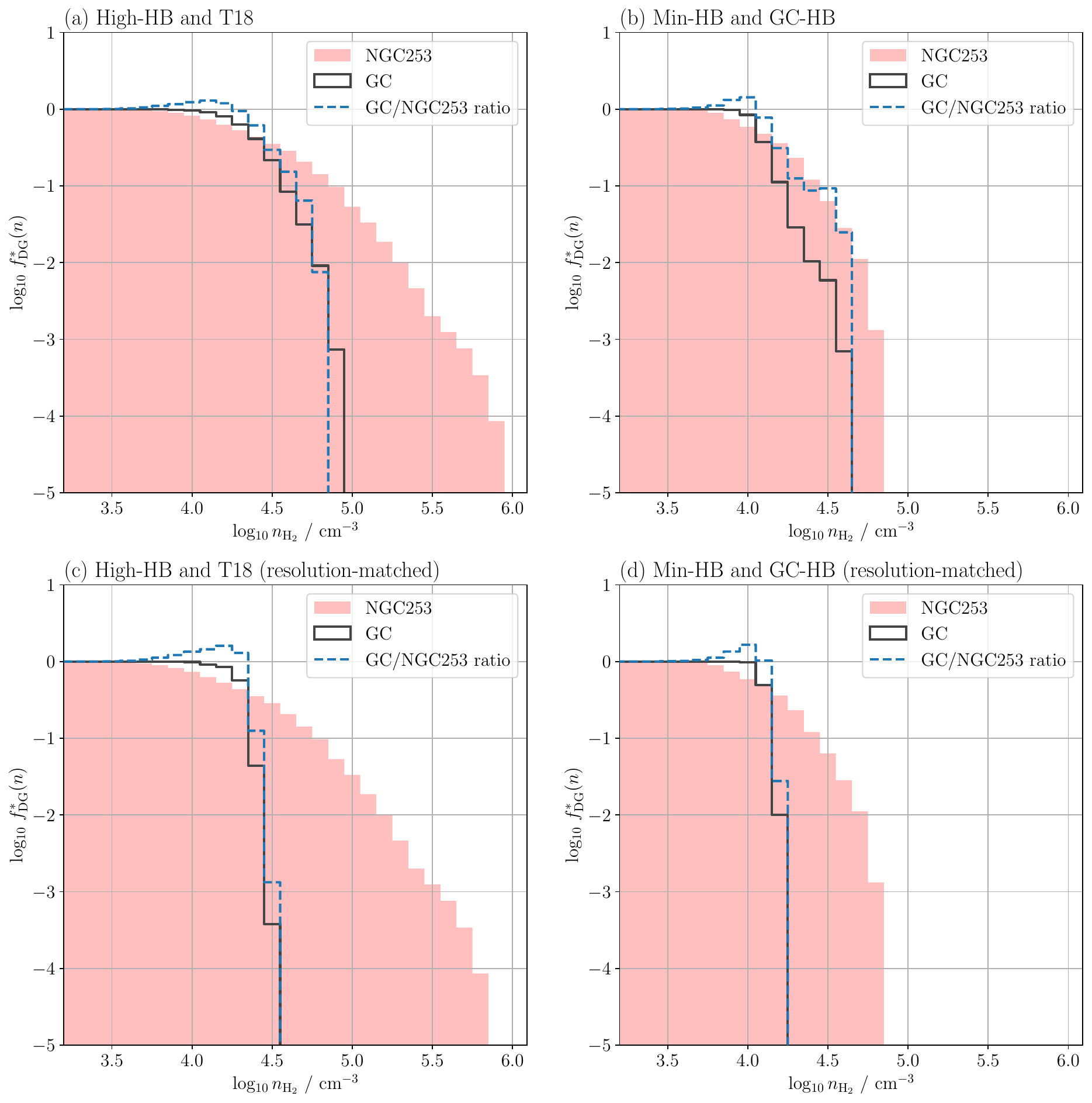}
\caption{(a) Inverse cumulative histograms of \nHH\ for NGC~253 (red hatched) and the GC (black solid) using the High-HB and \citetalias{Tanaka2018b} results, respectively.  The dashed line shows the frequency ratio of the GC to NGC~253.  (b) same as a, but using the T18 results smoothed to the same \PPV\ resolution as the High-HB result.  (c) same as a, but using the Min-HB and GC-HB result. (d) same as c, but using the GC-HB results smoothed to the same \PPV\ resolution as the Min-HB result. }
\label{fig:densegasmass}
\end{figure*}

The difference between the \nHH\ histograms of the NGC~253 and the GC seen in \S\ref{subsubsection:discussion:NnThistograms} could be translated into the difference in the dense gas mass fractions with varying threshold of the definition of ``dense gas''. 
We show the inverse cumulative form of the \nHH\ histograms in Figure \ref{fig:densegasmass}.  
Note that this inverse cumulative function represents the dense-gas mass fraction with varying threshold \nHH\ (except for the contribution from the low-density component), which we define as the modified dense-gas mass fraction \fDGn; i.e.,
\begin{eqnarray}
\fDGn\equiv\frac{\displaystyle\sum_{\nHH_i > n} \NHHbeam_i}{\displaystyle\sum {\NHHbeam}_{,i}} ,
\end{eqnarray}
where the subscript $i$ indexes the spaxels in the high-density component.  
In the \fDGn\ plot with the High-HB and \citetalias{Tanaka2018b} results (panel a), the NGC~253 and the GC show similar \fDGn\ profile in $\nHH \lesssim 10^{4.5}$ \pcc. However, \fDGn\ in the GC shows a steep decline in $\gtrsim 10^{4.5}$ \pcc, reaching 0 at $\nHH\sim 10^5$\ \pcc, while the NGC~253 CMZ plot maintains an approximately constant shallow slope in the range of $\nHH = 10^{4.5\mbox{--}6.0}$\ \pcc.
The \fDGn\ ratio between the GC to the NGC~253 CMZ reaches 0.1, i.e., the same as their $\mathrm{SFE}_\mathrm{DG}$ ratio,   at $\nHH = 10^{4.6}\ \pcc$.
Comparison using the Min-HB and GC-HB results (panel b) leads to a similar result, except for the threshold \nHH\ at which  the \fDGn\ ratio reaches 0.1 lies within the range of $10^{4.3\mbox{--}4.5}\ \pcc$.
Hence, if there exists a threshold \nHH\ that defines the dense gas consumed for SF, it is likely in $\sim 10^{4.3\mbox{--}4.6}\ \pcc$.
The rich abundance of the gas exceeding this threshold \nHH\ in the NGC~253 CMZ, or its scarcity in the GC, could be the critical difference characterizing the centers of the two galaxies.

We caution that the above analysis compares results that differ in spatial resolution by an order of magnitude.
Figure \ref{fig:densegasmass} also show the \fDGn\ plots where the GC results are smoothed to the working resolution for the NGC~253 CMZ, i.e., 27 pc and 20~\kmps\ in the space and velocity.  
The drop of the resolution-matched \fDGn\ in the GC data both for the \citetalias{Tanaka2018b} and GC-HB results (panels c and d, respectively) are steeper than in the non-smoothed data.
The threshold \nHH\ is estimated to be $\sim 10^{4.4}\ \pcc$ from the High-HB vs. \citetalias{Tanaka2018b} comparison, and $\sim 10^{4.2}\ \pcc$ from the Min-HB vs. GC-HB comparison.    
They are $\sim 0.2$--0.3 dex lower than the values without resolution-matching, demonstrating that the \nHH\ estimate is sensitive to the working resolution of the analysis.
Conversely, the \fDGn\ plot for the NGC~253 CMZ is expected to become shallower if measured at the resolution of the \citetalias{Tanaka2018b}, i.e., 2.4 pc and 10~\kmps\ in the space and velocity, which would yield a higher threshold \nHH\ than that without resolution matching.
By taking this resolution dependence into account, we conservatively conclude that the threshold density is $\sim 10^{4.2\mbox{--}4.6}\ \pcc$.

\subsection{Origin of the Contrasting SF Activities in the NGC~253 CMZ and the GC \label{subsection:discussion:originOfContrastingSF}}
\begin{figure}[ttt]
\plotone{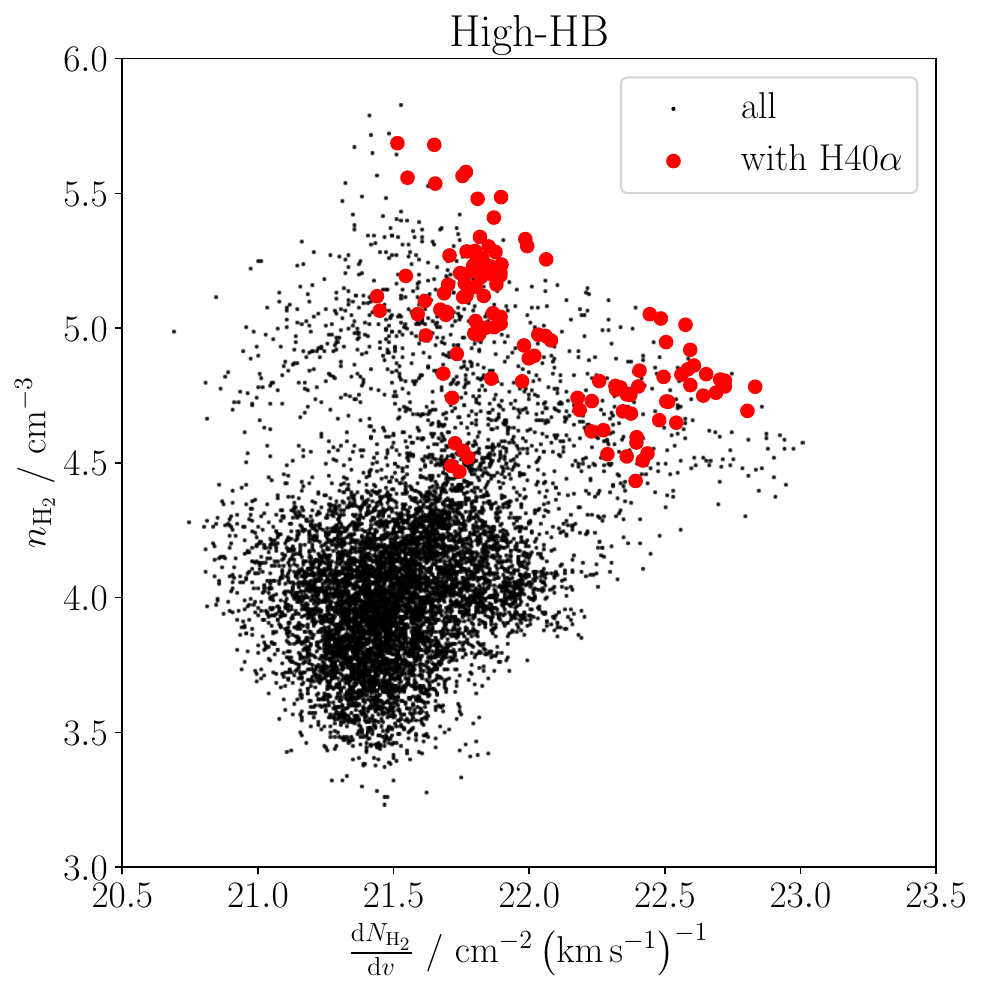}
\caption{Scatter plot of \nHH\ vs. \dNHHdv\ from the High-HB results.   The spaxels associated with compact H40$\alpha$ emission features \citep{Mills2021} are colored red. \label{fig:n_SF}}
\end{figure}

The above comparison of \fDGn\ between the NGC~253 CMZ and the GC (\S\ref{subsubsection:discussion:fDGx}) suggests that the \nHH\ defining the dense-gas consumed for SF be 0.4--0.8 dex higher than the boundary separating the low- and high-density components that are traced by low-$J$ CO and HCN transitions.
This is consistent with the actual distribution of embedded SF in the NGC~253 CMZ and the GC.
Figure \ref{fig:n_SF} shows \NHH--\nHH\ scatter plots for the high-density component of NGC~253 using the High-HB result, where the spaxels associated with H40$\alpha$ features \citep{Mills2021} are colored red.
Figure \ref{fig:n_SF} shows that the H40$\alpha$ features appear exclusively in $\nHH \gtrsim 10^{4.5}\ \pcc$ regions, which is approximately consistent with the above threshold \nHH\ obtained from the \fDGn\ plots of the High-HB and T18 result.
A similar result was obtained for the GC;
\cite{Tanaka2020} found that SF signatures (extended green objects, SF masers, and compact infrared sources) are primarily associated with dens-gas clumps with low virial parameter of $< 6$, whose typical density is $\nHH > 10^{4.6}\ \pcc$.  
This comparison may not be a rigorous one, as we apply different SF tracers and different spatial and velocity resolutions for NGC~253 and the GC, and hence it does not necessarily indicate the presence of a common threshold \nHH\ of $10^{4.5\mbox{--}4.6}\ \pcc$ in the centers of the two galaxies.
Nonetheless, the above results show that embedded star-forming regions are not uniformly distributed throughout the entire high-density components.
Instead, they are confined to the highest density portion of the high-density component, which qualitatively corroborates
our findings.

We note that it is not obvious that we can assume a common threshold \nHH\ for the centers of the two galaxies with different physical environment.
The turbulent cloud model predicts that the critical \nHH\ to initiate gravitational collapse is not universal but depends on the degree of turbulence and magnetic field strength \citep{Krumholz2005,Padoan2011}.  
\cite{Krieger2020} showed that the degree of turbulence measured by velocity dispersion and virial parameter in NGC~253 is similar to those in the GC when compared with high-\NHH\ regions ($\NHH\gtrsim 10^{22}\ \pcc$), where SF takes place.
However, the magnetic field strength in the literature is slightly higher for the NGC~253 CMZ ($160\pm20\ \mu$G; \citealt{Heesen2011AAp}) than the GC ($\ge 50\ \mu$G; \citealt{Crocker2010}), potentially resulting in an elevated critical \nHH\ in the NGC~253 CMZ.
In addition, the CR ionization rates reported for the dense clouds in the NGC~253 CMZ, $\sim 10^{-14\sim-12}\ \mathrm{s}^{-1}$ \citep{Harada2021ApJ,Holdship2021AAp,Holdship2022ApJ,Behrens2022ApJ}, tend to be higher than those reported for the GC, $10^{-16\sim-14}\ \mathrm{s}^{-1}$ \citep{OkaTakeshi2005,OkaTakeshi2019,Willis2020}, also suggesting a more hostile SF environment in the NGC~253 CMZ.
Therefore, it is not evident whether our results indicate a common SF threshold \nHH\ of $\sim 10^{4.2\mbox{--}4.6}$ \pcc\ in the NGC~253 CMZ and the GC.
Nevertheless, we could safely consider that the higher \fDGn\ in \nHH\ above this density range in the NGC~253 CMZ plays a critical role to overcome the disadvantageous conditions and promote starburst.
We emphasize that this difference in the \fDGn\ profile between the NGC~253 CMZ and the GC was not measurable from analysis of the luminosities without spatial information, and it provides a new clue to understand the low $\mathrm{SFE}_\mathrm{dense}$ in the GC, and more generally, the  origins of the dispersion in the SFR/HCN ratio among galaxies \citep{Usero2015,Bigiel2016a,Neumann2023}.

The origin of the broad \fDGn\ profile in the NGC~253 CMZ could be caused by the higher turbulent pressure within the GMCs.
\cite{Krieger2020} show the velocity dispersion \sigmav\ within the NGC~253 CMZ clouds is by a factor of 2--3 larger than that within the GC clouds over a wide range of spatial scale from 1 to 100 pc except for the highest \NHH\ region.
The turbulent cloud model predicts that such elevated degree of turbulence broadens the PDF of \nHH\ \citep{Krumholz2005,Neumann2023} as we observed in the NGC~253 CMZ.   
However, the turbulent cloud model also predicts that this broadening of the \nHH\ PDF is counteracted by an increase in the critical \nHH\ for gravitational instability, resulting in a net decrease in SFE.
The high $\mathrm{SFE_{dense}}$ in the NGC~253 CMZ contradicts this model prediction and the actually observed anti-correlation between \sigmav\ and the SFR/HCN ratio among galaxies \citep[e.g.][]{Neumann2023}.
Alternatively, we may attribute the different \fDGn\ in the NGC~253 CMZ and the GC to their different SF phases \citep{Callanan2021}.
In this line of theory, the GC could be considered to have undergone a past active phase suggested by several observations \citep[e.g.][]{Tanaka2007,Yusef-Zadeh2009}, during which dense gas has been depleted.
Meanwhile, observations suggest that the formation of new star-forming dense clouds is hindered by frequent destructive cloud--cloud collisions \citep{Tanaka2018b,Enokiya2019}.
The depletion of dense gas in the present-day GC could be attributed to a combination of these two effects.

\section{Summary and Conclusions\label{section:summary}}

We have performed a spatially-resolved excitation analysis of the CMZ of the starburst galaxy NGC~253 by exploiting the rich data set of the ALCHEMI line survey.  
On the basis of the comparison with the similar analysis performed for the quiescent CMZ of the Milky Way, we attempted to find the difference in the physical condition parameters of the molecular clouds that distinguish the starburst and quiescent galactic centers.
We applied the non-LTE analysis to three different sets of input lines (High-density, Low-density, and Minimal data sets) from the ALCHEMI data by employing two different statistical modeling approaches, namely, the hierarchical Bayesian (HB) and non-hierarchical Bayesian (NHB) models.  These modeling processes allowed the derivation of the full \PPV\ distributions of the fundamental physical condition parameters, including \NHH, \Tkin, and \nHH, delineated at spatial and velocity resolutions of 27 pc and 20 \kmps, respectively.
The main results are summarized below.

\begin{itemize}
    \item The main HB analyses using the Low- and High-density data sets (Low-HB and High-HB analyses) confirmed that the NGC~253 CMZ contains of at least two molecular cloud components: the low-density component (\Tkin = 85 K,  \nHH = $10^{3.3}$ \pcc) mainly traced by low-$J$ ($J\leq 3$) transitions of CO and its isotopologues, and the high-density component (\Tkin = 110 K,  \nHH = $10^{4.5}$ \pcc)  traced by low-excitation ($E_\mathrm{u}/k_\mathrm{B} \lesssim 70\ \mathrm{K}$) transitions  of high-density tracer species.   The low- and high-density components correspond to the first and second excitation components detected in previous dust SED and CO SLED studies, respectively (\S\ref{subsection:method:twoComponentModel},\ref{subsubsection:discussion:prevAnalysis:co}).
    The two components are found to be spatially mixed in the 27 pc resolution, exhibiting a strong spaxel-to-spaxel correlation in the \NHH\ distribution across the entire analyzed region  (\S\ref{subsection:results:LowVsHigh}).
    
    \item The physically unreasonable parameter distributions due to the severe parameter degeneracy in the NHB analysis, such as the anti-correlations in \NHH\ vs. \Tkin\ and \nHH\ vs. \Tkin, were successfully suppressed in the HB analysis for both the \myrev{Low-density and High-density data sets}.  The NHB analysis also yielded \NHH/\nHH\ values that were too low to fit the $r$-\sigmav\ relationship for the NGC~253 CMZ.
    Hence, the HB analysis was able to provide more realistic \PPV\ distributions of the physical condition parameters than the standard NHB analysis (\S\ref{subsection:results:HBvsNHB}).

    \item Molecular gas with high \Tkin\ and \nHH\ is concentrated in the starburst region within the innermost $\sim 10''$ radius, where typical (\Tkin, \nHH) are (100--200 K, $10^{3.0\mbox{--}3.6}\ \pcc$) and (100--150 K, $10^{4.5\mbox{--}5.5}\ \pcc$) in the low- and high-density components, respectively (\S\ref{subsection:results:maps}).  High-velocity warm gas was found associated with this region at velocities separated from the main velocity component by $\sim\pm 100\ \kmps$, indicating the presence of SF-driven shocks.      A few high-\Tkin\ and high-\nHH\ spots were detected outside the central starburst region, which may indicate interaction with shocks of non-SF origins, such as the large-scale molecular outflows and cloud--cloud collisions (\S\ref{subsection:results:high-nT}).

   \item The fractional abundances of tracers of shocks and high temperature molecular gas (SiO, $\mathrm{HC_3N}$, and \pformaldehyde) show outwardly increasing distributions from the central $\pm 10''$ region to the outskirts of the CMZ (\S \ref{subsection:results:xmols}).  This indicates that the low-excitation transitions of these species trace shocks of non-SF origin (\S\ref{subsubsection:discussion:prevAnalysis:otherALCHEMI}).

    \item \myrev{The [$^{12}\mathrm{C}$]/[$^{13}\mathrm{C}$] isotopic abundance ($R_{13}$) was fixed at 21 in the Low-HB analysis, based on the $\mathrm{^{12}C^{18}O/^{13}C^{18}O}$ measurement by \cite{Martin2019}.   However, the High-HB analysis yielded $R_{13}$ ranging from $\sim$45 to $\sim$60, which is 2--3 times that the Low-HB value and the values estimated from the ALCHEMI data by \cite{Martin2021A&A}. This difference with the previous results is caused by the one-zone approximation adopted in our analysis, which lead higher optical depths for the HCN, \HCOp, and CS transitions than \cite{Martin2021A&A}, who performed independent fittings for each species (\S\ref{subsection:results:xmols})}. 
   
    \item  The Low-HB analysis possibly overestimates the correlation coefficient between \NHH\ and \nHH, due to insufficient resolution of their degeneracy in the excitation analysis (\S\ref{subsection:results:hyperPrior}). The difference among the physical conditions traced by different molecular species in the High-density data set was confirmed to be insignificant, though HCN lines are somewhat biased to high-\Tkin\ and/or high-\nHH\ regions compared to other species (\S \ref{subsection:results:line2lineDifference}). Other caveats of the Low-HB and High-HB results are summarized in \S\ref{subsection:results:caveats}.
 
   \item The \nHH\ and \Tkin\ values from the Low-HB and High-HB results are confirmed to be approximately consistent with the previous measurement in literature (\S\ref{subsection:discussion:prevAnalysis}).  The difference from the CO SLED results  (\citetalias{Rosenberg2014} and \citetalias{Perez-Beaupuits2018a}) can be reasonably explained by the parameter degeneracy in the CO SLED analyses (\S\ref{subsubsection:discussion:prevAnalysis:co}).   The shocked gas of SF origin in the central starburst and a few cold gas features detected in the previous ALCHEMI paper and the \ammonia\ inversion transition studies are possibly missed in our analysis, due to the selection of the input data set and the one-zone LVG assumption employed in the HB analysis (\S\ref{subsubsection:discussion:prevAnalysis:nh3},\ref{subsubsection:discussion:prevAnalysis:otherALCHEMI}).

    \item The masses of the low-density and high-density components were estimated to be $1.6\times10^8\ \Msun$ and $0.50\times10^8\ \Msun$, respectively.  The dense gas mass fraction (\fDGx) of 0.24 is approximately twice that of the GC with the \nHH\ separating the high- and low-density components being $\sim 10^{3.5\mbox{--}3.8}\ \pcc$.  
    Their values depend on the assumed abundances of CO, $\mathrm{^{13}CO}$, and HNC, but are consistent with previous results from the CO SLED and dust SED analyses (\S\ref{subsection:results:LowVsHigh}).
    
    \item The Low-HB result yields the CO-to-$\mathrm{H_2}$ conversion factor for the NGC~253 CMZ is $X(\mathrm{CO})$ to be $0.7\times10^{20}\ \psc/\left(\mathrm{K}\,\kmps\right)$ for all of the 1--0, 2--1, and 3--2 transitions.  The HCN~1--0-to-$\mathrm{H_2}$ conversion factor for the high-density component is $2\times10^{20}\ \psc/\left(\mathrm{K}\,\kmps\right)$.   They are consistent with those calculated for the GC, confirming that the dense gas fraction (\fDG) defined as the HCN~1--0/CO~1--0 flux ratio serves as a good measure of \fDGx\ based on the mass estimate.  However, $\gtrsim$70\%\ of the mass of the high-density component is constituted by the spaxels with $\nHH < 3\times10^4$ \pcc, being consistent with the arguments against the reliability of the HCN~1--0 as an accurate measure of the dense-gas mass consumed for SF  (\S\ref{subsubsection:discussion:conversionFactor}).
        
    \item The star formation efficiency for the high-density component (SFE$_\mathrm{DG} \equiv$ SFR/high-density gas mass) is approximately 10 times that of the GC.  This difference is approximately an order of magnitude larger than their difference in \fDGx, confirming that SFE cannot be solely determined by \fDGx\ or \fDG\ (\S\ref{subsubsection:discussion:SFRandSFE}). 

    \item The \nHH\ frequency histogram of the NGC~253 CMZ derived from the High-HB result exhibits a 2.5 times broader profile toward he higher \nHH\ values than that of the GC using the \citetalias{Tanaka2018b} result, though they have similar mean \nHH\ of $10^{4.1\mbox{--}4.2}$\ \pcc\ (\S\ref{subsubsection:discussion:NnThistograms}).  The modified dense-gas mass fraction \fDGn, which is \fDGx\ redefined as a function of the threshold \nHH\ defining the ``dense gas'', declines with an approximately constant shallow slope in the range of $\nHH\sim 10^{4.5\mbox{--}6.0}$~\pcc\ for the NGC~253 CMZ, whereas that of the GC steeply declines at $\nHH \gtrsim 10^{4.5}$\ \pcc, reaching 0 at $\nHH\sim10^{5.0}$~\pcc.  When aligning the resolution of the GC data to that of the NGC~253~CMZ data, \fDGn\ drops more steeply at a lower \nHH\ of $\sim10^{4.4}$ \pcc (\S\ref{subsubsection:discussion:fDGx}).

    \item The above results were confirmed to be qualitatively unchanged when we used the results of the supplementary analysis with the transitions commonly included in the data set of the ALCHEMI and \citetalias{Tanaka2018b} (the Minimal data set) for both the NGC~253 and the GC, where bias from the selection of the tracer lines is absent (\S\ref{subsubsection:discussion:NnThistograms},\ref{subsubsection:discussion:fDGx}).  

    \item The \fDGn\ ratio between the GC to the NGC~253 CMZ reaches 0.1, i.e., the same as their $\mathrm{SFE}_\mathrm{DG}$ ratio, at $\nHH = 10^{4.2\mbox{--}4.6}\ \pcc$, where the threshold \nHH\ value depends on the resolution and the selection of the data set (\S\ref{subsubsection:discussion:fDGx}).      
    Signatures of massive star formation are associated with regions above this threshold \nHH\ both in NGC~253 and the GC. 
    It is not evident whether this threshold \nHH\ immediately translates into the universal \nHH\ threshold assumed in the scaling relation between the dense-gas mass and SFR. Nevertheless, the rich abundance of dense gas with $\nHH\gtrsim10^{4.2\mbox{--}4.6}\ \pcc$ in the NGC~253 CMZ is likely to play a critical role to promote starburst (\S\ref{subsection:discussion:originOfContrastingSF}). 

    \item The origin of the broader profile of \fDGn\ in the NGC~253 than the GC is potentially linked to the higher degree of turbulence in the former. However, the elevated $\mathrm{SFE_{dense}}$ in the NGC~253 CMZ \myrev{contradicts} the prediction of the turbulent cloud model.  Alternatively, we may speculate that the GC has consumed the dense gas during a past active SF phase, while the formation of new dense clouds is hindered by frequent destructive cloud--cloud collisions, resulting in the depletion of the dense gas.
    (\S\ref{subsection:discussion:originOfContrastingSF}).

\end{itemize}

\myrev{
The authors are grateful to the anonymous referee for valuable comments that helped to improve the quality of this paper.
}
This paper makes use of the following ALMA data: ADS/JAO.ALMA\#2017.1.00161.L.
\myrev{
ALMA is a partnership of ESO (representing its member states), NSF (USA) and NINS (Japan), together with NRC (Canada), MOST and ASIAA (Taiwan), and KASI (Republic of Korea), in cooperation with the Republic of Chile. The Joint ALMA Observatory is operated by ESO, AUI/NRAO and NAOJ.
}
S.V., M.B., K.-Y.H., and J.B. acknowledge support from  the European Research Council (ERC) under  the European Union's Horizon 2020 research and innovation program MOPPEX 833460.
K.S. acknowledges the support from the Ministry of Science and Technology (MOST) of Taiwan through the grant MOST 111-2112-M-001-039.
V.M.R. has received support from the project RYC2020-029387-I funded by MCIN/AEI /10.13039/501100011033.
H.K and T.T. were supported by JSPS KAKENHI Grant Number 20H00172 and the NAOJ ALMA Scientific Research Grant Number 2020-15A.
L.C. acknowledges financial support through the Spanish grant PID2019-105552RB-C41 funded by MCIN/AEI/10.13039/501100011033.

\facility{ALMA}

\appendix
\section{Probability Functions in the HB analysis\label{appendix:A}}
The likelihood function \prob{\mathcal{V}|\mathcal{P},\mathcal{E}} is 
\begin{eqnarray}
\prob{\mathcal{V}|\mathcal{P},\mathcal{E}} &=& \prod_{i,k} \exp\left[-\frac{1}{2}\left(\frac{\epsilon_{ik}\cdot F_k\left(\vec{p}_i\right) - v_{ik}}{\delta_k}\right)^2\right], \label{eq:lh}
\end{eqnarray}
where $F_k\left(\vec{p}\right)$ is the model intensity/ratio of the $k$th line calculated using the parameter $\vec{p}$\ and $\delta_k$ is the rms noise level of the $k$th line intensity/ratio.  Equation \ref{eq:lh} implicitly assumes that $v_{ik}$ includes additive uncertainty of $\sim \mathcal{N}\left(0, \delta_k^2\right)$ and $\delta_k$ is uniform across all spaxels.

The prior function \prob{\mathcal{P},\mathcal{E}|\theta} gives {\it a priori} PDF of $\mathcal{P}$ and $\mathcal{E}$, which is
parameterized by $\theta = \{\Vec{\mu}, \Sigma, \Vec{\sigma}\}$;
\begin{eqnarray}
\prob{\mathcal{P},\mathcal{E}|\theta} &=& \prod_i \prob{\vec{p}_i|\vec{\mu},\Sigma}\,\prob{\vec{\epsilon}_i|\vec{\sigma}}. \label{eq:priorHB}
\end{eqnarray}
We adopt the following truncated multivariate student prior and log-normal prior for $\mathcal{P}$ and $\mathcal{E}$, respectively; 
\begin{eqnarray}
\prob{\vec{p}_i|\vec{\mu},\Sigma} &=& T_\nu\left(\vec{p}_i|\vec{\mu}, \Sigma\right) R\left(\Vec{p}_i\right), \label{eq:priorP}\\
\prob{\vec{\epsilon}_i|\vec{\sigma}} &=& \prod_k L\left(\epsilon_{ik}|0, {\sigma_k}^2\right)\cdot R\left(\Vec{\epsilon}_i\right) \label{eq:priorE}.
\end{eqnarray}
where $T_\nu\left(\cdot|\vec{\mu},\Sigma\right)$ denotes the multi-variate student function with shape parameter $\nu$, location vector $\vec{\mu}$, and scale matrix $\Sigma$, and $L\left(\cdot|0, {\sigma}^2\right)$ the log-normal function with location 0 and dispersion $\sigma^2$.  We fix $\nu$ at 2 in the present analysis.
The function $R\left(\cdot\right)$ is a product of logistic functions for individual elements of the parameters, adopted to prevent them from diverging to extremely large or small values.  
The details of the logistic priors are provided in Appendix~\ref{subsection:method:rangelimits}.

The hyperprior function \prob{\theta} gives the prior probability of $\theta$.
We employ mainly non-informative functions for the hyperprior and additional functions to limit the variable ranges of a few specific elements of $\theta$; 
\begin{eqnarray}
\prob{\theta} & = & F_\mathrm{ss}\left(S,R\right)\cdot R\left(\theta\right), \label{eq:hyperprior}
\end{eqnarray}		
where $F_\mathrm{ss}\left(S,R\right)$ is the separate-strategy prior \citep{Barnard2000}, which gives a uniform PDF for a symmetric non-negative definite matrix. 
Matrices $S$ and $R$ are the scaling diagonal matrix and correlation matrix obtained from the decomposition $\Sigma = SRS$, respectively, i.e, $R_{ij}=\Sigma_{ij}/\sqrt{\Sigma_{ii}\Sigma_{jj}}$ and $S_{ij} = \delta_{ij}\sqrt{\Sigma_{ij}}$, with $\delta_{ij}$ being the Kronecker delta.
The function $R\left(\theta\right)$ represents the logistic hyperpriors introduced to explicitly forbid possible artificial correlations among the parameters.

Function $R\left(\cdot\right)$ in Equations \ref{eq:priorP}, \ref{eq:priorE}, and  \ref{eq:hyperprior} is a product of logistic functions.
Let $x$ be an element of $\mathcal{P}$, $\mathcal{E}$, or $\theta$, and 
the upper and lower limits can be set by logistic PDFs,
\begin{eqnarray}
R_{x,\mathrm{upper}}\left(x\right) & = & 
\left[1+\exp\left(\frac{x - x_\mathrm{max}}{k}\right)\right]^{-1} \label{eq:logMax}
\end{eqnarray}
and 
\begin{eqnarray}
R_{x,\mathrm{lower}}\left(x\right) & = & 
\left[1+\exp\left(-\frac{x - x_\mathrm{min}}{k}\right)\right]^{-1}, \label{eq:logMin}
\end{eqnarray}
respectively, where the scale parameter $k>0$ is taken to be sufficiently small.
We use these logistic (hyper-)prior functions to limit the variable ranges of the (hyper-)parameters to physically reasonable values or to prevent the prior functions (Equations \ref{eq:priorP} and \ref{eq:priorE}) from being too steep.
Details of the range limits of the (hyper-)parameters are given in Appendix~\ref{subsection:method:rangelimits}.

\section{Probability Functions in the Non-hierarchical analysis\label{appendix:B}}
The NHB analysis uses the same log-normal and logistic priors as the HB analysis for $\mathcal{E}$ and $\mathcal{P}$, respectively;
\begin{eqnarray}
\prob{\mathcal{P},\mathcal{E}} &=& 
\prod_{i,k} L\left(\epsilon_{ik} | 0, \sigma_\mathrm{c}^2\right)\cdot 
\prod_i R\left(\vec{p}_i\right). \label{eq:priorNonH}
\end{eqnarray}
The difference from the hierarchical prior (Equation \ref{eq:priorHB}) is that $\prob{\mathcal{P},\mathcal{E}}$ is not parameterized by variable hyperparameters.
We fix $\sigma_c = 0.1$ for all $i,j$, i.e, assume relative uncertainties of $\sim1\pm\sigma_c$ for all line intensities.
The variable ranges of $\mathcal{P}$ are chosen to be sufficiently wide to cover reasonable parameter values observed in molecular clouds, within which we obtain the marginal posterior function for $\mathcal{P}$ as
\begin{eqnarray}
\prob{\mathcal{P}|\mathcal{V}} & = & \int{\mathrm{d}\mathcal{E}}\cdot\prob{\mathcal{P},\mathcal{E}|\mathcal{V}}\nonumber \\ 
& \sim & \int{\mathrm{d}\mathcal{E}}\cdot\prob{\mathcal{V}|\mathcal{P},\mathcal{E}}\prod_{i,k} L\left(\epsilon_{ik} | 0, \sigma_\mathrm{c}^2\right), \label{eq:marginalposteriorNonH}
\end{eqnarray}
which is identical to the likelihood function where input line intensities have relative uncertainties.
Hence, the NHB analysis is equivalent to the standard maximum likelihood analysis with finite parameter ranges.

\section{Range Limits\label{subsection:method:rangelimits}}
We impose limitations to the variable ranges of several (hyper-)parameters using the logistic (hyper-)priors (Equations \ref{eq:logMax} and \ref{eq:logMin}).
Table \ref{table:rangeLimits} shows the upper/lower-limit values applied in the present analysis.
The limits to the $\vec{p}$ vector are introduced to prevent the parameters from diverging to physically unrealistic values in usual molecular clouds.   These limits are of a significant importance in the NHB analysis, but have almost no effect on the hierarchical analysis, in which the variable ranges of $\vec{p}$ are controlled by the prior and hyper-prior functions.

The limits of the off-diagonal elements of the $R$ matrix are interpreted as the prior correlation coefficients among the parameters.
We introduced the limits to $R$ for two purposes.
One is to forbid artificial anti-correlations between $\phi$ and $\NHH$, $\nHH$ and $\NHH$, and $\nHH$ and $\Tkin$.  These parameters are strongly degenerate in the excitation equations, and the NHB analysis tends to create anti-correlations among them.
We limit the $R$ elements for these parameter pairs to be non-negative.   
The assumption of non-negative $\phi$--$\NHH$ and $\nHH$--$\NHH$ correlations is physically reasonable, as it is likely that high-$\NHH$ clouds consist of high-\nHH\ and are in a crowded regions rather than isolated from extended molecular gas. 
The non-negative correlation between \nHH--\Tkin\ may not be obvious, but we adopted this  assumption based on the results toward the GC.
The comparison between the results by \cite{Nagai2007} and \citetalias{Tanaka2018b} have shown that both \Tkin\ and \nHH\ in the high-excitation component are higher than in the low-excitation component.  
\citetalias{Tanaka2018b} has also shown the absence of strong anti-correlation between \Tkin\ and \nHH\ on 2-pc scales. 
However, this does not necessarily indicate that the same applies to the low-excitation component of the CMZ of NGC~253. 
It should be noted that this range-limiting hyperprior might overcompensate for the correlations that may be present in real molecular clouds.

The other purpose to limit the variable ranges of the hyperparameters is to prevent the parameters from having too strong (anti-)correlation,
since strong (anti-)correlation makes the log-student prior of $\vec{p}$ extremely narrow, resulting in highly inefficient MCMC even though the hybrid MCMC technique is employed.
We limit $|R_{ij}| < 0.9$ for all parameter pairs to prevent the prior function from being too narrow.

\begin{deluxetable}{llcc}
\tablecaption{Range Limits of Parameters and Hyperparameters \label{table:rangeLimits}}
\tablecolumns{4}
\tablehead{
\colhead{} & \colhead{} & \colhead{minimum} & \colhead{maximum}
}
\startdata
(parameter\tablenotemark{$\dag$})
& $\dNHHdv$ & 19 & 23 \\
& $\Tkin$ & 1 & 3 \\
& $\nHH$ & 1 & 7 \\
& $\xmol{X}$ & $-10$ & $-5$ \\
& $R_{13}$ & 1 & 2\\
& $\phi$ & $-3$ & 2\\
(hyperparameter\tablenotemark{$*$})
& $R_{i,j}$   & $-0.9$ & 0.9 \\
& $R_{N,n}$   &  0 & \nodata \\
& $R_{N,\phi}$   &  0& \nodata \\  
& $R_{T,n}$   &  0 & \nodata
\enddata
\tablenotetext{\dag}{All parameters are in base-10 logarithms. }
\tablenotetext{*}{Subscript indices $N$, $n$, and $T$ denotes elements for \NHH, \nHH, and \Tkin, respectively.}
\end{deluxetable}

\section{\myrev{Integrated Intensity Images} \label{appendix:integratedIntensityMaps}}
\myrev{
Figure \ref{fig:integratedIntensityMaps} shows the integrated intensity images of the transitions used as the inputs for the analysis. 
All the images are regredded to a \ang{;;0.45} grid spacing and integrated over the \vlsr\ range of [0 500] \kmps\ after filtering of low S/N spaxels described in \S\ref{subsection:data:lowdensity},\ref{subsection:data:highdensity}.   
We refer readers to \S\ref{section:data} and \cite{Martin2021A&A} for more detailed data description.
}

\begin{figure*}[p]
\plotone{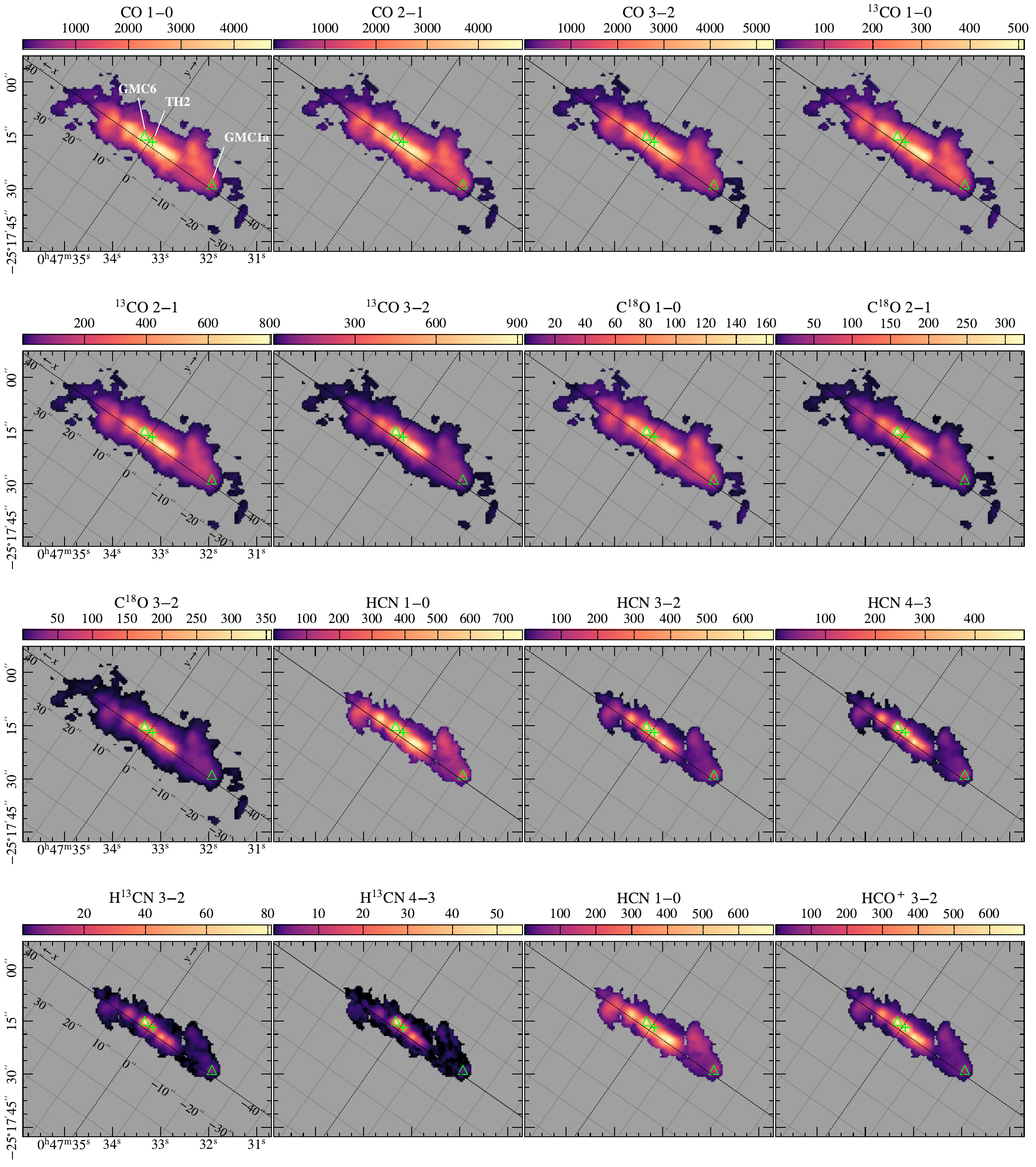}
\caption{\myrev{Integrated intensity images of the input transitions in the scale of $\int T_\mathrm{MB} \mathrm{d}v$ ($\mathrm{K}\,\kmps$).  The major and minor axis coordinates ($x$--$y$ coordinates) and the positions of TH2, GMC6, and GMC1a are shown for convenience in comparison with Figures \ref{fig:mapLowH}--\ref{fig:mapMinH} and \ref{fig:xmol}}. \label{fig:integratedIntensityMaps}}
\end{figure*}
\begin{figure*}[p]
\plotone{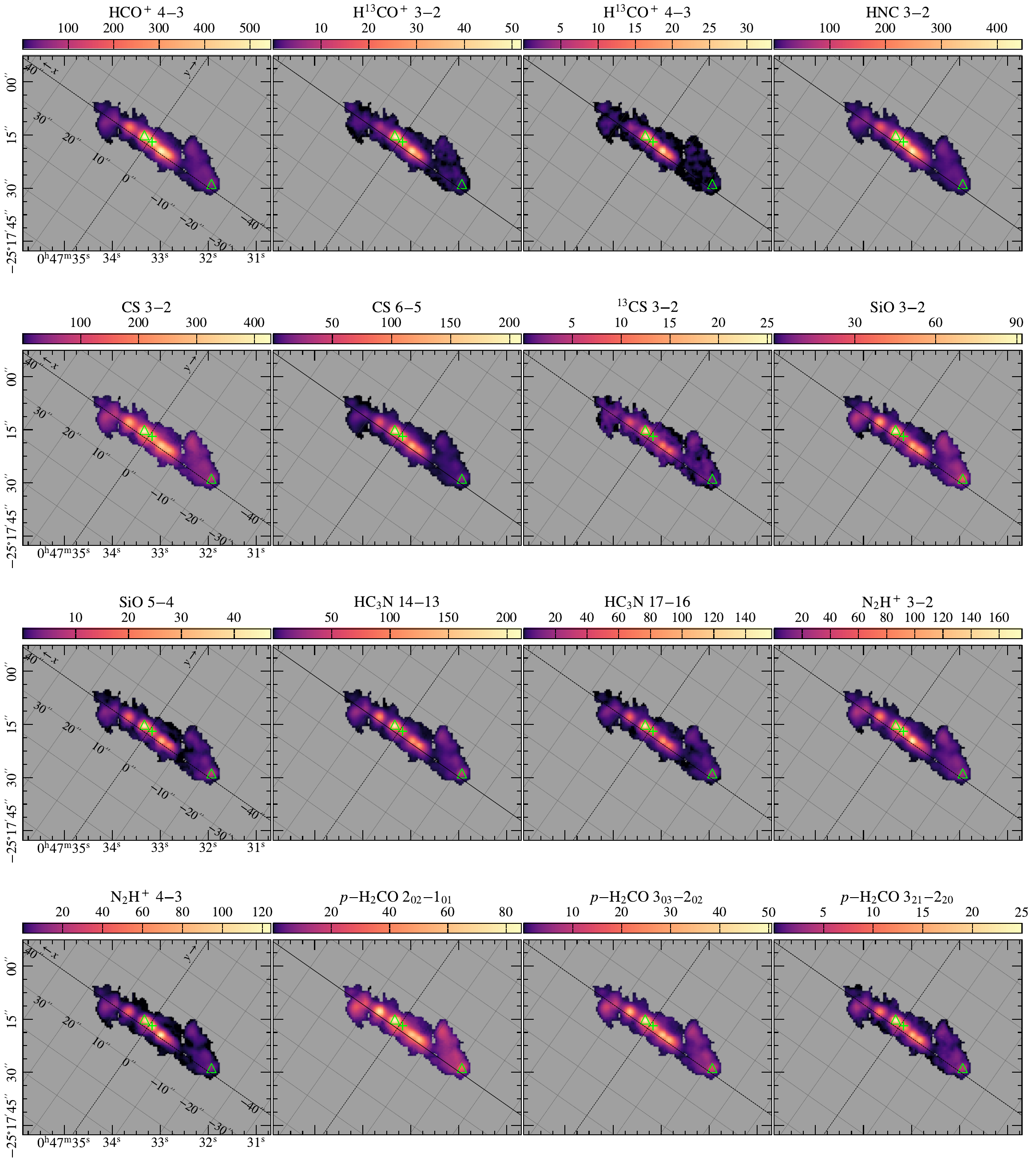}
\addtocounter{figure}{-1}
\caption{Continued.}
\end{figure*}

\section{MCMC results \label{appendix:C}}

\begin{figure*}[p]
\begin{tabular}{lr}
\epsscale{0.52}
\raisebox{27.5mm}{\plotone{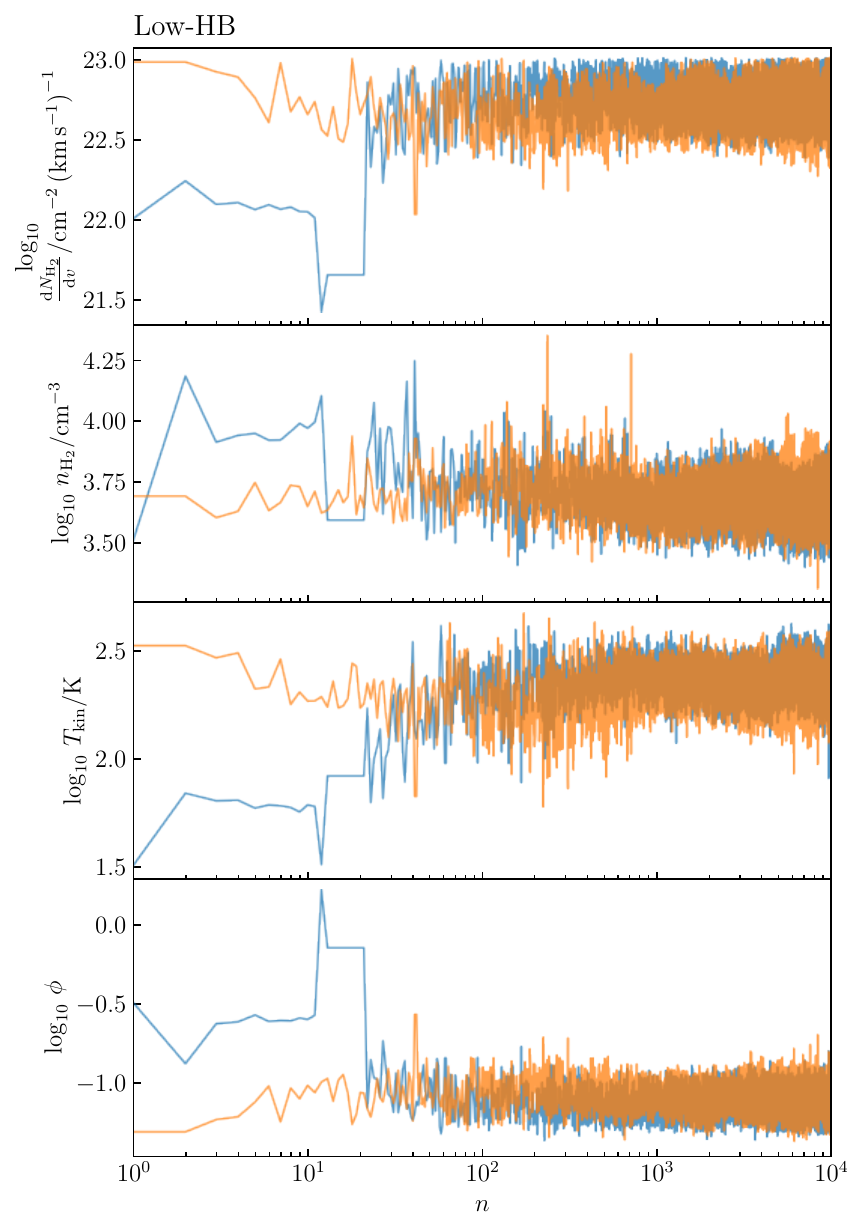}} &
\epsscale{0.52}
\plotone{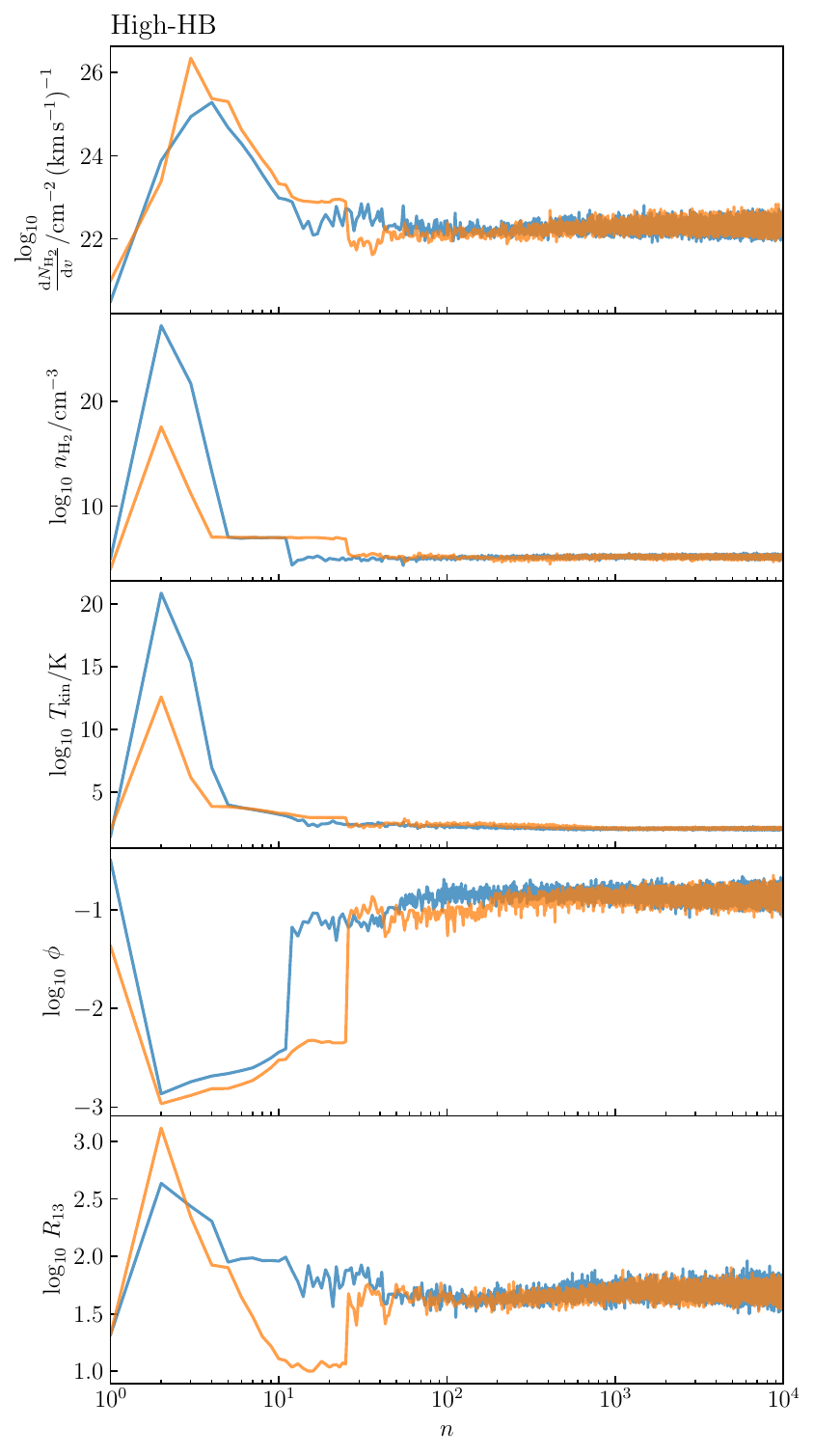}
\end{tabular}
\caption{MCMC trace plots of the Low-HB (left) and High-HB (right) runs at GMC6, showing parameter values (\dNHHdv, \nHH, \Tkin, $\phi$, and $R_{13}$) against MCMC step number, $n$, for first $10^4$ steps.
Results with two different initial parameters are overlaid on each panel. The Low-HB plot lacks $R_{13}$, whose value is constant in the analysis. \label{fig:traceplot}}
\end{figure*}

Figure \ref{fig:traceplot} shows MCMC trace plots of \dNHHdv, \nHH, \Tkin, $\phi$, and $R_{13}$ at GMC6 for the first $10^4$ MCMC steps in the Low-HB and High-HB runs. 
Each panel shows two different MCMC runs starting from different initial parameters.
In these runs, approximately $10^3$ burn-in steps were necessary before the convergence; after that, the MCMC reaches convergence with sufficient parameter coverage and density.  The convergent values are consistent irrespective of the initial parameters.  We ran more than a few $10^4$ steps after convergence in each run.

\begin{figure*}[p]
\begin{center}
\begin{tabular}{l}
{\epsscale{0.55}
\plotone{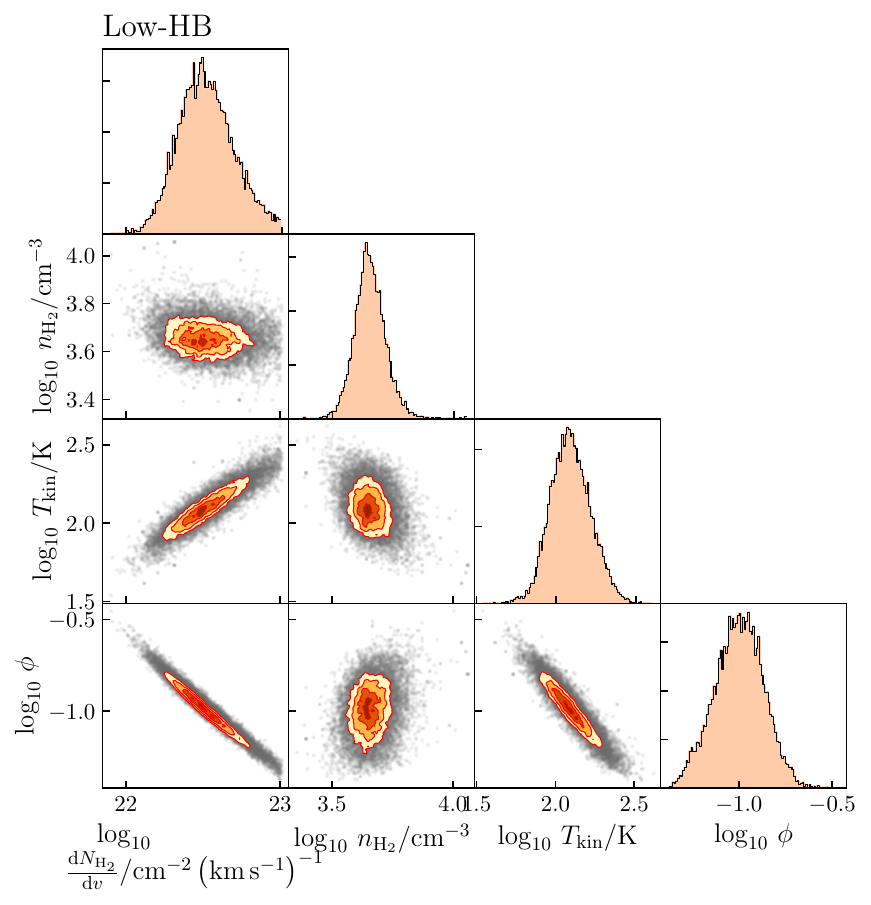}}
\\
{
\epsscale{0.65}
\plotone{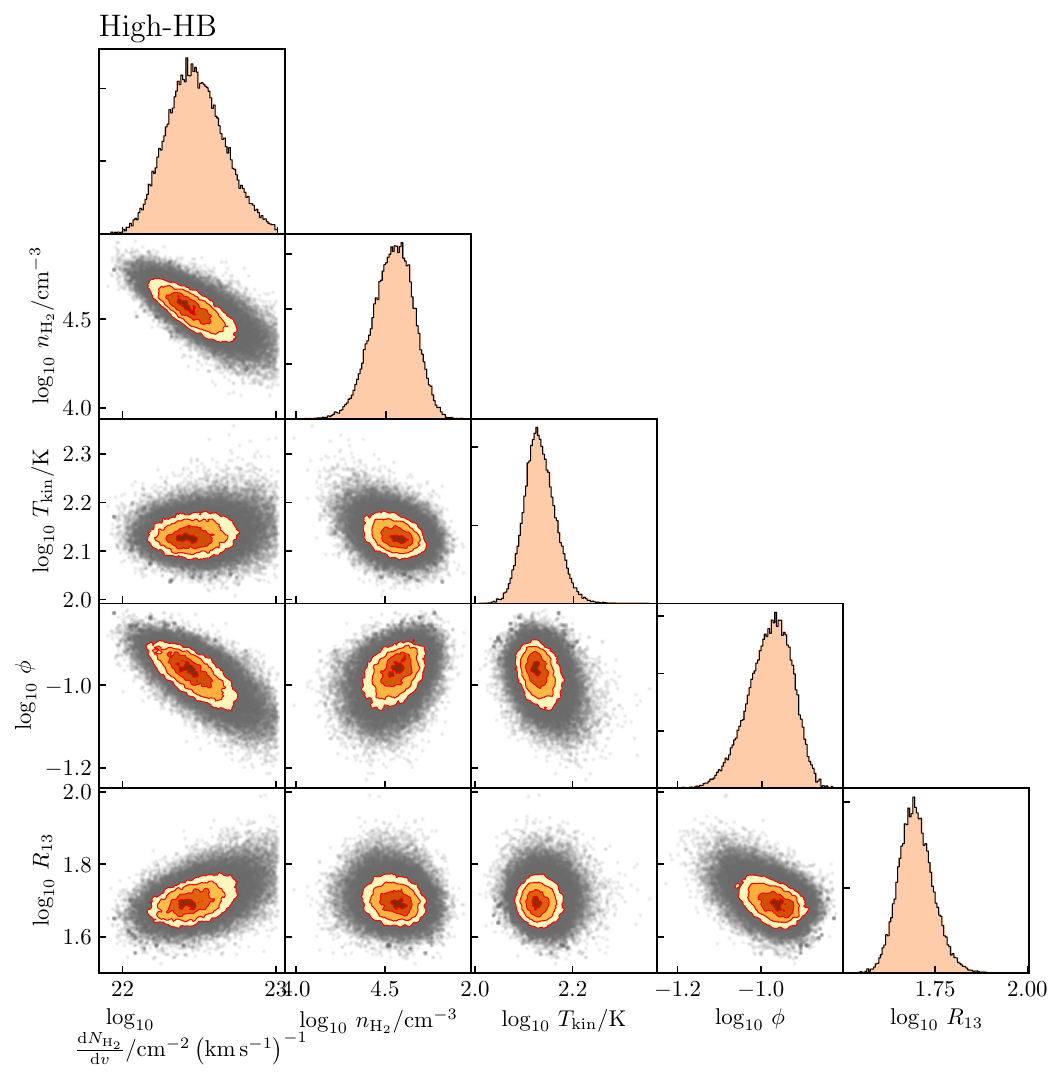}
}
\end{tabular}
\end{center}
\caption{MCMC corner plots of the Low-HB (upper panel) and High-HB (lower panel) runs at GMC6. Each panel shows 1-D (diagonal panels)  or 2-D histograms (off-diagonal panels) of the marginal posterior PDF.  The colored contours overlaid on the scatter plots are drawn at 30\%, 50\%, 80\%, and 95\% credible intervals. \label{fig:cornerplot}}
\end{figure*}

Figure \ref{fig:cornerplot} shows an example of the marginal posteriors of \dNHHdv, \nHH, \Tkin, $\phi$, and $R_{13}$ at GMC6 for the Low-HB and High-HB runs.

\section{\myrev{Optical Depths \label{appendix:tau}}}

\begin{figure*}[p]
\epsscale{0.9}
\plotone{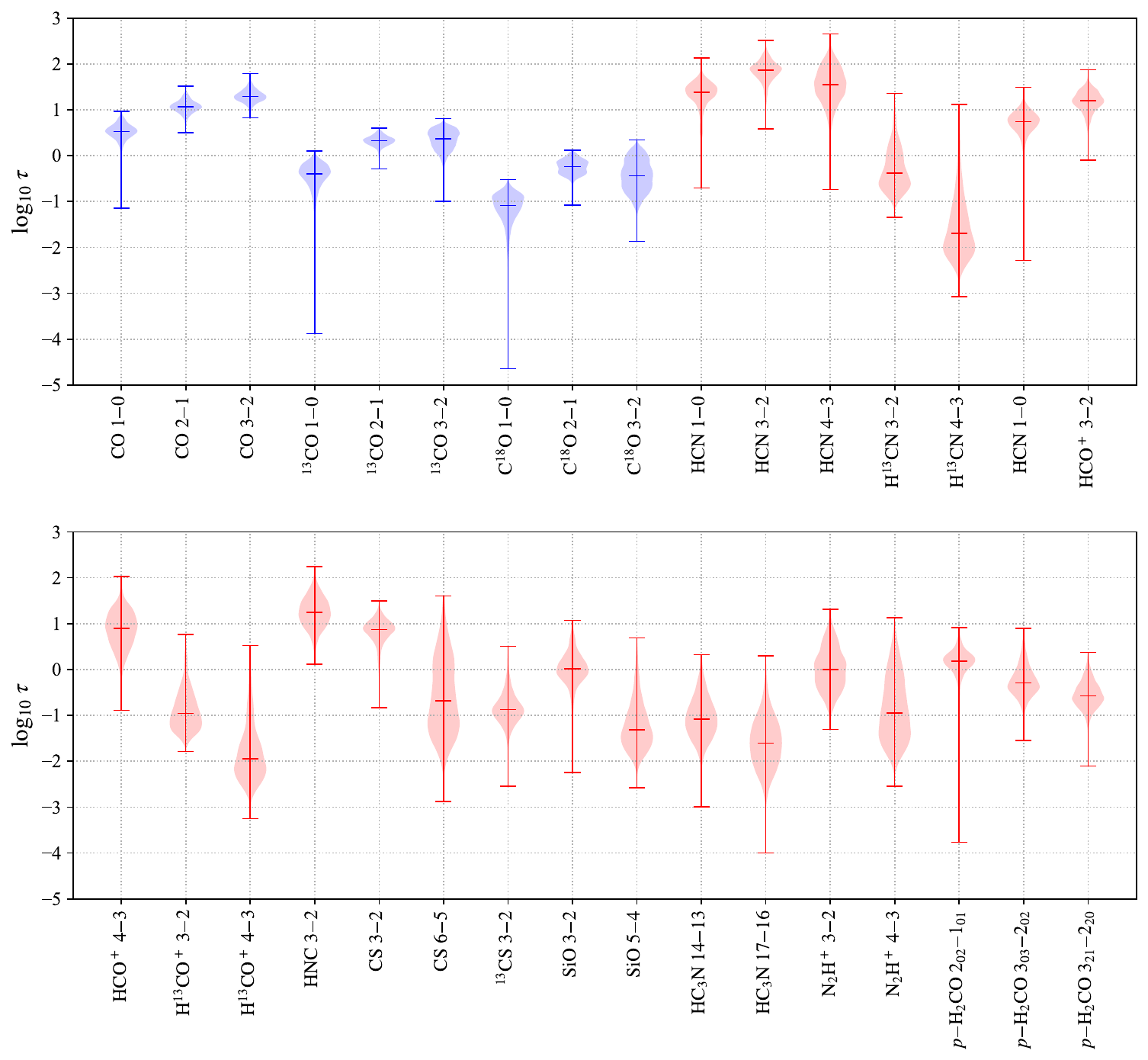}
\caption{\myrev{Optical depths evaluated from the parameters obtained from the Low-HB and High-HB analyses.  The violin plots show  spaxel-based frequency distributions, with bars indicating the frequency minima, medians, and maxima.  Transitions in the Low-density and High-density data sets are colored blue and red, respectively.} \label{fig:taus}}
\end{figure*}

\myrev{
We performed pixel-wise calculation of the optical depths of the transitions based on the LVG method, using the paramters from the Low-HB and High-HB analyses.
The calculation was performed for spaxels at which all of the \NHH, \Tkin, \nHH, $R_{13}$, and $x_\mathrm{mol}$ values were determined with uncertainties less than 0.3 dex. 
Figure \ref{fig:taus} shows the spaxel-based frequency distributions of the calculated optical depths.
}

\section{\myrev{Results of the non-hierarchical analyses and the hierarchical analysis for the GC \label{appendix:D}}}

\begin{figure*}[p]
\epsscale{1.0}
\plotone{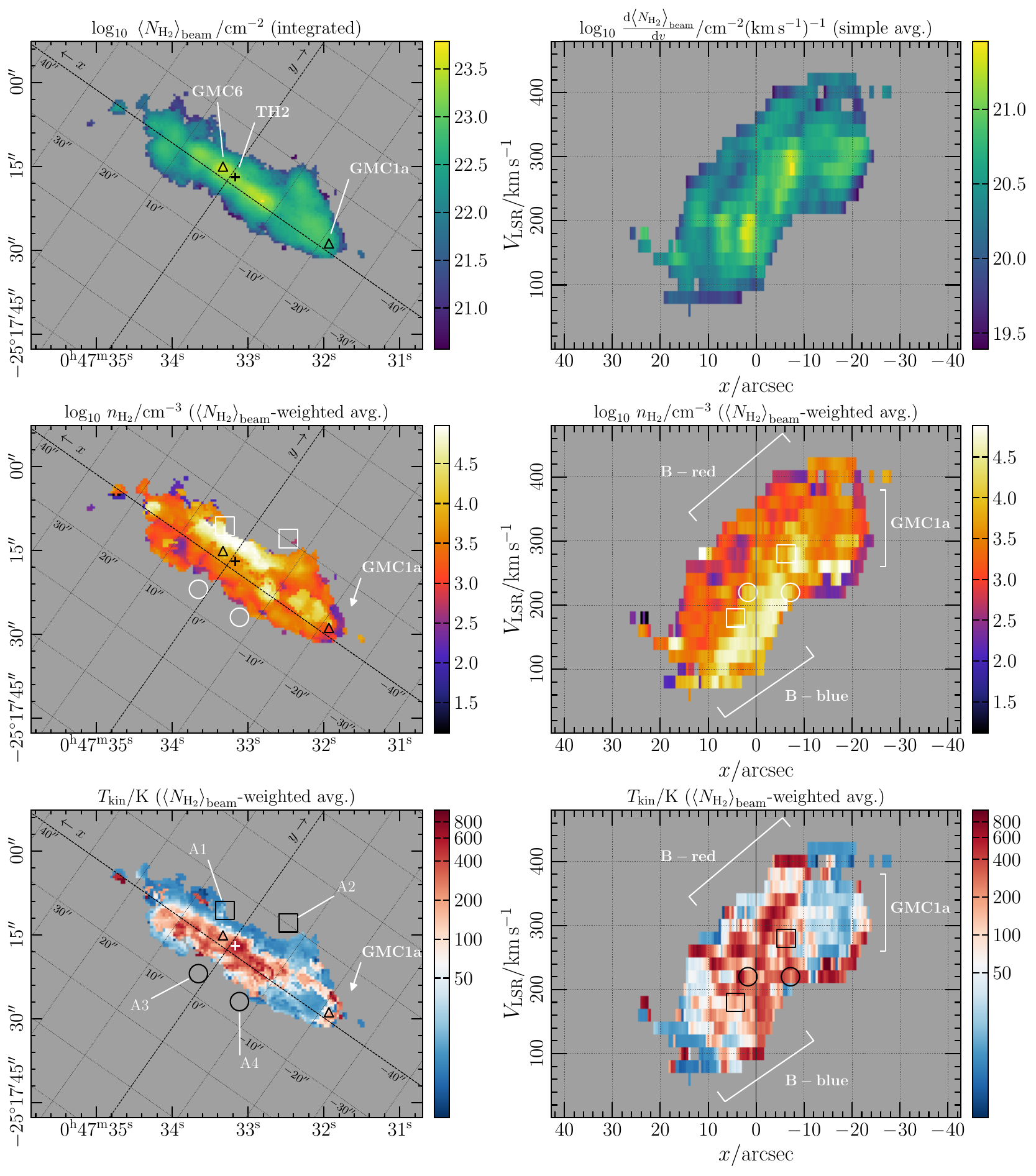}
\caption{Same as Figure \ref{fig:mapLowH}, but for the Low-NHB result. \label{fig:mapLowNHB}}
\end{figure*}
\begin{figure*}[p]
\epsscale{1.1}
\plotone{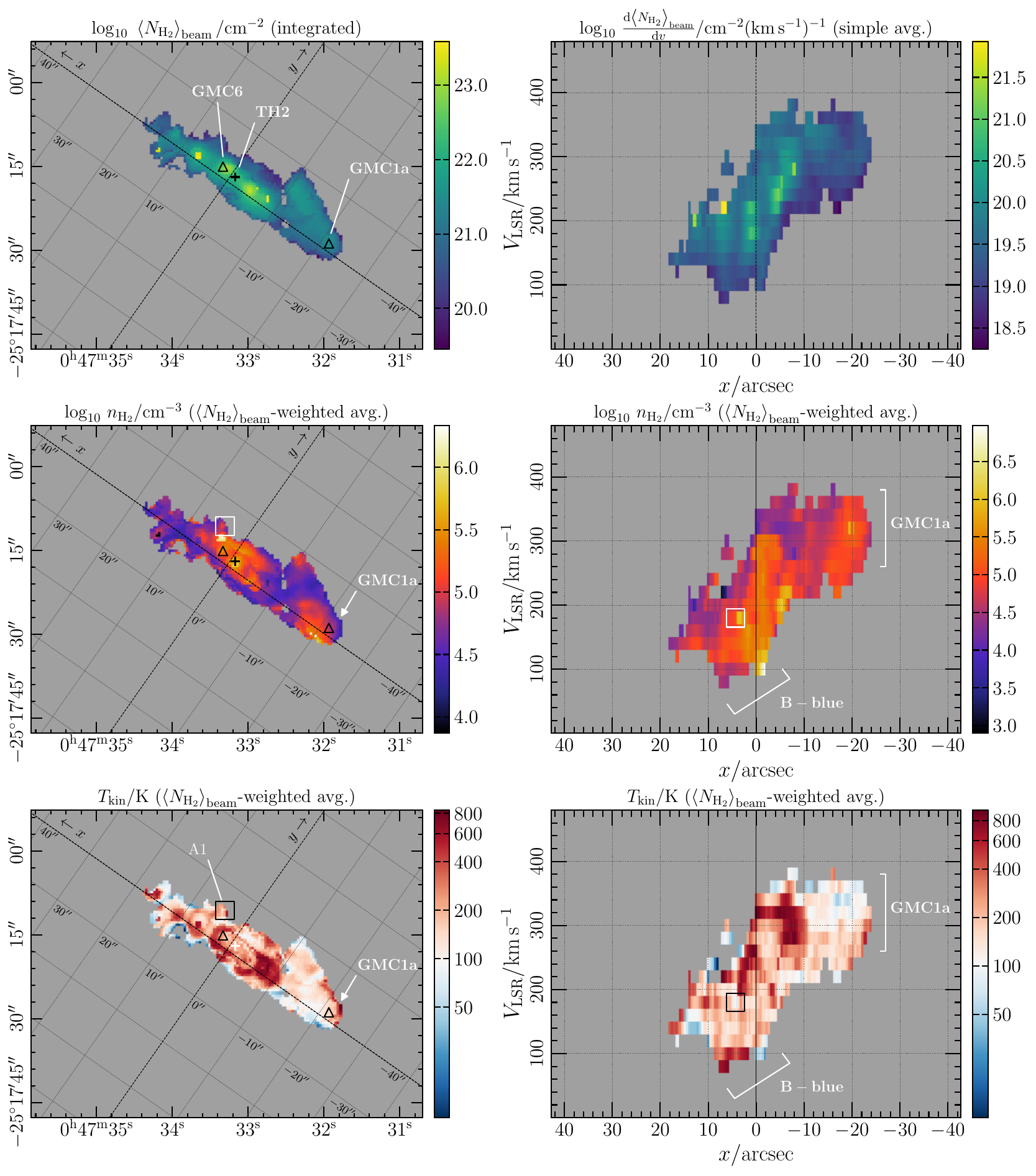}
\caption{Same as Figure \ref{fig:mapLowH}, but for the High-NHB result. \label{fig:mapHighNHB}}
\end{figure*}

The \PP\ and \PPV\ distribution of \NHHbeam, \nHH, and \Tkin\ for the Low-NHB, High-NHB, and the GC-HB analyses are shown in Figure \ref{fig:mapLowNHB}, \ref{fig:mapHighNHB}, and \ref{fig:mapGC} respectively.

Figure \ref{fig:T18GCComparison} shows spaxel-by-spaxel scatter plots of the \NHHbeam, \nHH, and \Tkin\ of the GC-HB results vs. the T18 results. 

\begin{figure*}[p]
\epsscale{0.85}
\plotone{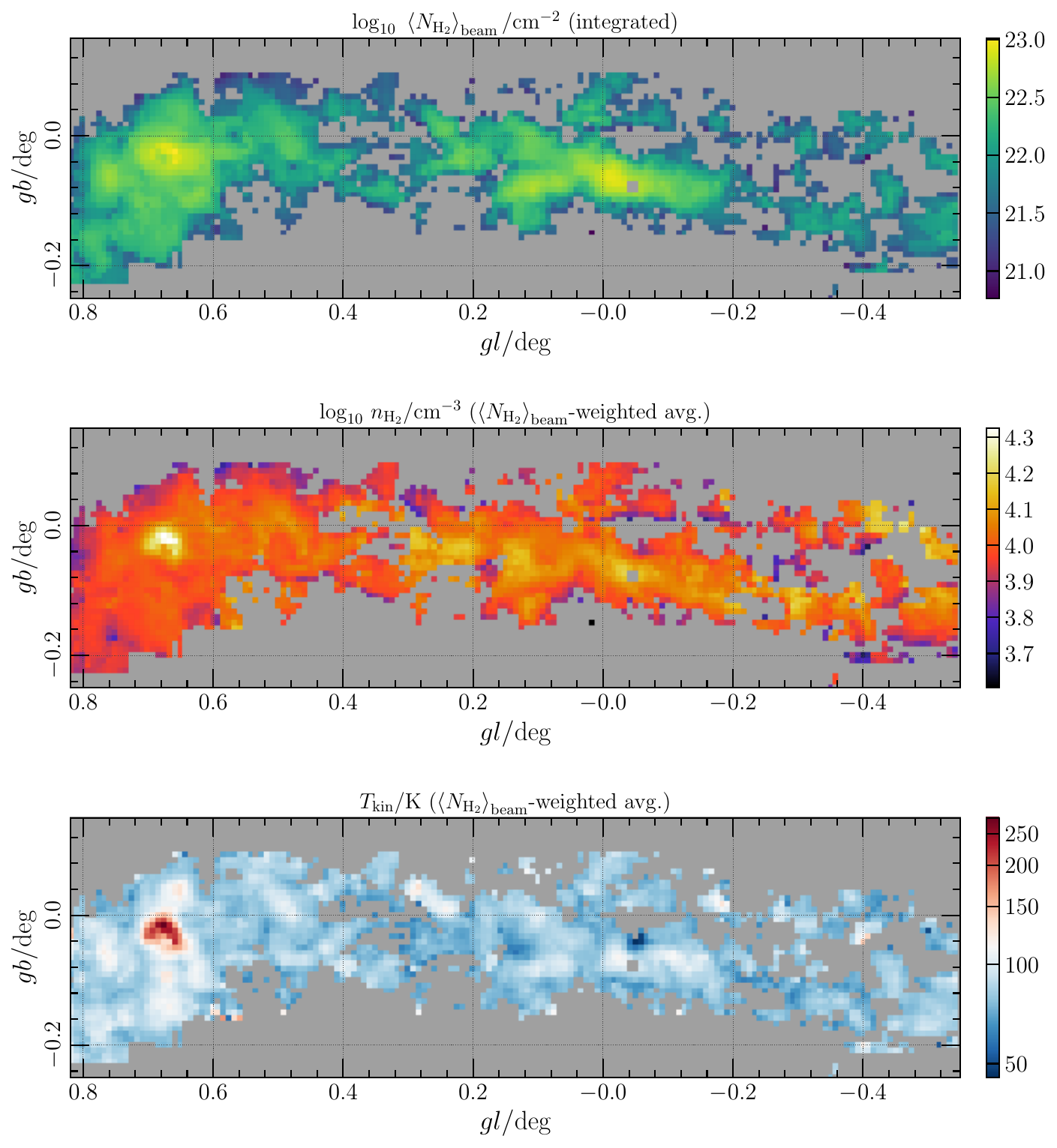}
\caption{Same as Figure \ref{fig:mapLowH}, but for the GC-HB result. \label{fig:mapGC}}
\end{figure*}

\begin{figure*}[p]
\epsscale{0.85}
\plotone{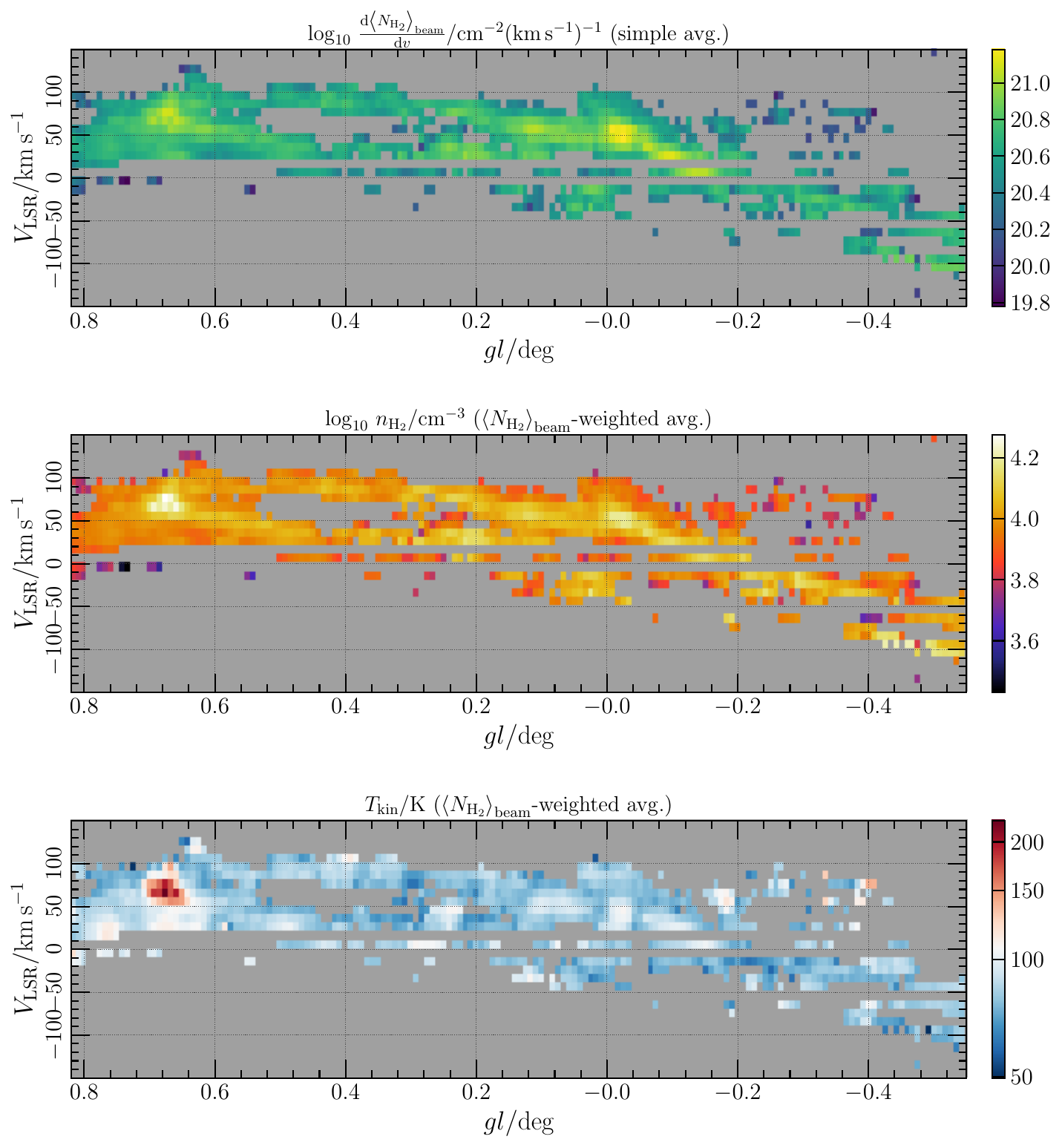}
\addtocounter{figure}{-1}
\caption{Continued.}
\end{figure*}

\begin{figure*}[p]
\plotone{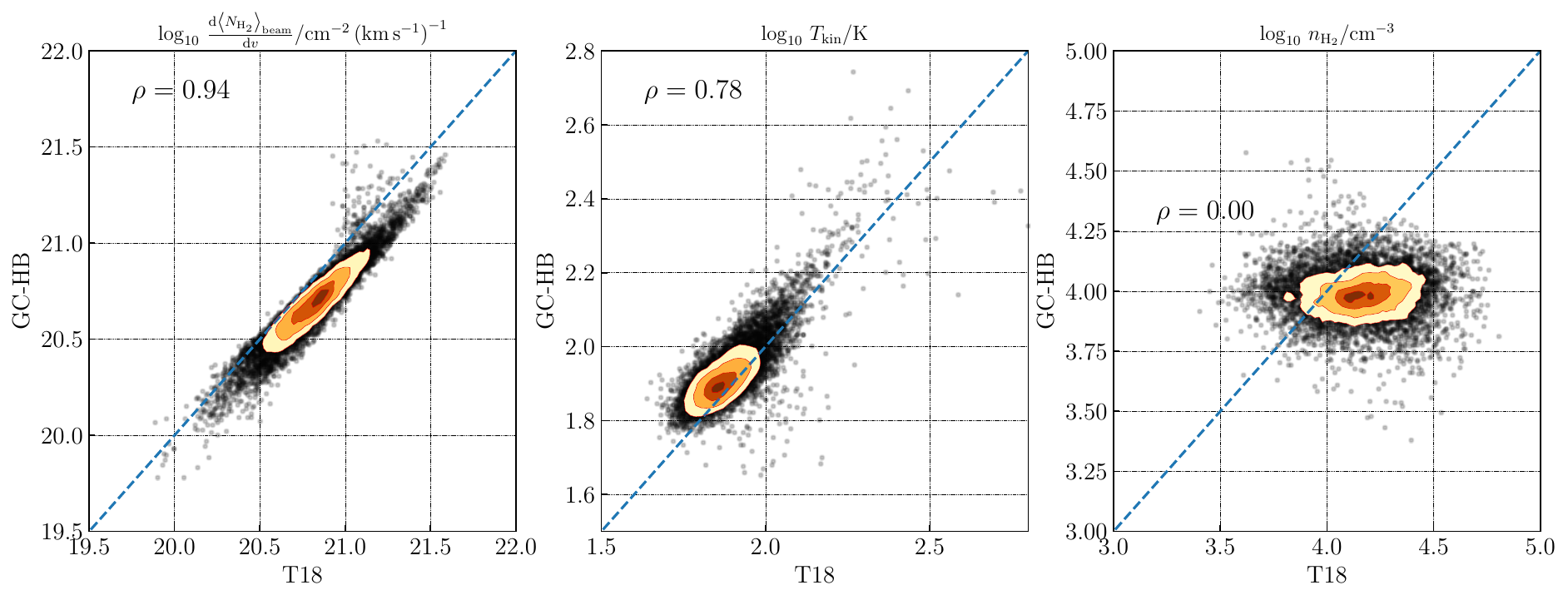}
\caption{Same as Figure \ref{fig:HighLowComparison} but for the T18 results vs. GC-HB results  .\label{fig:T18GCComparison}}
\end{figure*}

\newpage
\bibliographystyle{apj}
\bibliography{mendeley,local,lamda}

\end{document}